\documentclass[a4paper,fleqn,usenatbib]{mnras}

\usepackage{newtxtext,newtxmath}

\usepackage[T1]{fontenc}
\usepackage{ae,aecompl}

\usepackage{graphicx}	
\usepackage{epstopdf}	

\usepackage{soul}
\usepackage{pifont}

\newcommand{\cmark}{\ding{51}}%
\newcommand{\xmark}{\ding{55}}%
\newcommand{\LCDM}{$\Lambda$CDM }%

\usepackage[normalem]{ulem}

\newcommand{\DwarfBarLegend}{\vspace{0.0cm}\includegraphics[width=2\columnwidth]{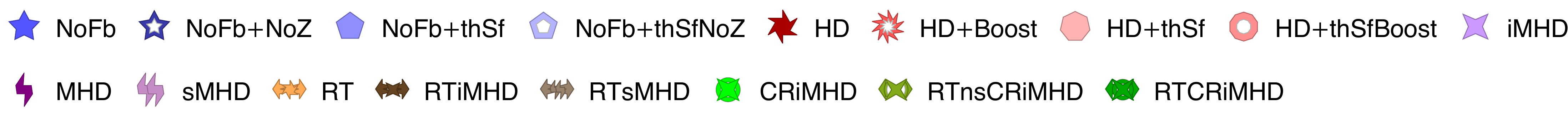}\vspace{-0.2cm}}
\newcommand{\DwarfBarLegendRT}{\vspace{0.0cm}\includegraphics[width=\columnwidth]{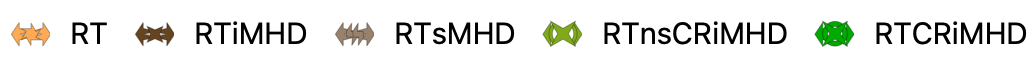}\vspace{-0.2cm}}
\newcommand{\HalfDwarfBarLegend}{\vspace{0.0cm}\includegraphics[width=\columnwidth]{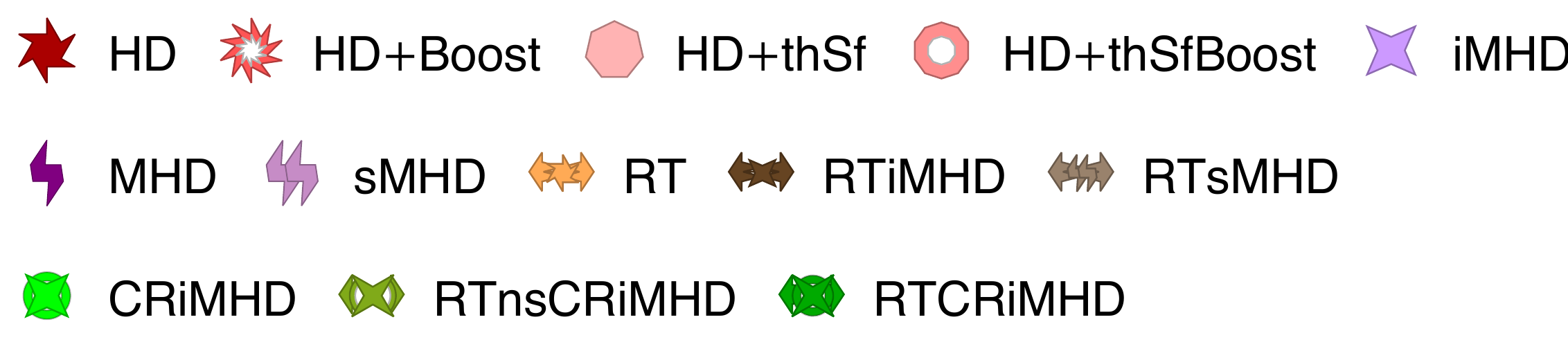}\vspace{-0.2cm}}

\newcommand{\DwarfBarLegendFrom}[1]{\vspace{0.0cm}\includegraphics[width=2\columnwidth]{#1}\vspace{-0.2cm}}
\newcommand{\HalfDwarfBarFrom}[1]{\vspace{0.0cm}\includegraphics[width=\columnwidth]{#1}\vspace{-0.2cm}}

\newcommand{\Mst}{M_*}%
\newcommand{\Mdynst}{M_{\text{dyn},*}}%
\newcommand{\Rhalfst}{R_\text{1/2,*}}%
\newcommand{\Mgas}{M_\text{gas}}%
\newcommand{\Mdyngas}{M_{\text{dyn},\text{gas}}}%
\newcommand{\Rhalfgas}{R_\text{1/2,gas}}%
\newcommand{\MHI}{M_\text{HI}}%
\newcommand{\MdynHI}{M_{\text{dyn},\text{HI}}}%
\newcommand{\RhalfHI}{R_\text{1/2,HI}}%
\newcommand{\sv}[1]{\sigma_{\text{v},\text{#1}}}%
\newcommand{\kappaCR}{\kappa_\parallel}%

\newcommand{\tablesymbol}[1]{\hspace{-0.3cm}\includegraphics[height=0.3cm]{Images/Symbols/#1.pdf}\hspace{-0.5cm}}
\newcommand{\HDthSf}{NoFb+thSfNoZ}
\newcommand{\HDthSfNoFb}{NoFb+thSf}
\newcommand{\HDthSfFb}{HD+thSf}
\newcommand{\HDthSfFbBoost}{HD+thSfBoost}
\newcommand{\HDSf}{NoFb+NoZ}
\newcommand{\HDSfNoFb}{NoFb}
\newcommand{\HDSfFb}{HD}
\newcommand{\HDSfFbBoost}{HD+Boost}
\newcommand{\MHDSfFb}{MHD}
\newcommand{\sMHDSfFb}{sMHD}
\newcommand{\iMHDSfFb}{iMHD}
\newcommand{\RTSfFb}{RT}
\newcommand{\RTsMHDSfFb}{RTsMHD}
\newcommand{\RTiMHDSfFb}{RTiMHD}
\newcommand{\CRiMHDSfFb}{CRiMHD}
\newcommand{\RTnsCRiMHDSfFb}{RTnsCRiMHD}
\newcommand{\RTCRiMHDSfFb}{RTCRiMHD}

\newcommand{\mH}{\mathrm{m_H}}
\newcommand{\Zsun}{\,\mathrm{Z_\odot}}
\newcommand{\Msun}{\,\mathrm{M_\odot}}
\newcommand{\erg}{\,\mathrm{erg}}
\newcommand{\cm}{\,\mathrm{cm}}
\newcommand{\km}{\,\mathrm{km}}
\newcommand{\kpc}{\,\mathrm{kpc}}
\newcommand{\pc}{\,\mathrm{pc}}

\newcommand{\s}{\,\mathrm{s}}

\newcommand{\gram}{\,\mathrm{g}}
\newcommand{\muG}{\,\mu\mathrm{G}}
\newcommand{\kmps}{\,\km\,\s^{-1}}
\newcommand{\gcc}{\,\gram\,\cm^{-3}}

\title[MHD, radiation and cosmic rays in dwarf galaxies]{The Pandora project. I: the impact of radiation, magnetic fields and cosmic rays on the baryonic and dark matter properties of dwarf galaxies}

\author[Martin-Alvarez et al.]{Sergio Martin-Alvarez$^{1,2}$\thanks{E-mail: martin-alvarez@stanford.edu (SMA)}, Debora Sijacki$^{1}$, Martin G. Haehnelt$^{1}$, Marion Farcy$^{3}$,\newauthor
Yohan Dubois$^{4}$, Vasily Belokurov$^{1}$, Joakim Rosdahl$^{5}$ and Enrique Lopez-Rodriguez$^{2}$\\
$^{1}$Institute of Astronomy and Kavli Institute for Cosmology, University of Cambridge, Madingley Road, Cambridge CB3 0HA, UK\\
$^{2}$Kavli Institute for Particle Astrophysics and Cosmology (KIPAC), Stanford University, Stanford, CA 94305, USA\\
$^{3}$Institute for Physics, Laboratory for Galaxy Evolution, EPFL, Observatoire de Sauverny, Chemin Pegasi 51, 1290 Versoix, Switzerland\\
$^{4}$Institut d’Astrophysique de Paris, 75014 Paris, France\\
$^{5}$Univ Lyon, Univ Lyon1, Ens de Lyon, CNRS, Centre de Recherche Astrophysique de Lyon UMR5574, F-69230 Saint-Genis-Laval, France\\
}

\date{MNRAS, submitted}

\pubyear{2023}

\begin{document}

\newcommand{\papertitle}[2]{\textbf{#1 \citep{#2}}.\newline}

\label{firstpage}
\pagerange{\pageref{firstpage}--\pageref{lastpage}}
\maketitle

\begin{abstract}
Enshrouded in several well-known controversies, dwarf galaxies have been extensively studied to learn about the underlying cosmology, notwithstanding that physical processes regulating their properties are poorly understood. To shed light on these processes, we introduce the Pandora suite of 17 high-resolution ($3.5$ parsec half-cell side) dwarf galaxy formation cosmological simulations. Commencing with magneto-thermo-turbulent star formation and mechanical supernova feedback, we gradually increase the complexity of physics incorporated, ultimately leading to our {\it full-physics} models combining magnetism, on-the-fly radiative transfer and the corresponding stellar photoheating, and SN-accelerated cosmic rays. We investigate multiple combinations of these processes, comparing them with observations to constrain what are the main mechanisms determining dwarf galaxy properties. We find hydrodynamical `SN feedback-only' simulations struggle to produce realistic dwarf galaxies, leading either to overquenched or too centrally concentrated, dispersion dominated systems when compared to observed field dwarfs. Accounting for radiation with cosmic rays results in extended and rotationally-supported systems. Spatially `distributed' feedback leads to realistic stellar and HI masses, galaxy sizes and integrated kinematics. Furthermore, resolved kinematic maps of our {\it full-physics} models predict kinematically distinct clumps and kinematic misalignments of stars, HI and HII after star formation events. Episodic star formation combined with its associated feedback induces more core-like dark matter central profiles, which our `SN feedback-only' models struggle to achieve. Our results demonstrate the complexity of physical processes required to capture realistic dwarf galaxy properties, making tangible predictions for integral field unit surveys, radio synchrotron emission, and for galaxy and multi-phase interstellar medium properties that JWST will probe.
\end{abstract}

\definecolor{brown}{rgb}{0.5, 0.3, 0.0}
\definecolor{orange}{rgb}{0.8, 0.5, 0.0}

\section{Introduction}
\def\dobullet{1}
Dwarf galaxies are intriguing dark matter dominated systems, subject to some of the most persistent controversies in the theory of galaxy formation. Most notable are the well-known missing satellites, cusp-core and too-big-to-fail problems \citep{Garrison-Kimmel2014, Onorbe2015, Sawala2016, Bullock2017}, which have prompted many studies challenging the standard cosmological \LCDM model \citep{Spergel2000, Libeskind2013}. Some of these problems likely arise due to inaccurate treatment of the complex baryonic physics. `Missing' satellites have been largely accounted for by considering a combination of their disruption due to photoionization \citep{Efstathiou1992, Benson2002, Bose2018}, photoevaporation \citep{Bullock2000} linked to a filtering mass \citep{Gnedin2000b, Okamoto2008}, photoheating starvation \citep{Hoeft2006, Katz2020}, and the impact of stellar feedback \citep{Dekel2003, Garrison-Kimmel2019}. Furthermore, some problems may be alleviated by observational advances, which allow the detection of fainter previously `missing' systems \citep[e.g.][]{Belokurov2008, Koposov2015} and by accounting for observational biases \citep{Oman2016}. While dwarf galaxies are the most abundant type of galaxy, their low masses and luminosities reduce their detectability and study to relatively modest redshifts \citep{Tolstoy2009}. Detailed observations are mostly limited to those found within the Local Group or not far beyond \citep[e.g.][]{Read2017, Wheeler2017, Kirby2017}. However, excitingly well-resolved observations of dwarf galaxies at high redshift will very soon become available with JWST \citep{Jeon2019}.

Similarly, as for the missing satellites problem, baryonic physics may help to resolve the cusp-core controversy as well, with this line of research gaining traction over the years. The gravitational response of dwarf galaxy haloes to explosive supernova (SN) events is capable of producing cores in the central region of dwarf density profiles \citep{Navarro1996}. This process is particularly efficient when driven by resonantly cyclic SN bursts \citep{Pontzen2012, Governato2012}, where systems featuring younger stellar populations could produce more cored profiles than those in which star formation ceased rapidly early on \citep{Read2019a}. In fact, because of their shallow potential wells and low masses, dwarf galaxies are particularly sensitive probes of baryonic physics \citep{Bullock2017}. Furthermore, due to their rapid quenching at high redshifts \citep{Frebel2014, Rey2020, Chiti2021} and during the epoch of reionization \citep{Barkana1999}, dwarf galaxies are unique archaeological probes of physical processes shaping the formation of galaxies in the early Universe.

Various galaxy formation simulations generated with different codes and sub-grid prescriptions for star formation and stellar feedback have been able to reproduce the central scaling relations between the stellar component and the haloes of galaxies predicted by abundance matching \citep[see e.g. ][]{Dubois2014, Vogelsberger2014, Crain2015, Onorbe2015, Fattahi2016, Henden2018, Pillepich2018b, Hopkins2018c}. These scaling relations are rather uncertain in the dwarf regime, where very high resolution simulations are needed to resolve the internal processes and the ISM of dwarfs \citep[see e.g.][]{Wheeler2019, Smith2019, Agertz2020, Gutcke2021a}. While SN feedback has traditionally been invoked as the dominant physical process regulating the properties of dwarf galaxies \citep{White1978, Dekel1986, White1991, Efstathiou2000}, it is now understood that other physical processes beyond SN feedback are also required to reproduce realistic dwarf properties \citep[e.g.][]{Rosdahl2018, Smith2019}. Early stellar feedback in the form of e.g. winds from massive stars or stellar radiation provides a mechanism to rapidly suppress star formation, with the impact of photoheating shown particularly important within the mass regime of dwarf galaxies \citep{Rosdahl2015b, Emerick2018}. Alternatively, warm or self-interacting dark matter have been explored as a means to alleviate the cusp-core problem \citep{Villaescusa-Navarro2011, Lovell2014, Vogelsberger2016}. Additionally, recent observational evidence supporting the presence of active galactic nuclei (AGN) in dwarf galaxies has sparked interest in whether AGN feedback in these galaxies may have a significant effect on their evolution \citep[e.g.][]{Habouzit2017, Dashyan2018, Koudmani2019, Koudmani2021, Koudmani2022}. However, well-known baryonic physics, such as magnetic fields, stellar radiation and $\sim\!\text{GeV}$-energy cosmic rays still need to be studied in detail in full cosmological simulations of dwarf galaxies.

Estimates of the magnetic energy in galaxies reveal the importance of magnetic fields for the ISM, with observations suggesting equipartition between the thermal, turbulent and magnetic pressure components \citep{Basu2013, Beck2015}. Magnetic fields with strengths of $\gtrsim\!\!\muG$ have been detected in dwarf galaxies \citep{Chyzy2011}. Such magnetic fields can affect the galaxies in multiple ways. At sub-ISM scales, magnetic fields are well-known to be an important factor in regulating the dynamics of ISM turbulence \citep{Padoan2011}. They affect thermal instabilities and thus regulate the different phases of the ISM \citep{Iffrig2017, Kortgen2019} as well as influencing gas fragmentation \citep{Inoue2019}. On galactic scales, magnetic fields have the potential to impact galactic outflows \citep{Gronnow2018, Steinwandel2019}, halo gas mixing \citep{vandeVoort2021}, global galactic properties \citep{Pillepich2018a, Martin-Alvarez2020} and possibly play a role during galaxy merging \citep{Whittingham2021}. Magnetic fields are nonetheless not expected to have a major effect on fundamental properties such as the final stellar mass of galaxies \citep{Su2017, Pakmor2017, Martin-Alvarez2020}.

Even though SN explosions may be the most obvious form of stellar feedback, stellar radiation is widely recognised as an important agent in galaxy evolution. Stellar radiation is particularly fundamental for low mass, late-type dwarf galaxies \citep{Rosdahl2015b}. Accounting for radiative feedback leads to a gentler self-regulation of dwarf galaxies and reduces the importance of explosive SN events \citep{Agertz2020}. \citet{Katz2020} showed how photoheating by stellar radiation can lead to high redshift quenching of dwarf galaxies by evaporating the filaments responsible for their gas supply. Stellar radiation also augments the impact of individual SN events by pre-processing gas parcels where these explosions will take place \citep{Geen2015b}. Through this effect, stellar radiation has been argued to support the driving of galactic outflows \citep{Emerick2018}. However, other studies have claimed opposite effects \citep{Rosdahl2015b, Smith2021}, as photoionization feedback may disrupt local star forming clouds much more rapidly than the first SN exploding, hence leading to a less bursty and less effective SN feedback.

Finally, cosmic rays have gained a lot of attention in recent years due to their ability to efficiently drive continuous and colder galactic outflows, especially when compared with winds driven solely by SNe \citep[e.g.][]{Booth2013, Salem2014a, Girichidis2018, Samui2018, Dashyan2020, Buck2020, Farcy2022, Rodriguez-Montero2023}. Cosmic rays are able to reduce star formation rates in isolated \citep{Pakmor2016, Dashyan2020} and cosmological \citep{Jubelgas2008, Hopkins2020} galaxy formation simulations. Such reduction is found for most galaxy masses in isolated simulations. In a cosmological context, \citet{Hopkins2020} found this effect only for their larger galaxy masses, also varying based on the selected cosmic ray diffusion coefficient $\kappaCR$. This disparity points towards some dependence on the employed physics of cosmic ray transport (diffusion, streaming, etc.), as well as on the $\kappaCR$ coefficient, which remains poorly constrained. Depending on their implementation, cosmic rays may also alter the appearance of galaxies \citep{Buck2020} and their ISM \citep{Commercon2019, Dashyan2020, Nunez-Castineyra2022}. When converting a small fraction of SN energy into cosmic rays (typically $\sim\!\!10\%$; e.g. \citealt{Hopkins2020, Dashyan2020}), this non-thermal energy has the potential to enhance the deposition of momentum by SNe \citep{Diesing2018, Rodriguez-Montero2022}. Most importantly, once they escape SN remnants, cosmic rays can establish a `smooth' pressure gradient beyond galactic scales which is believed to accelerate gas at the edges of galactic discs \citep{Hopkins2021}. 

In addition to their individual effects, due to the highly non-linear nature of galaxy formation physics, these different processes have a complex interplay when combined in the same simulation. For example, in the presence of magnetic fields the efficiency of star formation ramps up in molecular clouds \citep{Federrath2012, Zamora-Aviles2018}, but these clouds will then be rapidly dissipated by the radiation produced by the newly formed stars \citep{Murray2010}. Likewise, radiation has the potential to puff up gas discs in galaxies \citep{Roskar2014} whereas cosmic rays are capable of accelerating the diffuse gas located at high altitudes above discs \citep{Breitschwerdt1991, Girichidis2016}. In combination, these effects may thus increase the amount of gas expelled by cosmic rays.

In this work, we aim to shed light on the role played by each of these additional physical processes, as well as on their interplay in the formation of dwarf galaxies. For this purpose, we perform a suite of cosmological zoom-in simulations of galaxy formation that, starting with our fiducial star formation and stellar feedback models, gradually includes additional and more sophisticated physics up to `full-physics' simulations featuring magneto-thermo-turbulent star formation and mechanical SN feedback with magnetic fields, on-the-fly radiative transfer, and cosmic rays. We investigate the evolution and resulting properties of the simulated dwarf galaxies and compare them with observations of similar galaxies in our local Universe to understand which physical processes are more likely to shape dwarf properties. To produce an agnostic and objective assessment of our different models, and to be able to better discriminate them against observations and empirical relations, we base the selection of all the model parameters in Pandora exclusively on physical considerations (see Section~\ref{s:Methods}, with a summary of our simulations in Table~\ref{table:setups}), without any aim to match any specific observables. That is, we do not attempt to tune our model parameters and wherever a specific model fails to match expected quantities, we focus on investigating the physical mechanisms at play causing this result.

This work is organized as follows. The numerical methodology to generate and evolve our simulations is described in Section~\ref{s:Methods}, with further descriptions of our magnetic fields (Section~\ref{ss:MHDmethods}), radiative transfer (Section~\ref{ss:RTmethods}), and cosmic rays (Section~\ref{ss:CRmethods}) models. Section~\ref{ss:SimsSuite} describes our simulations suite and Section~\ref{ss:ObsComparison} discusses which observed dwarf galaxies provide the best comparison to Pandora. Section~\ref{s:Results} reports our main results, commencing with the evolution of the stellar mass - halo mass relation (Section~\ref{ss:HMSM}), followed by the study of stellar and gas morphology of our galaxy (Section~\ref{ss:sizes}), its resolved and integrated kinematics (Section~\ref{ss:kinematics}), the colour-magnitude relation (Section~\ref{ss:colmagnitude}), magnetic field and synchrotron synthetic observations (Section~\ref{ss:magnetism}), and finally by analysing the impact of different physical processes on its dark matter distribution (Section~\ref{ss:CuspCore}). Finally, we conclude with a summary of our work and its main conclusions in Section~\ref{s:Conclusions}.

\section{Methods} 
\label{s:Methods}
\begin{figure*}
    \centering
    \includegraphics[width=2.1\columnwidth]{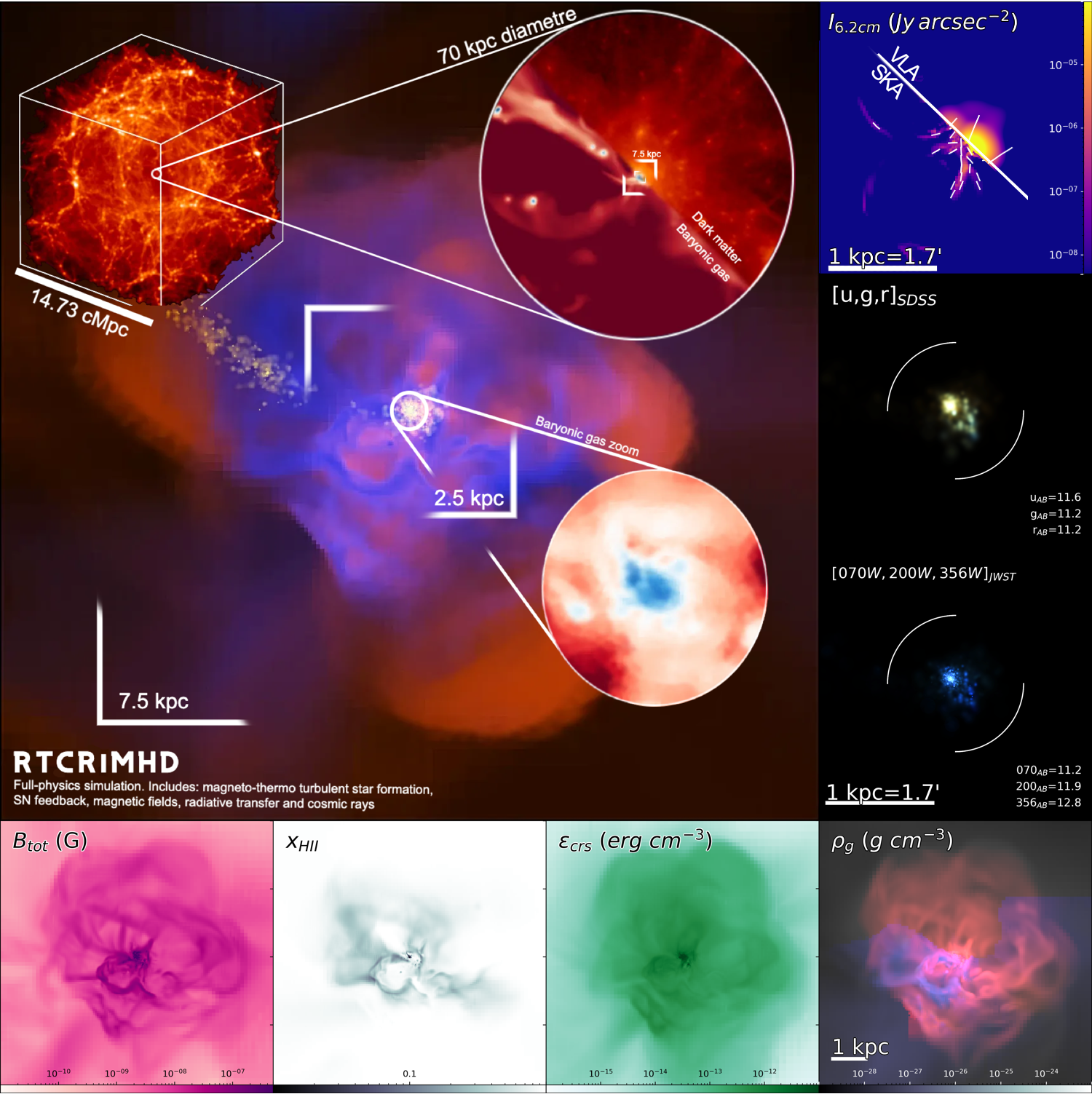}\\
    \caption{`Full-physics' simulation (\RTCRiMHDSfFb) projections centred on the dwarf galaxy at $z = 3.5$. {\bf (Inset box)} Mass projection of the entire simulated box, with the zoom-in circle encompassing approximately twice the virial radius of the galaxy studied. {\bf (Large panel)} RGB projections of a $(20\,\text{kpc})^3$ box showing gas density (blue), gas temperature (orange) and stellar density (yellow). Square labels indicate the projected regions shown in the bottom row (7.5 kpc) and the right-hand column (2.5 kpc). The circular panel in the central zoom region shows a gas density close-up view of the inner 500 pc of the dwarf galaxy. {\bf (Right-hand panels)} Synthetic observations of the galaxy in a $(2.5\,\text{kpc})^3$ box. The top panel shows radio synchrotron ($\lambda = 6.2\cm$, dashes show 90º-rotated polarisation to align with the magnetic field) as would be observed by VLA in its D configuration in the top right corner and for SKA-like resolutions at the bottom left. The middle and bottom panels, show respectively the optical emission as would be observed by SDSS, and the near infrared as would be observed by JWST. For these synthetic observations we artificially position the galaxy at a distance of 2 Mpc from Earth in order to emulate Local Group dwarf galaxies analogue distances. Each image is convolved with the PSF of the corresponding telescope. For the optical and near-IR panels, we provide the apparent magnitude within the circular aperture displayed by the white circle wedges. {\bf (Bottom row)} From left to right, these four panels are projections of a $(7.5\,\text{kpc})^3$ box showing magnetic field, hydrogen ionization fraction, cosmic ray energy density, and gas density (gray) separated into inflowing (blue gas; $v_\text{gas,radial} < -10\kmps$) and outflowing (red gas; $v_\text{gas,radial} > 10\kmps$) gas. The galaxy in our full-physics model appears particularly extended (compared to the fiducial hydrodynamical case, discussed in Fig.~\ref{fig:Galaxies1Close}), with an envelope of outflowing gas that correlates spatially with the high energy density region of cosmic rays and strong magnetic fields, extending to approximately 3 kpc.}
    \label{fig:FrontPanel}
\end{figure*}

All galaxy formation simulations studied in this work are new cosmological zoom-in simulations generated with our own modified version of the publicly available {\sc ramses} code \citep{Teyssier2002}. {\sc ramses} models the dark matter and stellar components as an ensemble of collisionless particles. These are coupled to each other and the baryonic gas through the gravity solver of the code. The evolution of the gaseous component is instead solved on an adaptive mesh refinement (AMR) grid that is recursively refined in regions of interest. We extend our version of {\sc ramses} to simultaneously and self-consistently model radiative transfer \citep{Rosdahl2013,Rosdahl2015a}, constrained transport (CT) magnetohydrodynamics \citep[MHD;][]{Teyssier2006,Fromang2006} and cosmic rays \citep{Dubois2016,Dubois2019}. Each of these physical components and the configurations we employ for them are described in Sections \ref{ss:MHDmethods} (MHD), \ref{ss:RTmethods} (radiative transfer), and \ref{ss:CRmethods} (cosmic rays).

We re-simulate one of the dwarf galaxies studied by \citet{Smith2019}, labelled as Dwarf1 in their work. Note that our simulations are generated with the {\sc ramses} code instead of {\sc arepo}, and that they feature different resolutions as well as physical models and implementations. An initial analysis of our setup showed good agreement of our halo and galaxy masses with those reported by \citet{Smith2019}. As we employ different sub-grid prescriptions for e.g., star formation and stellar feedback, we expect, however, multiple differences to emerge between our results and theirs. Our initial conditions are for a cubic box of $\sim 14.73$~comoving Mpc (cMpc) per side, initialised at $z = 127$ and discretised with a uniform grid of $256^3$ cells. At the centre of our computational domain a dwarf galaxy forms in a halo of virial mass $M_\text{vir} (z = 0) \sim 10^{10} \Msun$. We follow the formation of this dwarf galaxy using a convex hull zoom region. At $z = 127$, this region is approximately $2.5$~cMpc across, and it is evolved using a passive refinement scalar advected with the fluid and the high-resolution dark matter particles. The dark matter and stellar particle mass resolutions in this region are $m_\text{DM} = 1.5 \cdot 10^{3} \Msun$ and $m_\text{*} = 400 \Msun$, respectively. In the zoom region, we allow the AMR to refine the grid down to a resolution of $\Delta x \sim 7$~physical pc (or 3.5 pc radius/half-cell size). Cells are marked for refinement when their total contained mass $(\Omega_m / \Omega_b)\,m_\text{baryons} + m_\text{DM}$ surpasses $8 \times m_\text{DM}$ or when their size is $\Delta x > \lambda_J / 4$, with $\lambda_J$ being the local Jeans length. All our simulations employ the \citet{PlanckCollaboration2016b} cosmology. Furthermore, we impose an initial metallicity floor of $10^{-4} \Zsun$, corresponding to the critical metallicity required for gas fragmentation to allow PopII stellar clusters to be formed ($\Zsun = 0.012$; \citealt{Schneider2012}). Due to their computational cost, we evolve the majority of our simulations only to $z = 3.5$, when the mass of the studied halo is $M_\text{vir} (z = 3.5) \sim 5 \cdot 10^{9} \Msun$. We only continue some of our hydrodynamical runs (\HDSfFb, \HDSfFbBoost, \HDSfNoFb~and \HDSf; listed in Table~\ref{table:setups}) down to $z = 0.5$.

Unless indicated in Section \ref{ss:SimsSuite}, all our simulations include metal cooling above and below a threshold of $10^4$~K interpolating {\sc cloudy} cooling tables \citep{Ferland1998} and following \citet{Rosen1995}, respectively. We model the effects of ionizing radiation as an homogeneous ultraviolet (UV) background according to \citet{Haardt1996}, activated at $z < 9$. We always assume the baryonic gas to be monatomic and ideal (i.e. with a specific heat ratio $\gamma = 5/3$).

To determine the position of the main galaxy and its halo, we use {\sc halomaker} \citep{Tweed2009}, employing a shrinking spheres algorithm \citep{Power2003} to attain a higher centring accuracy. The application of the halo finder to the dark matter determines the location and properties of the studied dwarf galaxy halo. Whenever required, we obtain the centre of the galaxy and its angular momentum by applying {\sc halomaker} to the baryons (gas and stars), with the centre of the galaxy being located inside the central region of the halo ($r < 0.2\;r_\text{halo}$).

\subsection{Star formation and stellar feedback} 
\label{ss:StarsFormation}
The process of star formation in our simulations is modelled by converting a fraction of the gas mass contained within a given cell into a new stellar particle. For this, we investigate two different prescriptions. The first is our fiducial prescription: a magneto-thermo-turbulent (MTT) star formation model, presented in \citet{Kimm2017}, and \citet{Martin-Alvarez2020} in its MHD extension. In this model, we only allow cells at the highest level of refinement to form stars \citep{Rasera2006} as long as they fulfil the condition that the gravitational pull due to their gas content is higher than the support provided by the combined turbulent, thermal and magnetic pressure. Star forming cells convert gas into stars following a Schmidt law \citep{Schmidt1959},
\begin{equation}
    \dot{\rho}_\text{star} = \epsilon_\text{ff} \frac{\rho_g}{t_\text{ff}}\,,
    \label{eq:SchmidtLaw}
\end{equation}
with star formation efficiency, $\epsilon_\text{ff}$, and free-fall time, ${t_\text{ff}}$. In this model, $\epsilon_\text{ff}$ is a local parameter determined by the MTT properties of the local gas cells. For its computation, we follow the multi-free fall version of the \citet{Padoan2011} model as described by \citet{Federrath2012}. For further details, we refer to Appendix B of \citet{Martin-Alvarez2020}.

A small subset of our simulations assume an alternative star formation model, the more commonly used gas density threshold. This model assumes a fixed star formation efficiency of $\epsilon_\text{ff} = 0.015$ and allows star formation to proceed according to Equation~(\ref{eq:SchmidtLaw}) whenever a cell at the highest refinement level \citep{Rasera2006} has a gas density that exceeds $\rho_g > \rho_\text{th} = 10\; \mH / \cm^3$, with $\mH$ the hydrogen atom mass. These values follow \citet{Smith2019}, and are selected to produce a star formation model analogous to that by \citet{Krumholz2007}. These simulations using density threshold star formation are the most similar within this study to the simulations studied by \citet{Smith2019} in terms of configuration, but the caveats described above remain.

Whenever radiative transfer is included in our simulations, stellar particles emit radiation into specific energy bins. Radiative feedback from stars is described in more detail in Section~\ref{ss:RTmethods}.
The majority of our simulations also feature SN feedback, employing the mechanical SN feedback prescription (Mech) by \citet{Kimm2014} and \citet{Kimm2015}. To determine when SN events occur, each stellar particle has its initial mass function (IMF) stochastically sampled during the first 50 Myr after its formation. Each stellar particle undergoing a SN event injects into its hosting and neighbouring cells mass, momentum and energy, with a specific energy of $\varepsilon_\text{SN} = E_\text{SN} / M_\text{SN}$ (except for the boosted feedback simulations - as indicated in Section~\ref{ss:SimsSuite}), where $E_\text{SN} = 10^{51} \erg$ and $M_\text{SN} = 10\,\Msun$. We assume a Kroupa initial mass function \citep{Kroupa2001}, returning a fraction $\eta_\text{SN} = 0.213$ of the total SN exploding mass to the ISM gas. A further fraction $\eta_\text{metals} = 0.075$ of this total mass corresponds to the gas returned as metal mass. In some of our simulations our SN feedback also injects magnetic and cosmic ray energy back to the ISM. A detailed description of these injections by the SN feedback are provided in Sections~\ref{ss:MHDmethods} and \ref{ss:CRmethods}, respectively.

\subsection{Magnetohydrodynamics} 
\label{ss:MHDmethods}
We model magnetic fields and their coupling to the gas fluid with the CT ideal MHD implementation of {\sc ramses} by \citet{Fromang2006} and \citet{Teyssier2006}. This implementation models the magnetic field $\vec{B}$ as a face-centred quantity in each cell. The algorithm ensures that the divergence of the magnetic field fulfils the solenoidal constraint ($\vec{\nabla} \cdot \vec{B} = 0$) exactly to the numerical precision. This guarantees the absence of spurious MHD modifications and the preservation of conserved quantities, which is not ensured with other methods\footnote{See e.g. \citet{Hopkins2016} for a performance comparison of various MHD solver methods in simple and complex astrophysical problems.} \citep{Toth2000}. The time evolution of the magnetic field is computed solving the induction equation.

As magnetic diffusivity in most astrophysical environments is negligible, we set this quantity to zero in our simulations. This implies that all magnetic diffusive effects in our simulations will result from the numerical magnetic diffusivity emerging when the domain is discretised into a finite grid. 

In the absence of battery terms in the ideal MHD induction equation (such as the implementation of a Biermann battery, e.g. \citealt{Attia2021}), a magnetic seed has to be introduced to obtain any $\vec{B} \neq 0$ field. We investigate two different approaches: a) an ab-initio magnetic field, and b) a SN-injected magnetic field. In the first approach, we permeate the simulated domain with a uniform magnetic field along its $z$ axis with comoving strength $B_0$. This method is the most commonly employed in MHD simulations \citep[e.g.][]{Pakmor2016, Martin-Alvarez2018, Marinacci2018}, and can be interpreted as a magnetic field of primordial origin coherent on large scales. Galaxy formation simulations seeded with sufficiently small $B_0$ retain negligible or no memory of the initial seed \citep{Marinacci2015}, with only the strongest primordial magnetic fields being able to affect the properties of galaxies ($B_0 > 10^{-12}$ G; \citealt{Martin-Alvarez2020}). The second method of seeding injects small-scale circular loops of magnetic field around SNe as the SN events take place. This guarantees $\vec{\nabla} \cdot \vec{B} = 0$ when magnetising the SN ejecta in our mechanical feedback. As the ejecta expand, they advect the injected magnetic field to larger scales in the ISM. Each SN explosion is assumed to inject $E_\text{inj,mag} = 0.01 E_\text{SN} \sim 10^{49} \erg$. This corresponds to a magnetic field strength of $\gtrsim 10^{-5}$ G when injected at scales of $\sim\!\!10$~pc, in reasonable agreement with the observed high magnetisation of supernova remnants \citep{Parizot2006}. This magnetic injection model is capable of reproducing the magnetic fields observed in galaxies \citep{Martin-Alvarez2021}. Additional details of the magnetic field SN injection implementation can be found in Appendix A of \citet{Martin-Alvarez2021}.

\subsection{Radiative Transfer} 
\label{ss:RTmethods}
The implementation for ionizing emissivity in our simulations is similar to the one in the {\sc sphinx} simulations \citep{Rosdahl2018, Rosdahl2022}, which provide a good match to observational constraints on the reionization history of our Universe. We employ the {\sc ramses-rt} implementation by \citet{Rosdahl2013,Rosdahl2015a}. Radiative transfer is particularly sensitive to the distribution of multiphase gas within the ISM. \citet{Kimm2014} find that a resolution of $\sim\!\!5$~pc is required to have a converged escape of ionizing radiation into the ISM. Due to our similarly high spatial resolutions ($\Delta x \sim 7$~pc) we expect our radiative transfer approach within the ISM as well as its escape from the galaxy to be reasonably well resolved.

We separate the radiation into three spectral bins: HI ($13.6 \;\text{eV} \leq \epsilon_\text{photon} < 24.59 \;\text{eV}$), HeI ($24.59 \;\text{eV} \leq \epsilon_\text{photon} < 54.42 \;\text{eV}$), and HeII ($54.42 \;\text{eV} \leq \epsilon_\text{photon}$). We allow the radiation solver to subcycle over the hydrodynamical solver up to a maximum of 500 steps, and assume a reduced speed of light $0.01\, c$. This is sufficient for modelling the propagation of ionization fronts through the ISM of galaxies. In our radiative transfer simulations, stellar particles are the only source of ionizing radiation. Each stellar particle radiates energy into its hosting cell with a spectral energy distribution given by the {\sc bpass}v2.0 model \citep{Eldridge2008,Stanway2016} according to particle mass, metallicity and age.

\subsection{Cosmic rays} 
\label{ss:CRmethods}
Our simulations including cosmic rays use the cosmic ray implementation in {\sc ramses} by \citet{Dubois2016,Dubois2019}. This implementation assumes cosmic rays to behave as a fluid characterized by an energy density which is evolved by an implicit solver.
Unless explicitly indicated for a given model, our simulations with cosmic rays account both for anisotropic diffusion and streaming of cosmic rays. We assume the cosmic ray diffusion coefficient to be $\kappaCR = 3 \cdot 10^{28} \cm^2 \text{s}^{-1}$. This value has been found to be consistent with observations of $\gamma$-rays generated through cosmic ray hadronic losses \citep{Ackermann2012,Salem2016,Pfrommer2017a}. This value is also in agreement with estimates for the isotropic coefficient in the Milky Way \citep{Trotta2011,Cummings2016}. We assume that the streaming velocity is equal to the Alfv\'en velocity. In addition to streaming and adiabatic energy losses, the cosmic ray implementation in {\sc ramses} also accounts for hadronic and Coulumb cooling of cosmic rays \citep{Guo2008}. Cosmic rays are assumed to be generated by SN explosions. In our simulations with cosmic rays, each SN event injects a fraction of its total energy as cosmic rays to its hosting cell, where this energy follows $E_\text{CR} = f_\text{CR} E_\text{SN} = 10^{50} \erg$ and is extracted from the standard thermal injection. The selected fraction $f_\text{CR} = 0.1$ is in agreement with observational estimates for cosmic ray injection by SN remnants \citep{Morlino2012,Helder2013}. Finally, we note that the selected values for $\kappaCR$ and $f_\text{CR}$ have been frequently studied in the literature \citep[e.g.][]{Pfrommer2017b, Wiener2017, Butsky2018, Dashyan2020}.

\subsection{The simulation suite} 
\label{ss:SimsSuite}
Our suite of simulations builds up from a fiducial simulation employing hydrodynamics (HD) with {\it star formation} and {\it stellar feedback} physics up to a full-physics simulation with radiative transfer, cosmic rays and magnetic fields. A compilation of all the simulations studied in this work is presented in Table \ref{table:setups}. In this table, and throughout the rest of the manuscript, we label our simulations according to the physical processes included: magnetohydrodynamics ({\it MHD}), radiative transfer ({\it RT}) and cosmic rays ({\it CR}). According to the source of magnetic fields, we have simulations with a primordial magnetic field of a moderately high strength ($B_0 = 3 \cdot 10^{-13}$ G) simply labelled {\it MHD}; a strong primordial magnetic field ($B_0 = 3 \cdot 10^{-11}$ G) labelled {\it sMHD}, and magnetised SN seeding (note that these simulations still include a negligible primordial field, $B_0 = 3 \cdot 10^{-20}$ G) labelled {\it iMHD}. We do not need to discriminate different radiative transfer configurations, as all our simulations share the same radiative transfer implementation (Section \ref{ss:RTmethods}). For cosmic rays, our simulations contain both cosmic ray streaming and diffusion (simply identified as {\it CR}) or exclusively diffusion with deactivated streaming (in the simulation labelled \RTnsCRiMHDSfFb; identified by {\it nsCR}). Finally, the names of some of our simulations also contain information of their treatment for star formation and stellar feedback. Unless indicated otherwise, all our simulations include our fiducial MTT star formation implementation. The simulations using the standard density threshold star formation criterion are identified by adding {\it thSf} to their name. We also have some hydrodynamical simulations where SN feedback events do not return any energy to the ISM. These have {\it HD} replaced in their name by {\it NoFb}. Amongst these simulations, we distinguish those where SN feedback events do not take place at all ({\it NoZ}) from those where the SN feedback only returns mass (including metals) to the ISM. Finally, we include two simulations where we boost the specific energy of the mechanical feedback by a factor of 4 ({\it Boost}, with $\varepsilon_\text{SN} = 2\,E_\text{SN} /\,0.5\,M_\text{SN}$). This calibration is inspired by the {\sc sphinx} simulations \citep{Rosdahl2017}, which found that such a boost factor for the energy injection by SN feedback\footnote{Note that our boosted SN feedback models do not include radiative transfer, whereas the {\sc sphinx} simulations do.} reproduces well the observations by \citet{Read2017} and the stellar mass to halo mass ratio predicted by abundance matching \citep{Behroozi2013} in their low-mass galaxies at $z = 6$. With this calibration of the stellar mass-halo mass relation, {\sc sphinx} obtains a remarkable match to observations of the reionization history.

In summary, our simulations can be separated into four groups:
\begin{enumerate}
    \item {\bf HD} simulations, exploring different prescriptions for star formation and parameters for stellar feedback.
    \item {\bf MHD} simulations, exploring different sources of magnetic fields.
    \item {\bf RT}/RTMHD simulations, exploring the effects of stellar radiation and its combination with magnetic fields.
    \item {\bf CRMHD}/RTCRMHD simulations, exploring the impact of cosmic rays and their combination with radiative transfer and magnetic fields.
\end{enumerate}

Out of these, we focus most of our investigation on 7 simulations: \HDSfNoFb, \HDSfFb, \HDSfFbBoost, \RTSfFb, \RTiMHDSfFb, \RTnsCRiMHDSfFb, and \RTCRiMHDSfFb. This subsample builds up from no SN feedback up to `full-physics' simulations. We place special emphasis on three simulations that serve us to explore how standard modern simulations fare against the upcoming more complete models, and to study the importance of additional, well-known baryonic physics. These are stellar feedback \HDSfFb, calibrated stellar feedback \HDSfFbBoost, and our `full-physics' simulation \RTCRiMHDSfFb. A detailed view of the full-physics dwarf galaxy at $z = 3.5$ is shown in Fig.~\ref{fig:FrontPanel}. Gas density maps and synthetic optical views for other representative simulations are presented in Fig.~\ref{fig:Galaxies1Close}.

\begin{table*}
\centering
\caption{Suite of simulations studied in this work. Columns indicate from left to right the unique symbol and ID label of the simulation, the solver employed, the initial (uniform) magnetic field seed $B_0$, whether the simulation accounts for radiative transfer (Section~\ref{ss:RTmethods}) and cosmic rays (Section~\ref{ss:CRmethods}), the star formation prescription, the stellar feedback configuration, and further details regarding the configuration of the simulation. From top to bottom, simulations increase in complexity, accounting for additional baryonic physics.}
\label{table:setups}

\begin{tabular}{l l l l l l l l l l}
\hline
& Simulation & Solver & $B_0$ (G) & RT & CR & Stars & Stellar feedback & Further details \\
\hline

\tablesymbol{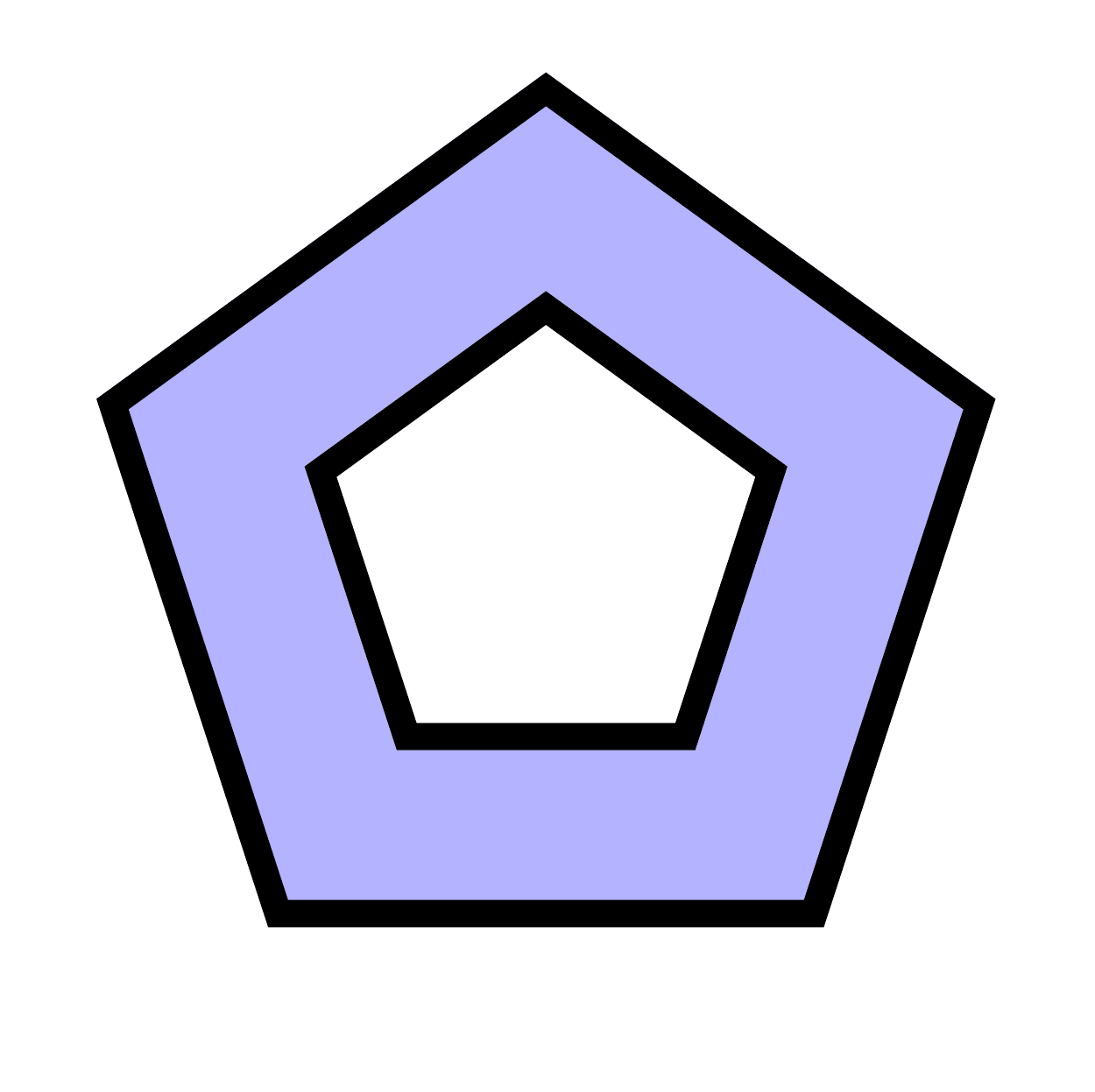} & \HDthSf & Hydro
&\xmark&\xmark&\xmark& thres & \xmark & Classic density threshold (thSf) star formation. \\

\tablesymbol{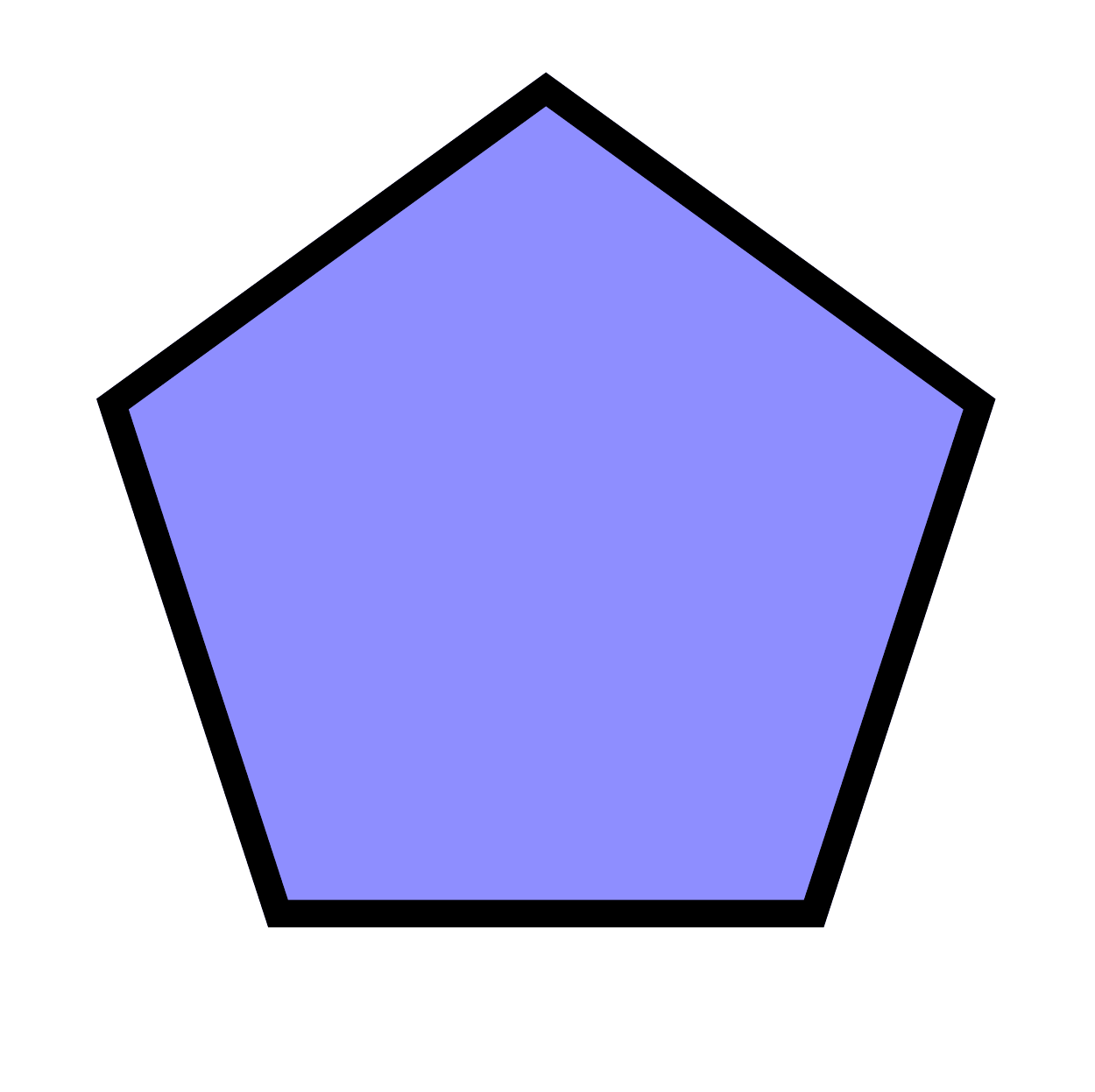} & \HDthSfNoFb & Hydro
&\xmark&\xmark&\xmark& thres & No energy/mom injection & Classic SF; SNe only inject gas and metal mass. \\

\tablesymbol{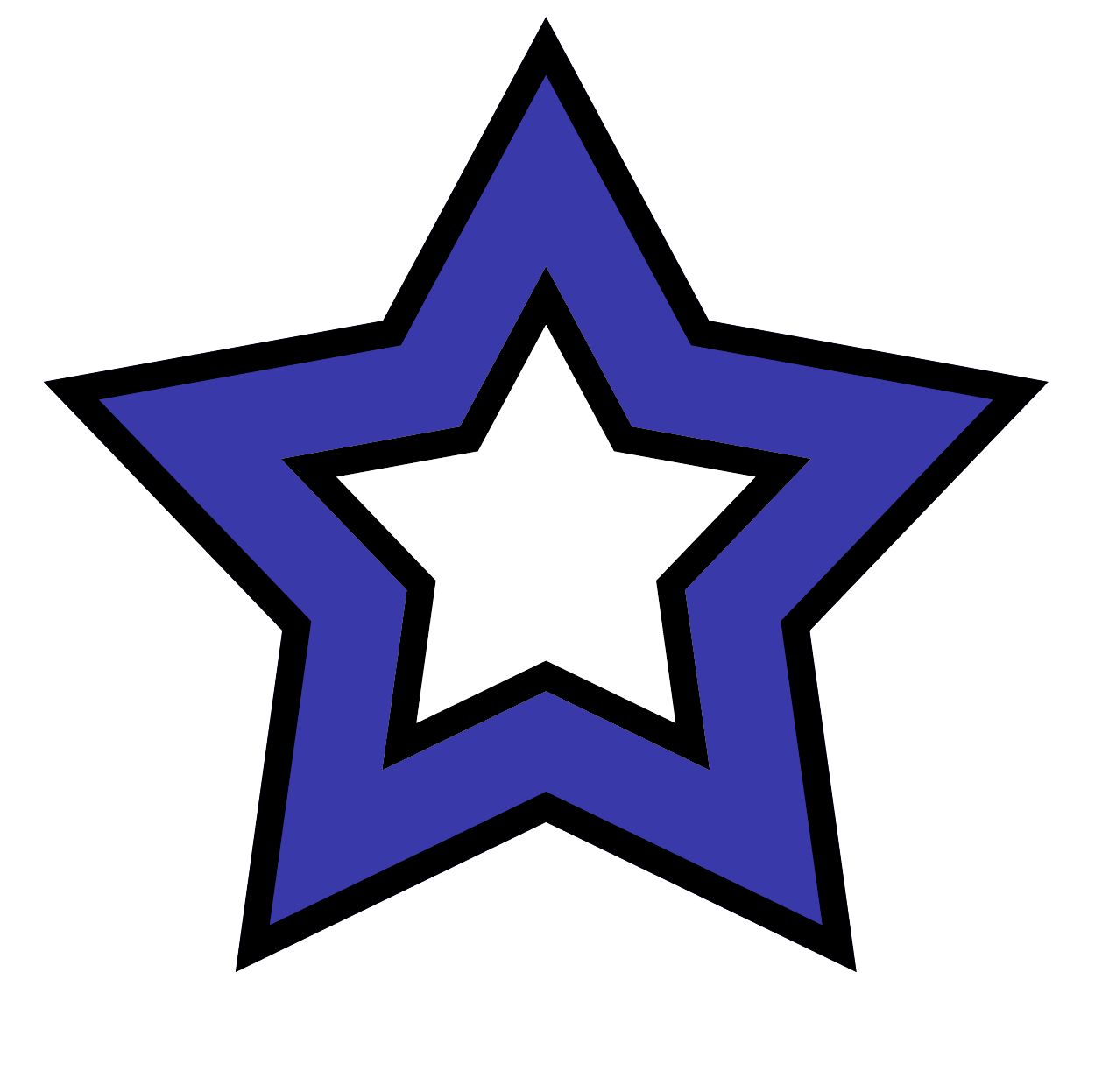} & \HDSf & Hydro 
&\xmark&\xmark&\xmark& MTT & \xmark & Magneto-thermoturbulent star formation (MTT).\\

\tablesymbol{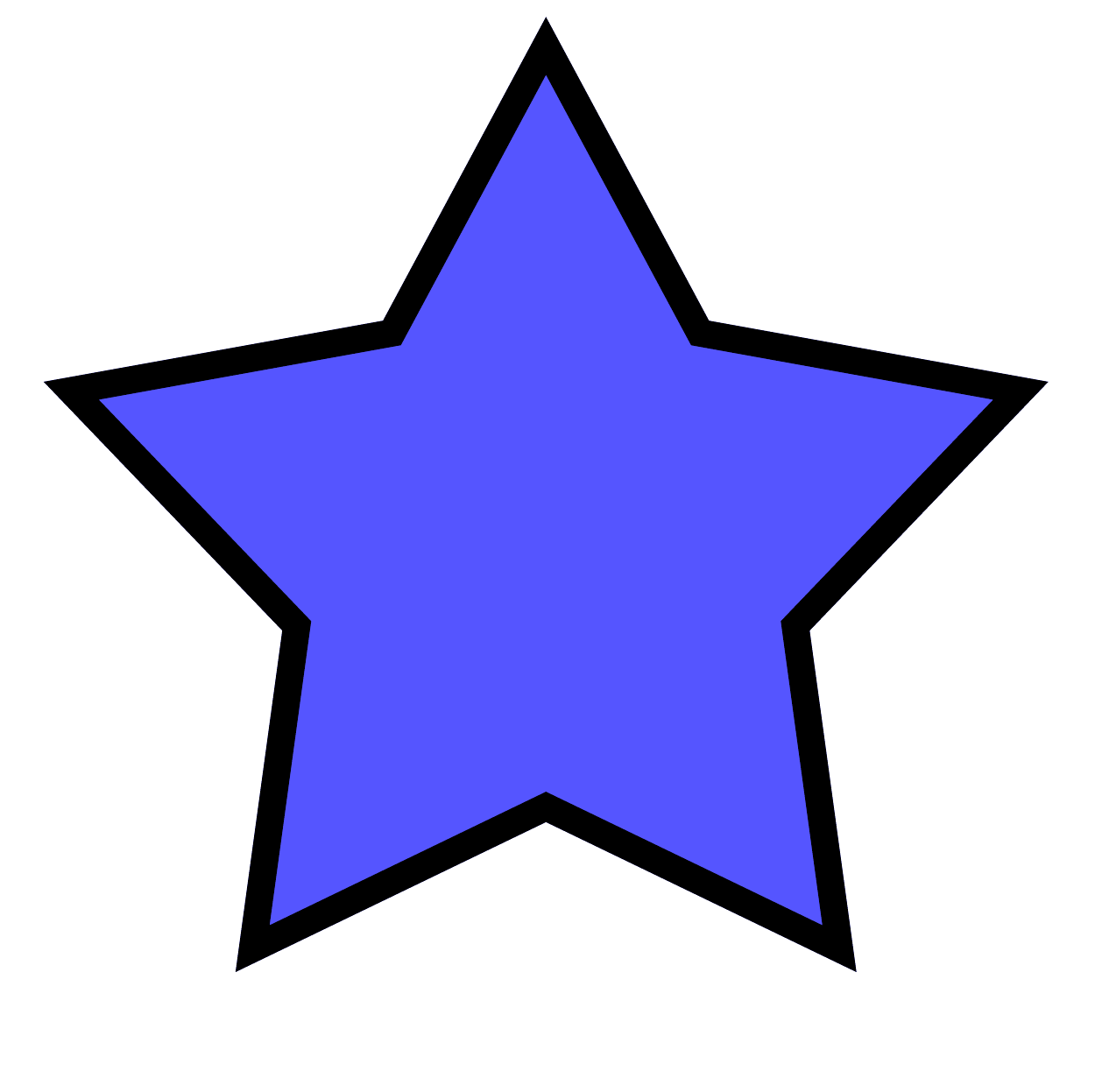} & \HDSfNoFb & Hydro 
&\xmark&\xmark&\xmark& MTT & No energy/mom injection & SNe only inject gas and metal mass. \\

\tablesymbol{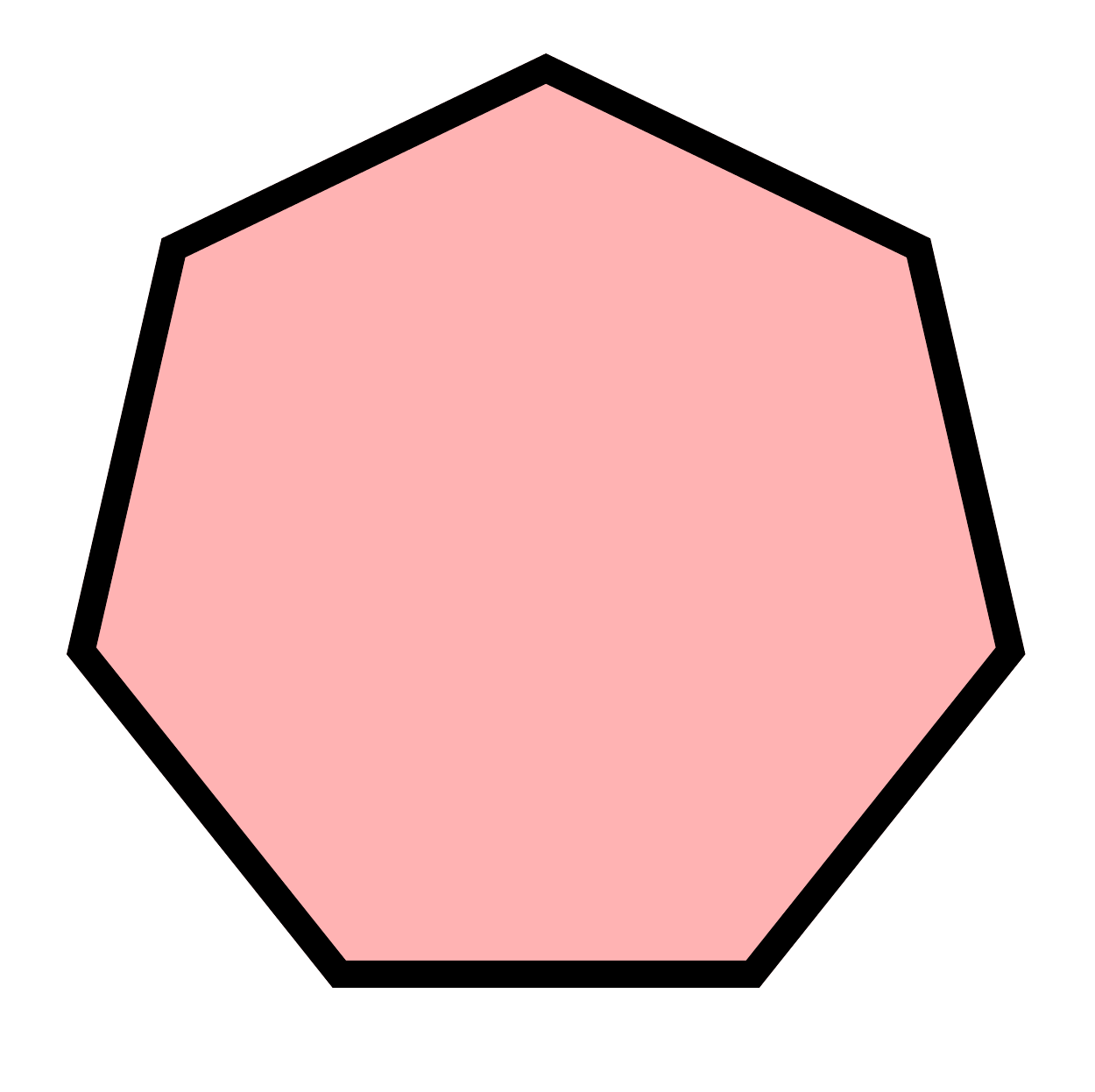} & \HDthSfFb & Hydro
& \xmark & \xmark & \xmark & thres & Mech & Classic SF. \\

\tablesymbol{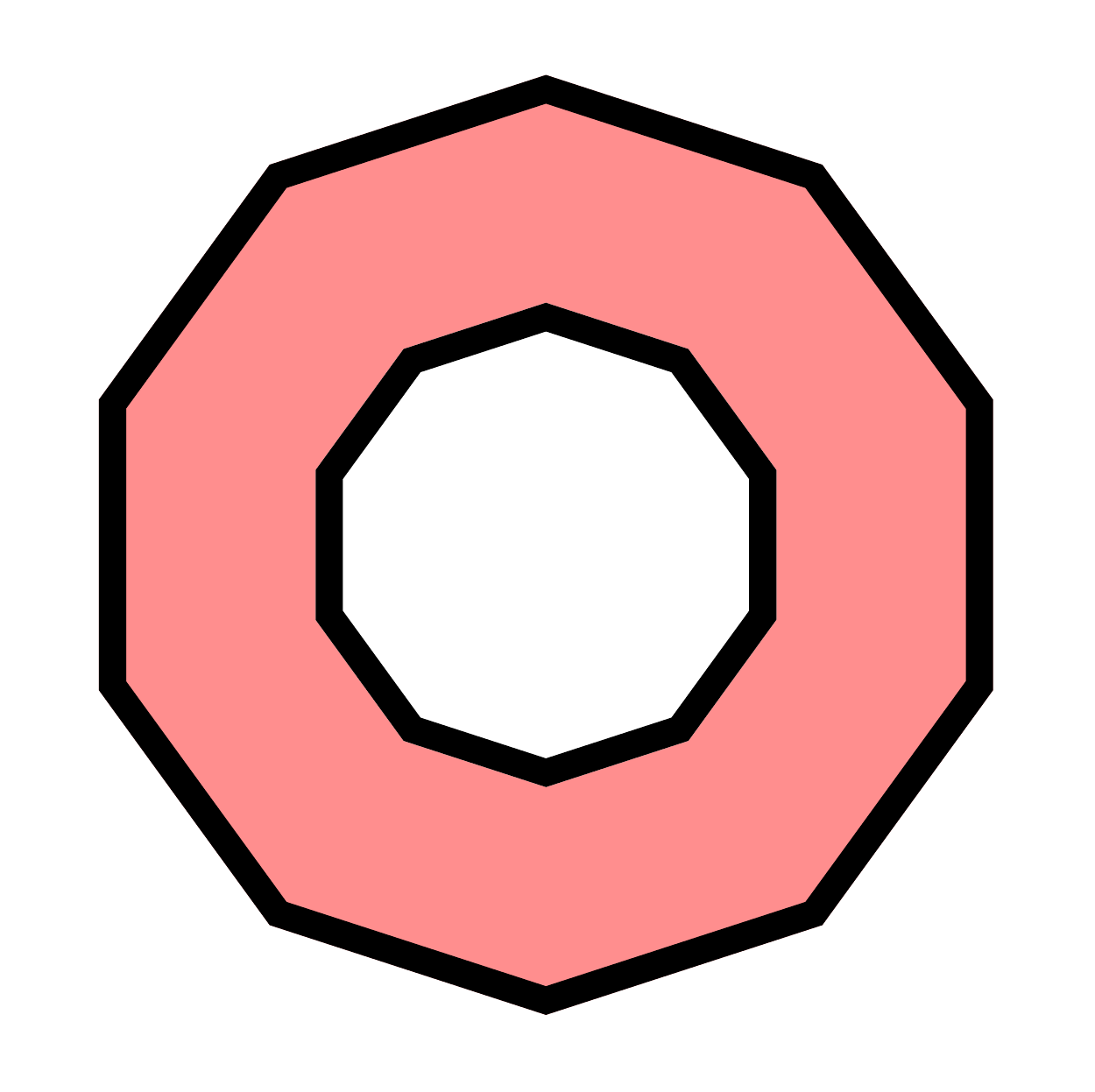} & \HDthSfFbBoost & Hydro 
& \xmark & \xmark & \xmark & thres & Boosted Mech & Classic SF; Boosted SN: $2 E_\text{SN}$, $0.5 M_\text{SN}$. \\

\tablesymbol{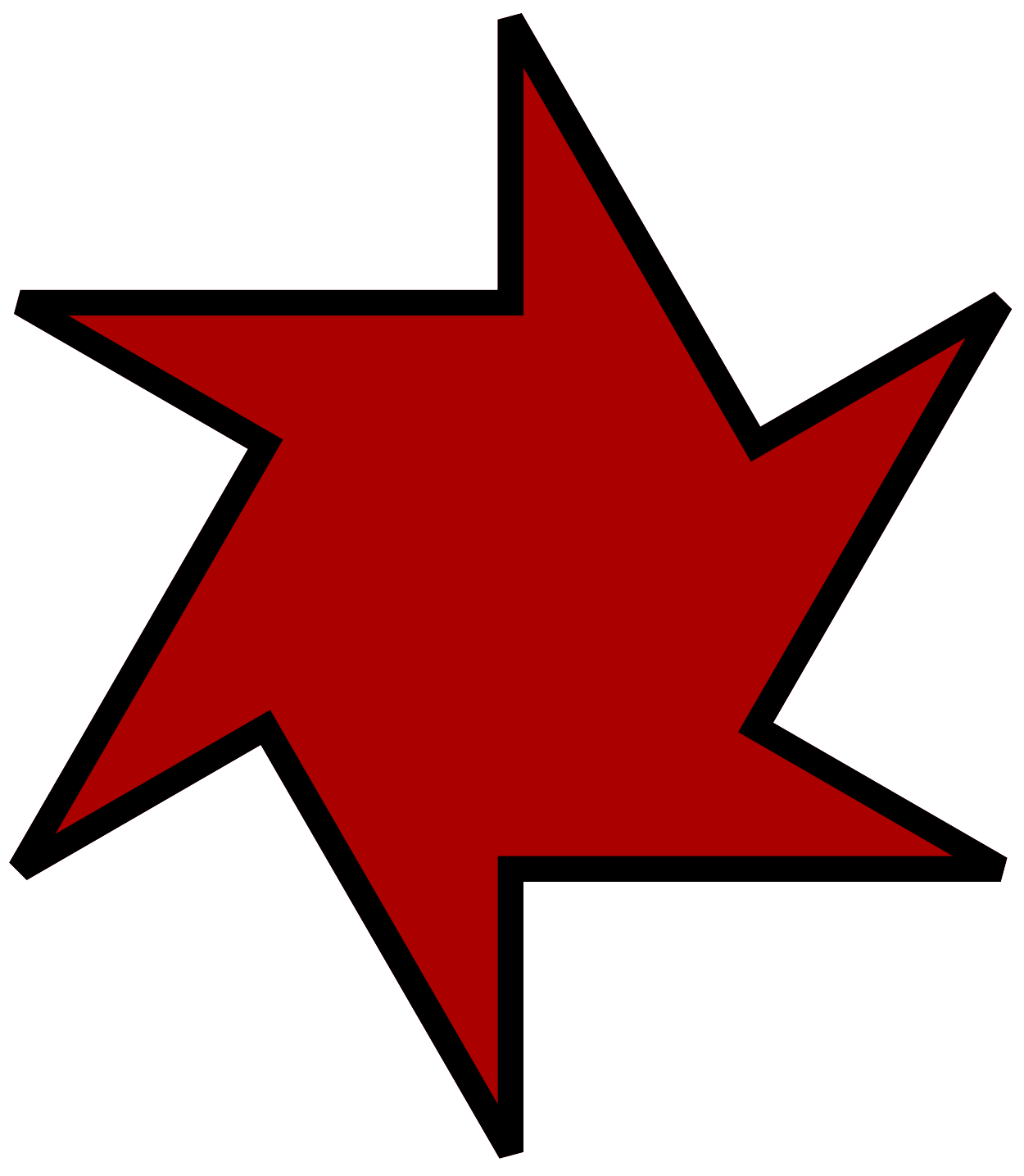} & \HDSfFb & Hydro
& \xmark & \xmark & \xmark & MTT & Mech & Fiducial physics simulation \\

\tablesymbol{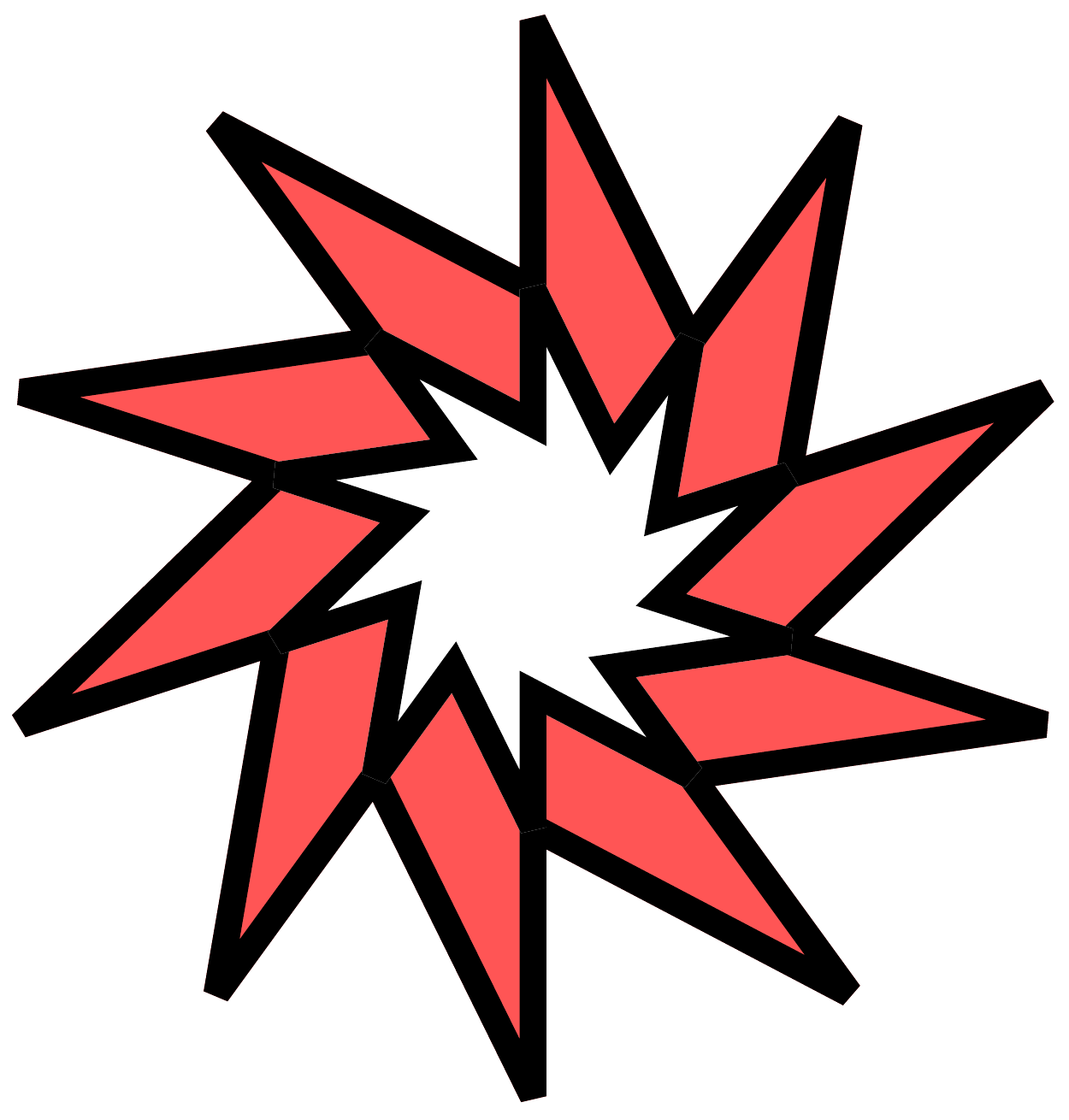} & \HDSfFbBoost & Hydro
& \xmark & \xmark & \xmark & MTT & Boosted Mech & Boosted SN: $2 E_\text{SN}$, $0.5 M_\text{SN}$. \\

\tablesymbol{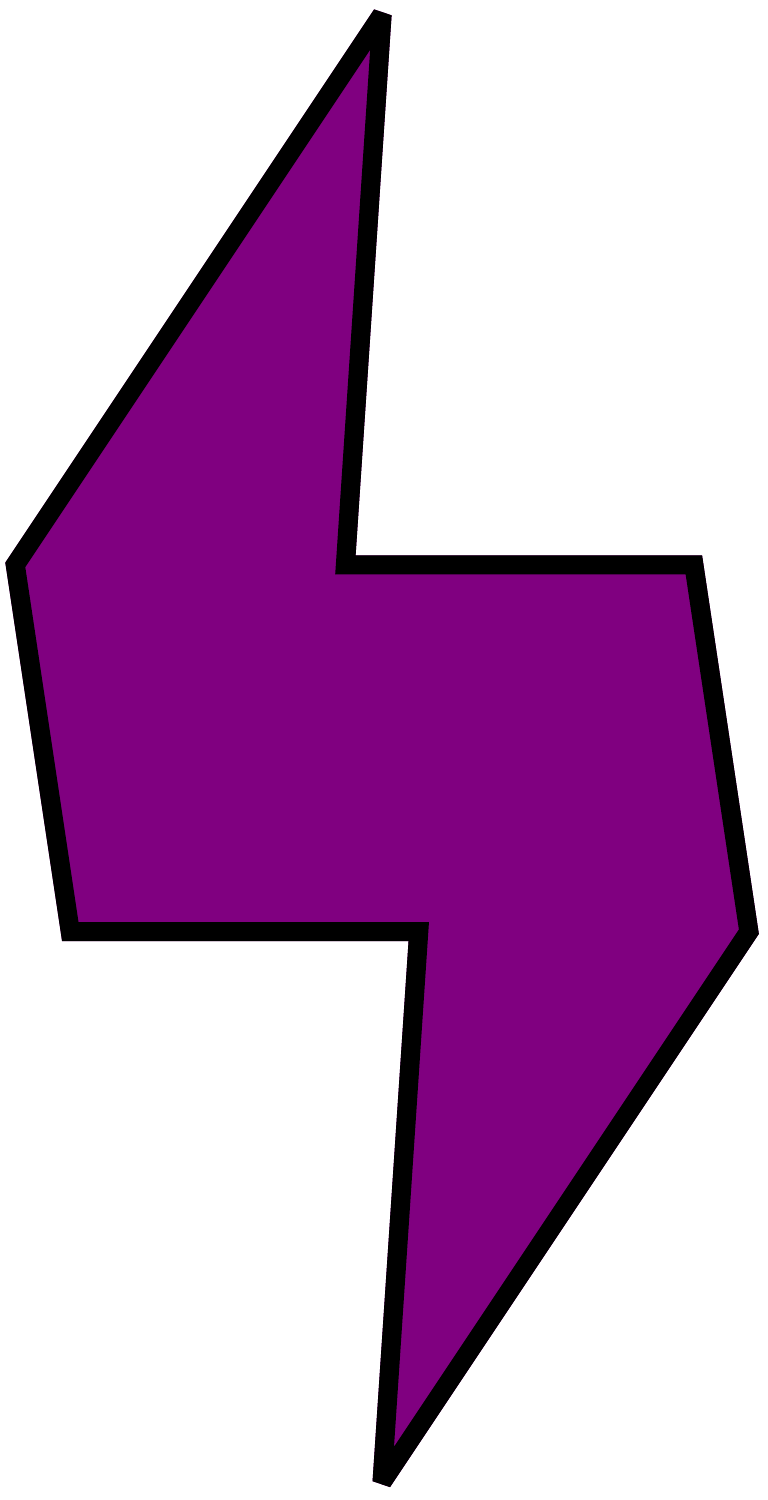} & \MHDSfFb & MHD
& $3 \cdot 10^{-13}$ & \xmark & \xmark & MTT & Mech & Intermediate primordial magnetic field\\

\tablesymbol{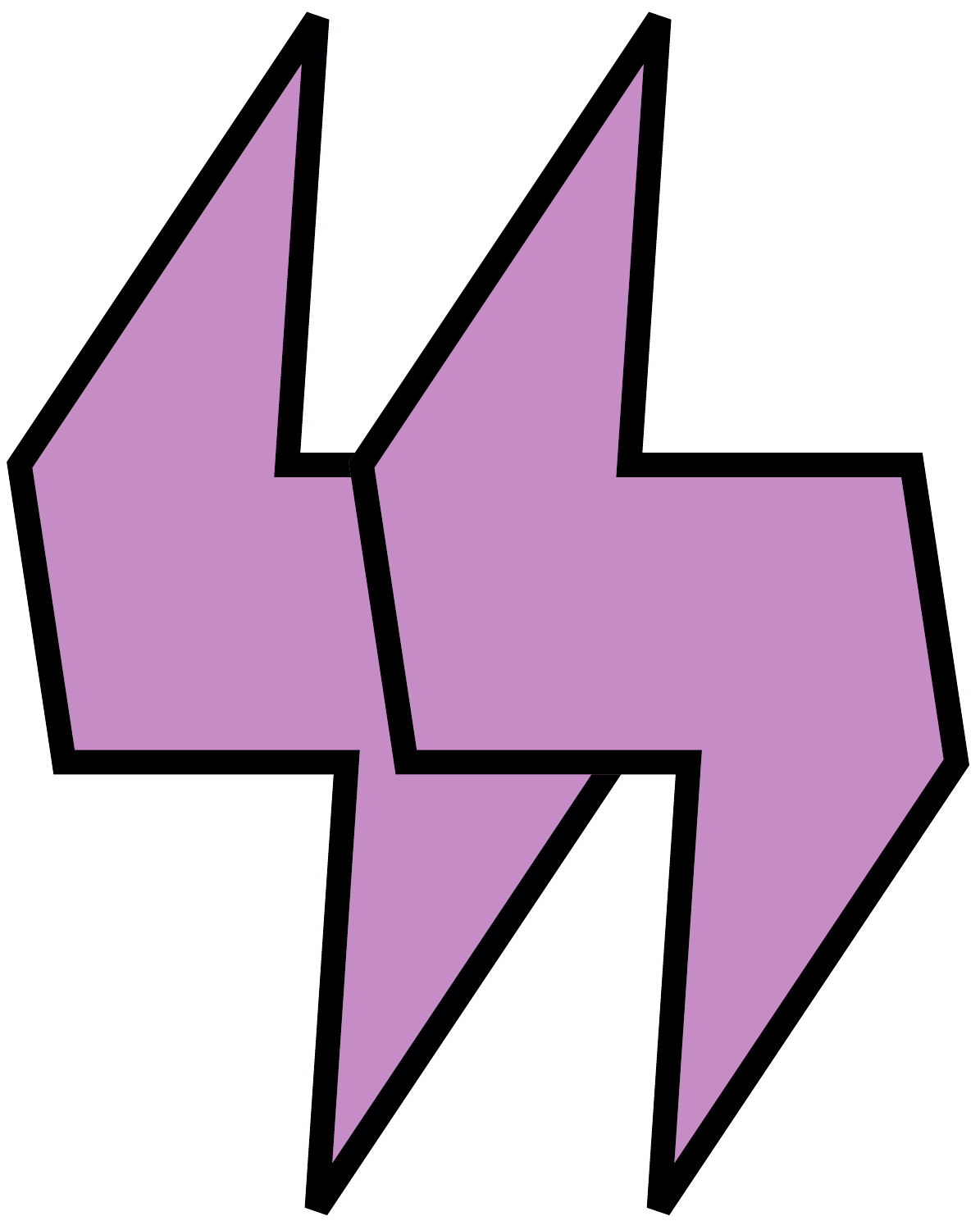} & \sMHDSfFb & MHD
& $3 \cdot 10^{-11}$ & \xmark & \xmark & MTT & Mech & Extreme primordial magnetic field\\

\tablesymbol{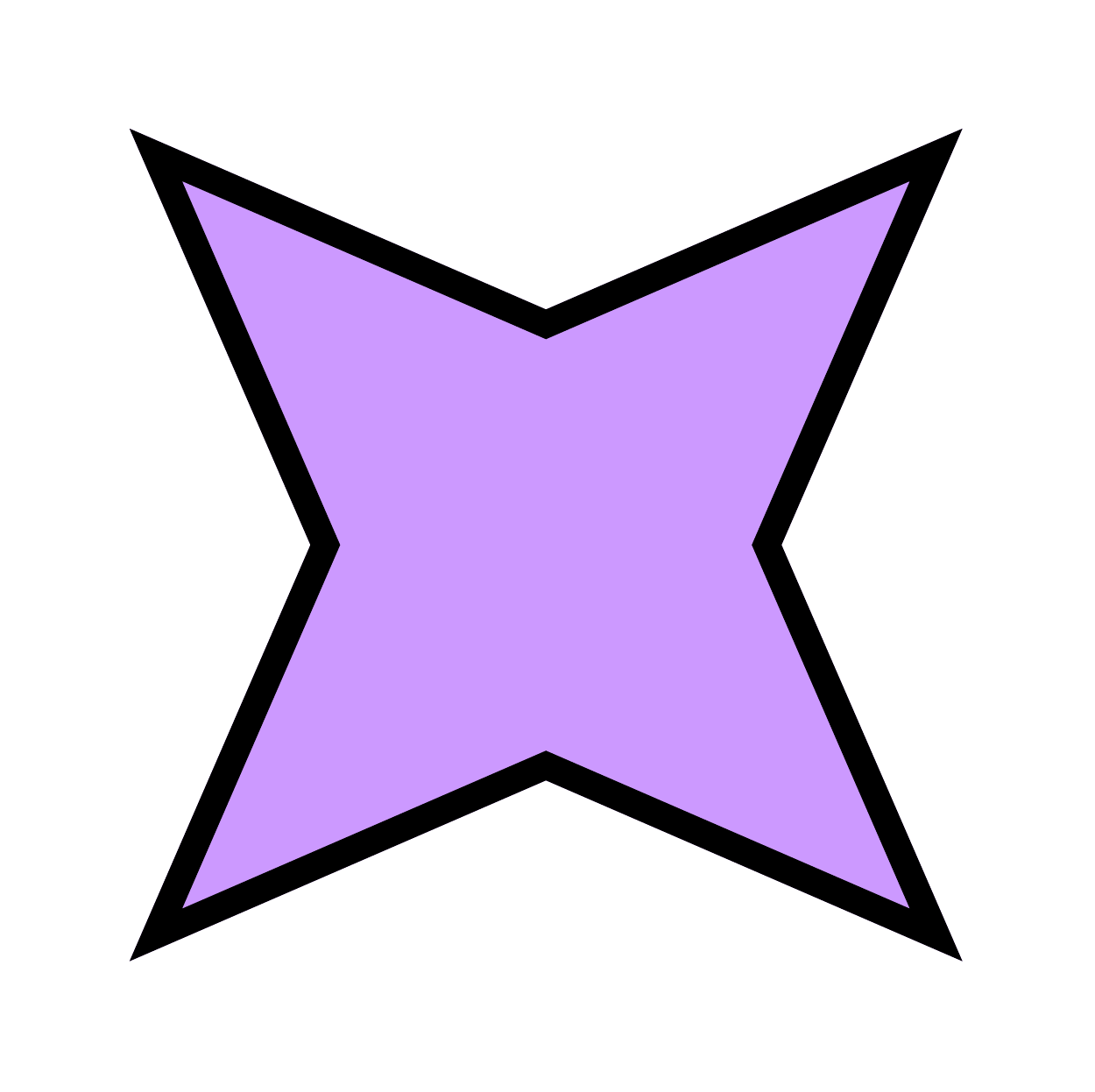} & \iMHDSfFb & MHD
& $3 \cdot 10^{-20}$ & \xmark & \xmark & MTT & MagMech & SN inject $E_\text{mag,SN}$ (MagMech). \\

\tablesymbol{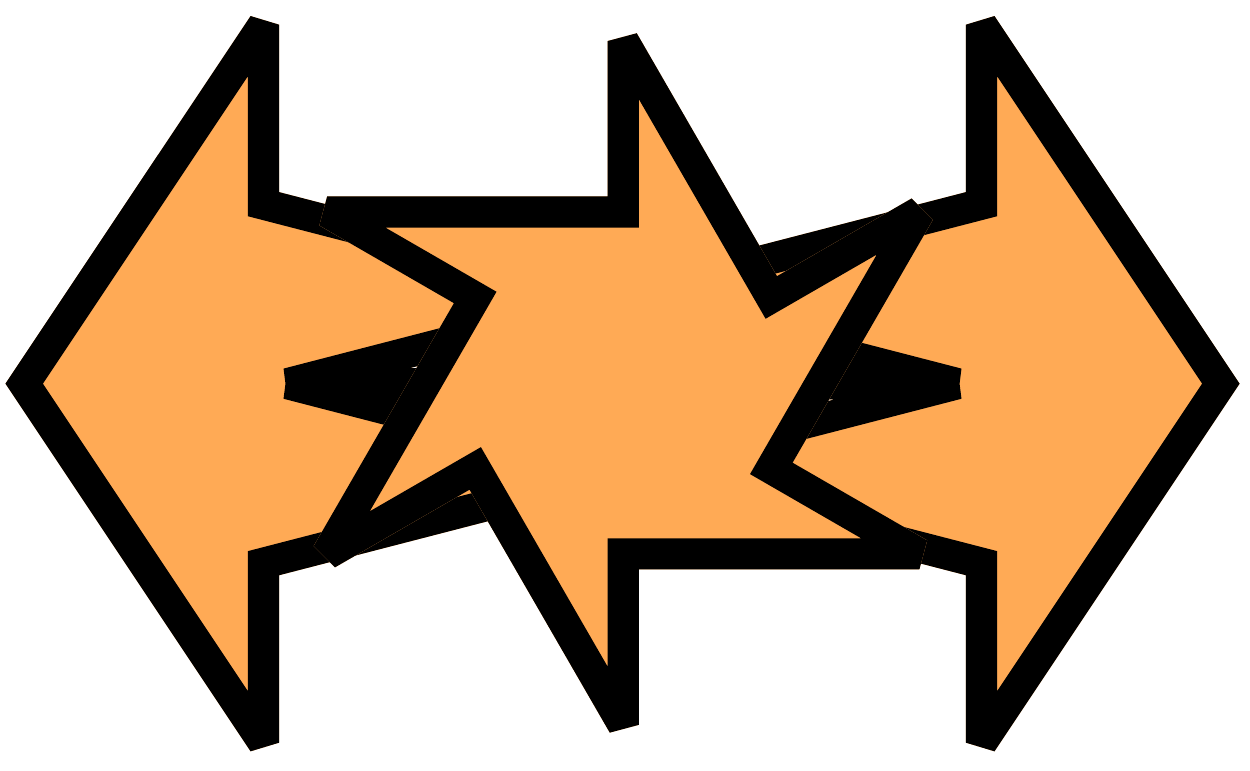} & \RTSfFb       & Hydro
& \xmark & \cmark & \xmark & MTT & Radiation + Mech & RT fiducial physics. \\

\tablesymbol{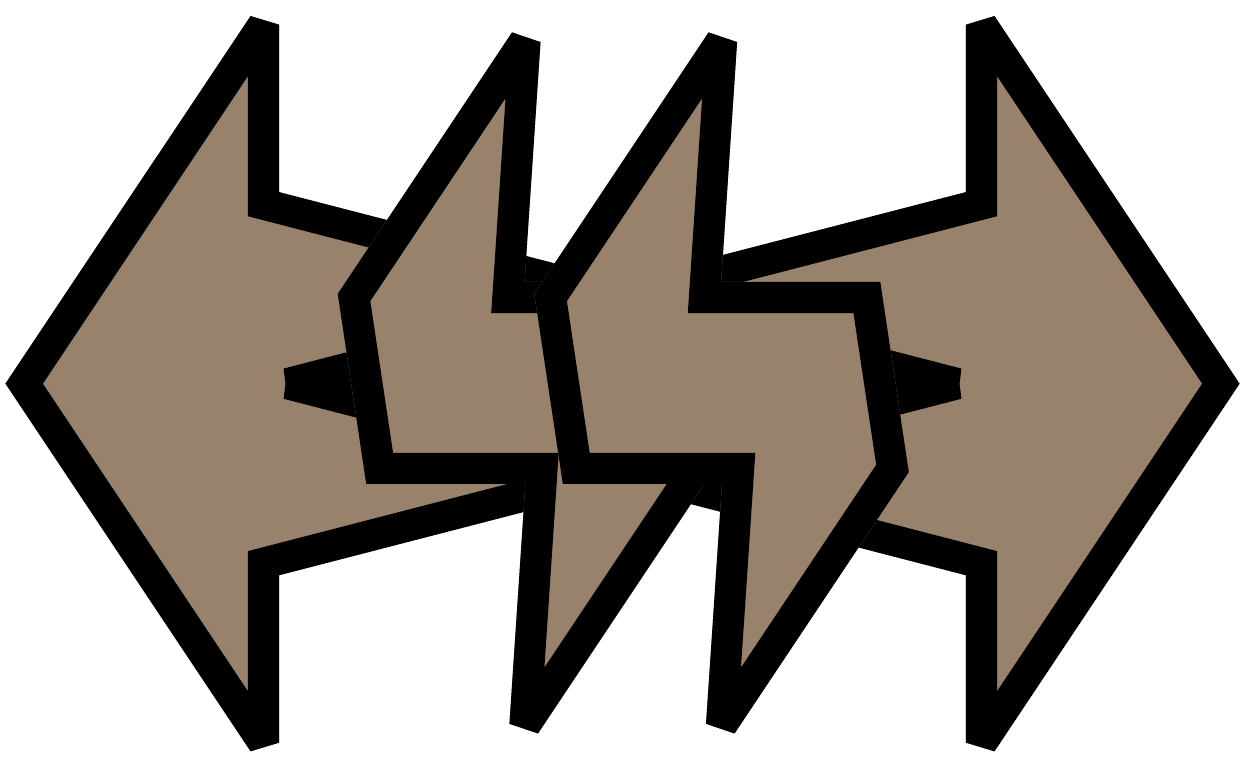} & \RTsMHDSfFb   & MHD
& $3 \cdot 10^{-11}$ & \cmark & \xmark & MTT & Radiation + Mech & RT extreme primordial magnetism.\\

\tablesymbol{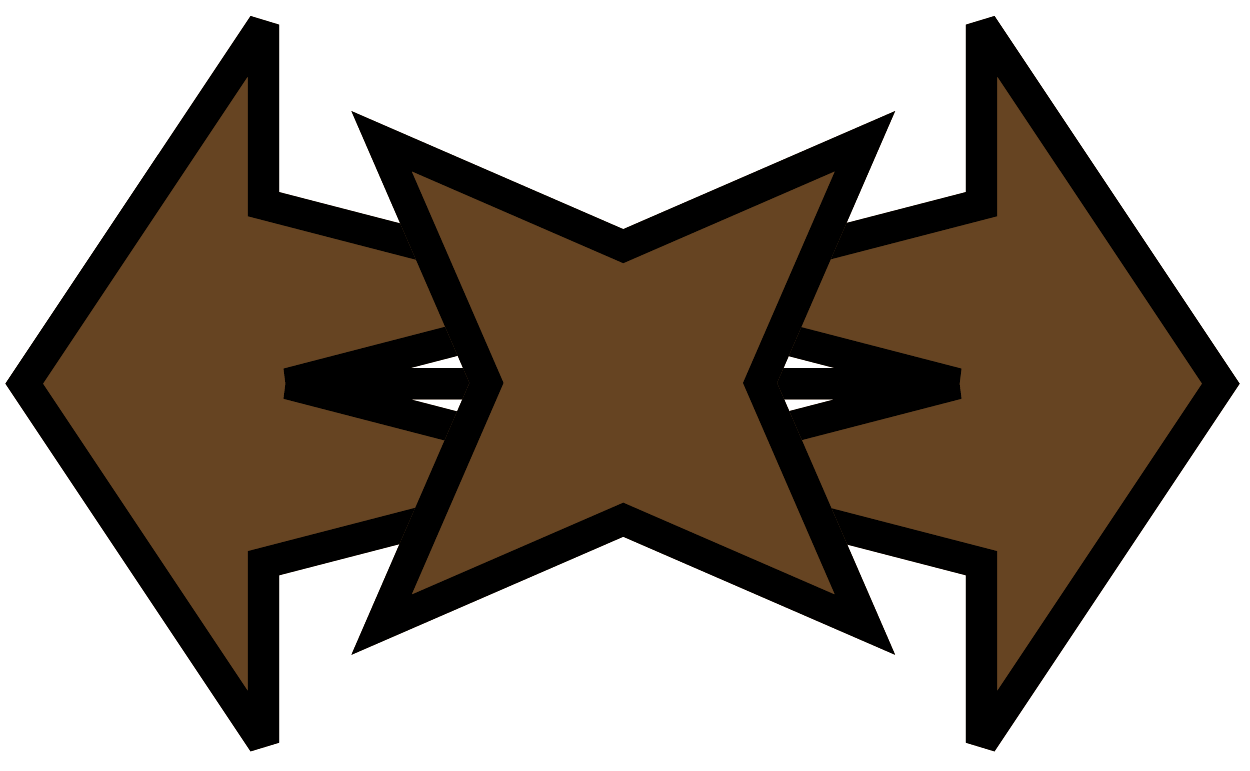} & \RTiMHDSfFb   & MHD
& $3 \cdot 10^{-20}$ & \cmark & \xmark & MTT & Radiation + MagMech & RT magnetism with SN inject $E_\text{mag,SN}$.\\

\tablesymbol{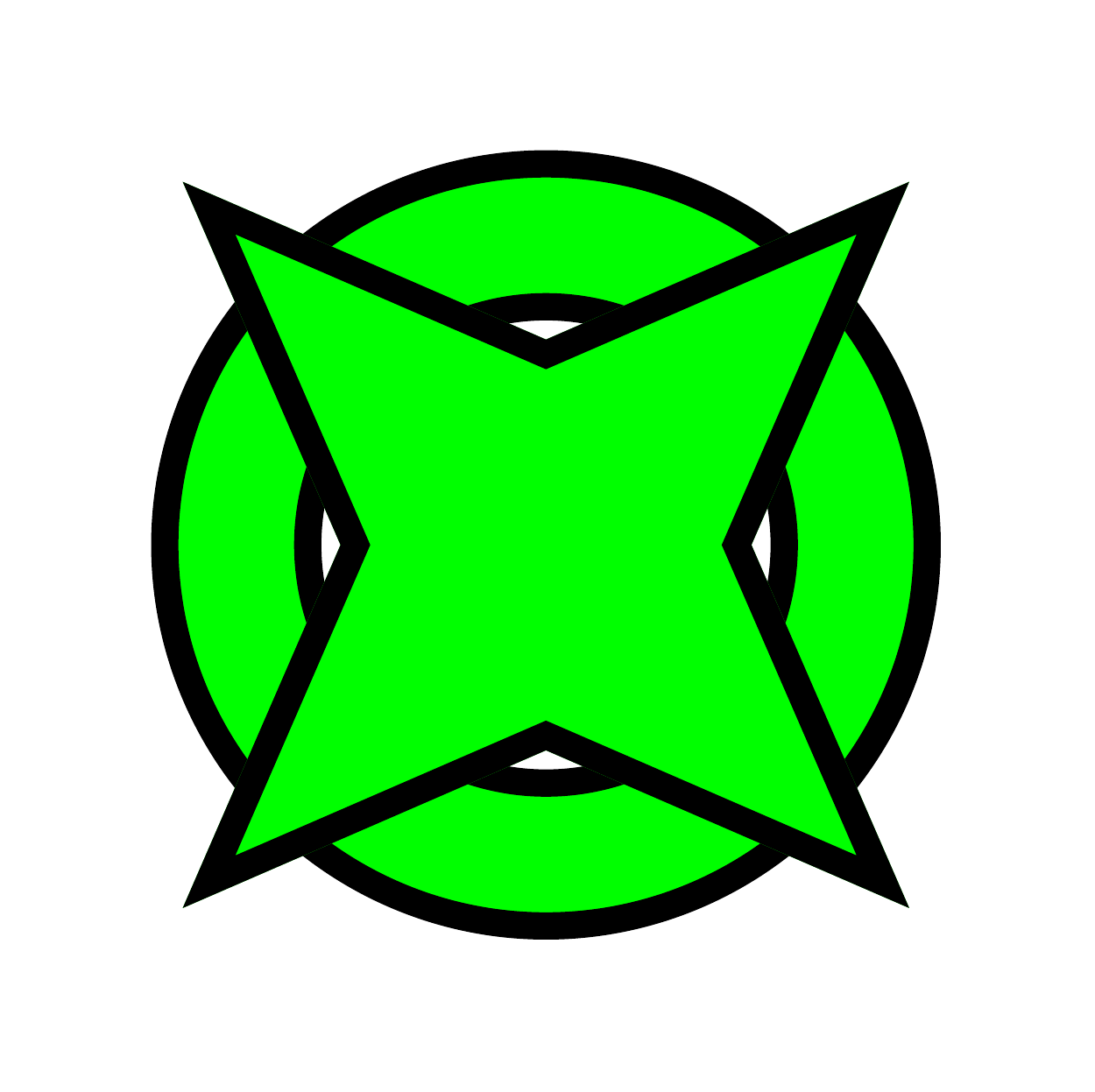} &\CRiMHDSfFb   & MHD
& $3 \cdot 10^{-20}$ & \xmark & \cmark & MTT & CRMagMech & SN inject $E_\text{mag,SN}$ + $E_\text{CR,SN}$ (CRMagMech). \\

\tablesymbol{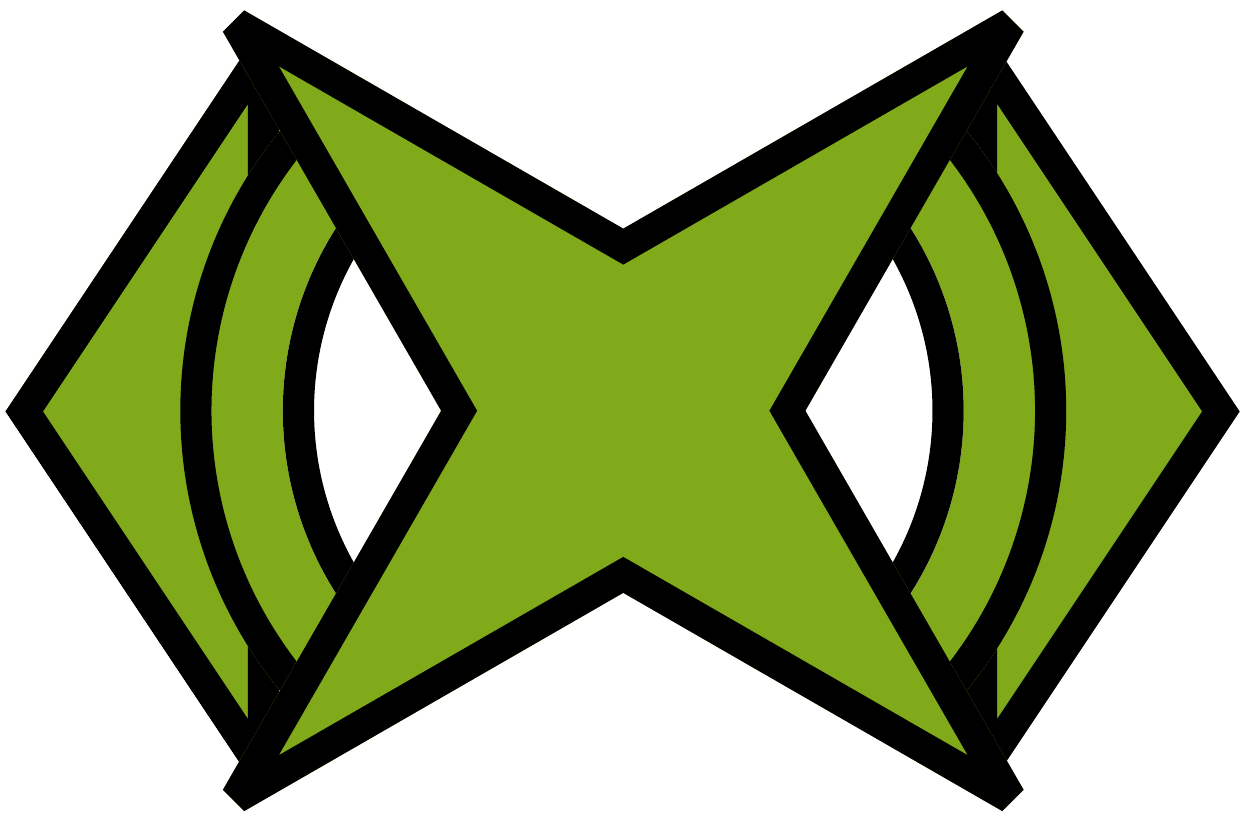} & \RTnsCRiMHDSfFb & MHD
& $3 \cdot 10^{-20}$ & \cmark & \cmark & MTT & Radiation + CRMagMech & Same as \RTCRiMHDSfFb~with no CR streaming.\\

\tablesymbol{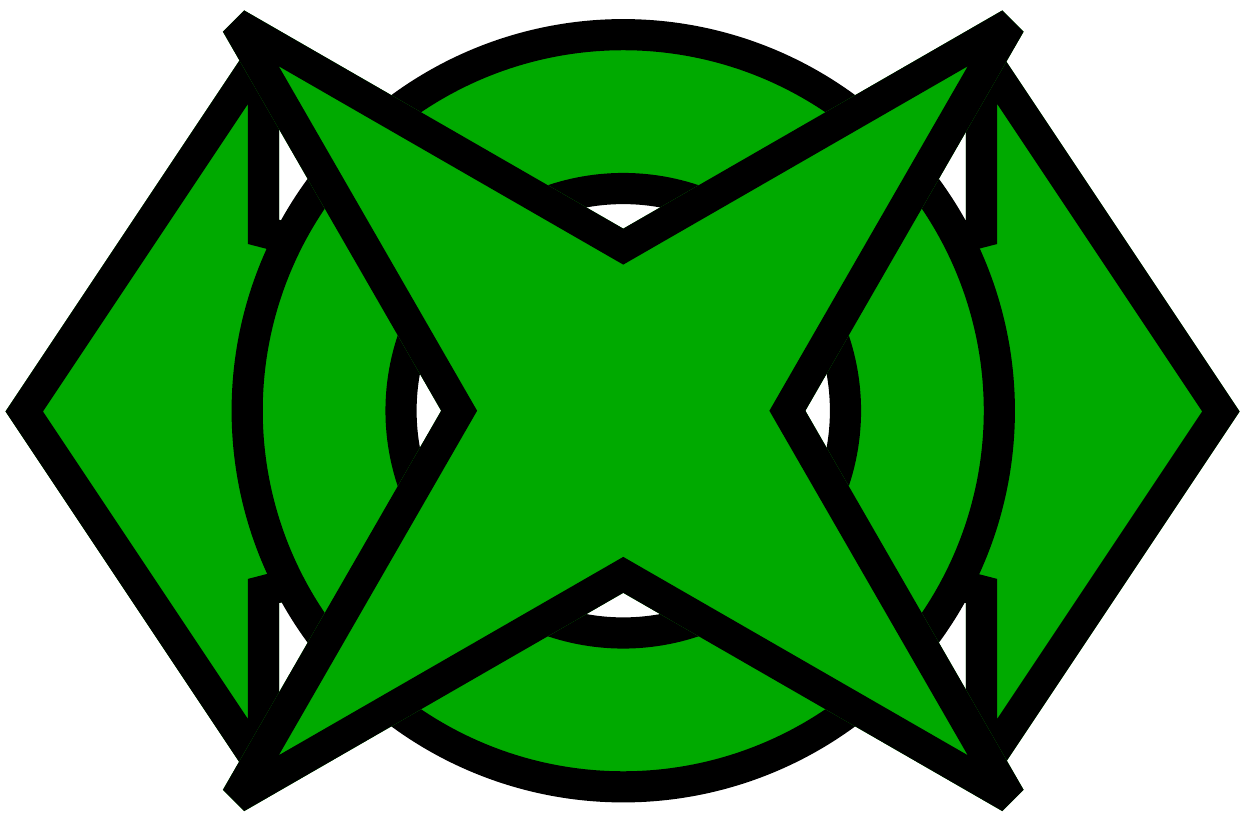} & \RTCRiMHDSfFb & MHD
& $3 \cdot 10^{-20}$ & \cmark & \cmark & MTT & Radiation + CRMagMech & Full physics; SN inject $E_\text{mag,SN}$ + $E_\text{CR,SN}$.\\
\hline
\end{tabular}
\end{table*}

\subsection{Comparable dwarf galaxies in observations} 
\label{ss:ObsComparison}
In order to aid in reviewing our results and providing some context for galaxy properties and relations, we include observational data in various figures. The observational data included features a broad population of low-mass and dwarf galaxies. While our simulated models may resemble the properties of many of these different systems across the multiple relations, we focus the comparison on those observed galaxies that have similar masses and are found in comparable environments.
Our simulated dwarf galaxy is an isolated system that forms in a relatively quiet environment. Consequently, we favour comparison with the most isolated dwarf galaxies in the Local Group, with similar halo masses ($M_\text{vir} (z = 0) \sim 10^{10} \Msun$) and within a comparable stellar mass range ($\Mst \sim 10^{6} - 5 \cdot 10^{7} \Msun$). The best analogues in the observational data are the dwarf galaxies WLM, LeoA, VV124 and SagDIG.
On the other hand, dwarf galaxies that have lower halo and stellar masses at $z = 0$ are candidates for reionization quenching of star formation. Some of these systems, such as LeoT ($M_\text{halo} (z = 0) \sim 5 \cdot 10^8 \Msun$) have only recently re-ignited their star formation \citep{Irwin2007, Rey2020}. As these correspond to a different category of dwarf galaxies undergoing different evolutionary processes, we avoid comparing our models with them.

\section{Results} 
\label{s:Results}

\subsection{Dwarf galaxy morphology with different ISM physics}

\def \thiswidth{0.42} 
\begin{figure*}
    \centering
    \includegraphics[width=2.1\columnwidth]{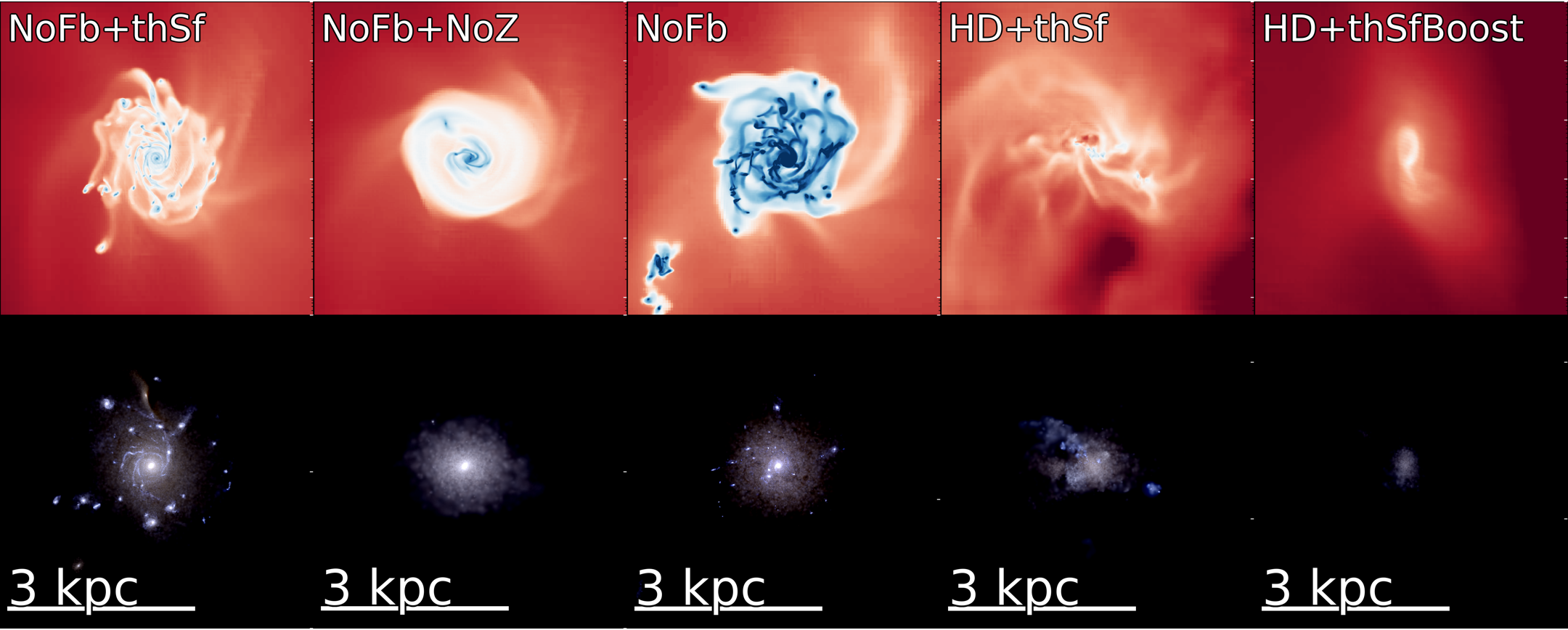}\\
    \includegraphics[width=2.1\columnwidth]{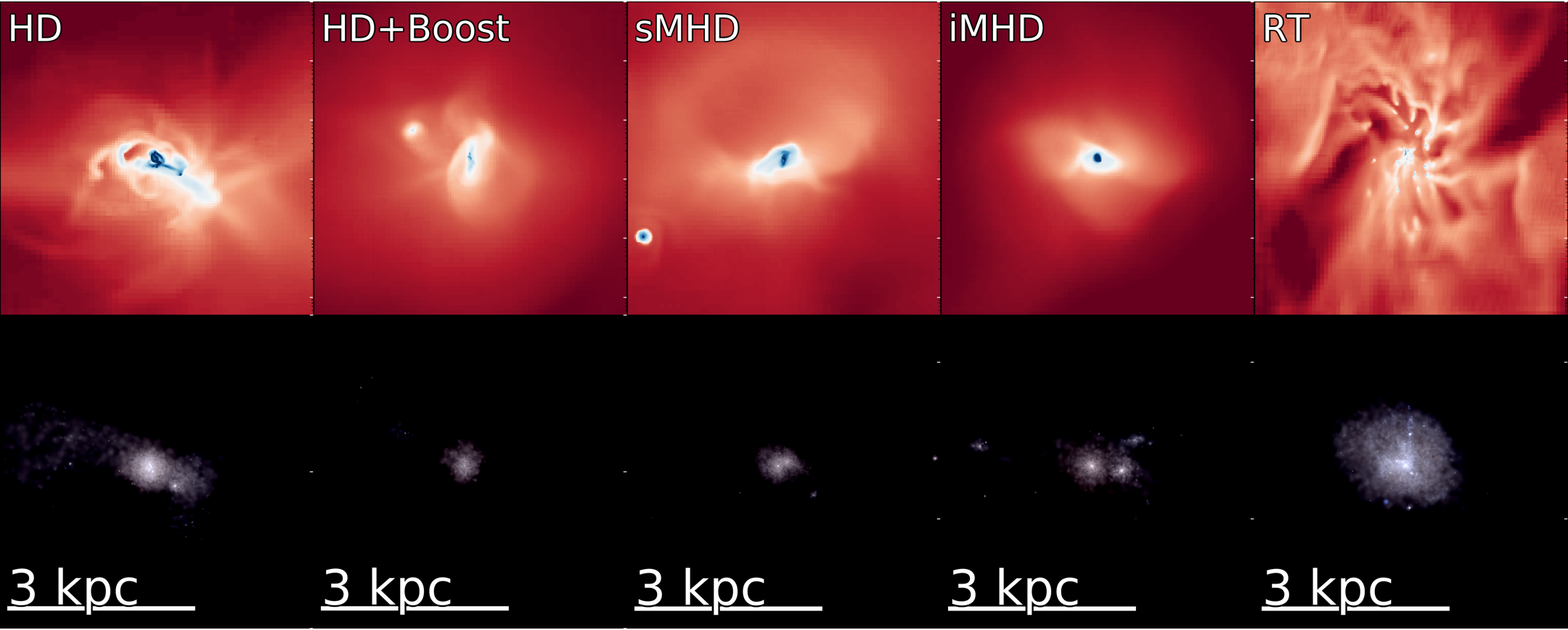}\\
    \includegraphics[width=2.1\columnwidth]{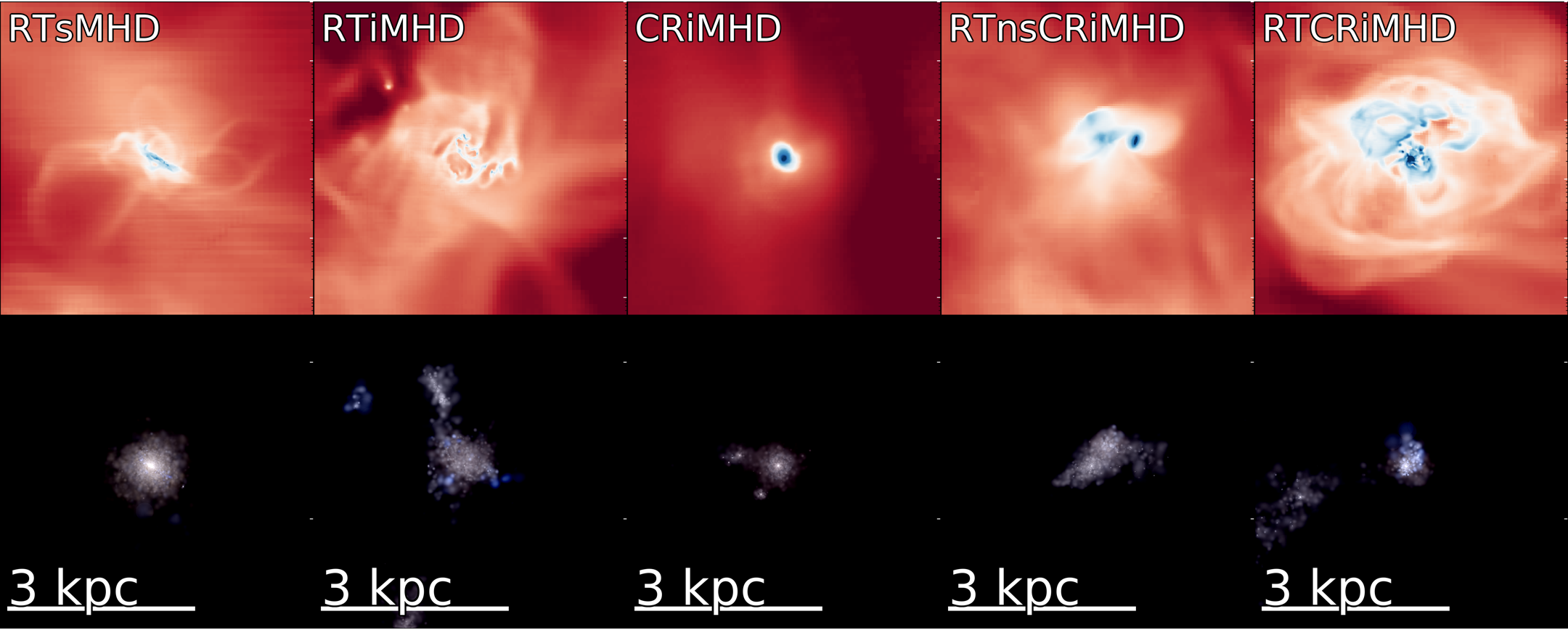}\\
    \caption{Face-on projection centred on the studied dwarf galaxy at $z = 3.5$ for different simulation models as indicated on the panels. Panels show in each column gas density (top) and a synthetic optical observation (bottom). The synthetic observations employ the SDSS [u,g,r] filters in the galaxy rest-frame, accounting for dust extinction but not convolved with any telescope PSF. We reduce the luminosity of the NoFb images by 1 dex to avoid saturated images.}
    \label{fig:Galaxies1Close}
\end{figure*}

We evolve all our simulations from $z = 127$ to $z = 3.5$. In Fig.~\ref{fig:Galaxies1Close} we present face-on views of the dwarf galaxy at this lower redshift for our most representative simulations. The panels are projections of $(5 \kpc)^3$ cubic boxes centred on the galaxy with their line-of-sight oriented along the total baryonic angular momentum of the system. For each simulation, the two rows represent gas density (top) and a synthetic optical RGB image for the SDSS [u,g,r] filters in rest frame (bottom). We assume each stellar particle to be a single stellar population, with emission as predicted by \citet{Bruzual2003}. All emission is integrated along the line-of-sight, accounting for dust obscuration by assuming a 3D absorption screen with an exponential extinction law of $R_v \sim 3.1$. We model the dust content of each cell employing a simple metal-to-dust ratio of $0.4$, found to be a reasonable approximation of a full post-processing radiative transfer treatment for this type of synthetic images by \citet{Kaviraj2017}. 

The physical processes included in the simulations and their different configurations significantly affect the properties and appearance of the galaxy. Focusing first on the simulations with no stellar feedback (top rows, 3 leftmost columns), they exhibit, unsurprisingly, extended gaseous discs and massive stellar components with an exceptionally pronounced stellar density peak (bulge) at their centre. Enabling the return of metals to the ISM has a substantial impact on the morphology of gaseous discs. They develop spiral structures and over-dense `knots', which correspond to the clearly visible stellar clusters in the optical emission, particularly prevalent in the simulation with the density threshold star formation model. By construction, the density threshold model allows for a greater fraction of cold, dense gas to become star forming, whereas with the MTT model, gas must reach significantly higher densities before becoming star forming. This leads to the density-threshold models converting a larger fraction of their gas into stars, consequently reaching higher stellar masses.

Including SN feedback (top rows, 2 rightmost columns and central rows, 2 leftmost columns) modifies the appearance of the dwarf galaxy considerably, becoming an irregular system with no clear disc. Furthermore, both the stellar and gas mass content is reduced, with the optical image showing a much fainter luminosity (note that we decreased the luminosity of the NoFb models in this figure by 1 dex to avoid image saturation). The galaxies produced with the density-threshold star formation model have notably more diffuse distribution of gas and stars, due to a more efficient impact from SN feedback. This is the result of gas being allowed to be converted into stars at earlier times in the density threshold models, as well as due to the formation of additional stars in diffuse regions. Hence, in the density threshold model, more SN events occur earlier than in the MTT model, and more events take place in regions of lower density, where SN feedback is more impactful. The simulations using the MTT model for star formation have a clumpier appearance, resembling local Universe dwarf irregulars as well as the more massive high redshift analogues observed by GOODS and GEMS \citep{Elmegreen2009b}. Finally, in addition to featuring even lower stellar masses and fainter luminosities, the boosted SN feedback simulations have depleted the system of the densest gas.

The remaining panels in Fig.~\ref{fig:Galaxies1Close} compare various of our simulations with additional physical processes. 
\if\dobullet1 \begin{itemize} \item {\bf Magnetic fields:} \fi
The simulation with a strong primordial magnetic field (central rows, central column, \sMHDSfFb) shows a higher central concentration of the baryonic component \citep{Martin-Alvarez2020} than the \HDSfFb~simulation. The presence of magnetic fields further produces a smoother gas structure of the ISM \citep{Kortgen2019, Martin-Alvarez2020}, also apparent in the \iMHDSfFb~simulation (central rows, fourth column). However, note that this latter simulation has had its star formation quenched for a considerable time, which is the aspect dominating its appearance.
\if\dobullet1 \item {\bf Radiative transfer:} \fi
The addition of radiative transfer (central rows, last column) generally leads to a higher fraction of gas in the circumgalactic medium (CGM). We will show later how radiative transfer leads to a higher gas mass being retained in the galaxy, as the photoheating due to the stellar radiation prevents a fraction of gas to form stars and leads to gentler SN feedback. The stellar component is similar to that in the \HDSfFb~simulation. However, the spatial distribution of the stars is more extended and diffuse as well as bluer due to some recent star formation. The combination of radiative transfer and magnetic fields (bottom rows, first two columns) is particularly interesting. We will explore the interplay of these two physical processes in more detail below. For now, we note two important aspects in these simulations. First, the appearance of the gas and stars is somewhere in between the MHD-only and radiative transfer-only runs. Second, it is worth noting that the simulation with the strong primordial magnetic field (\RTsMHDSfFb; bottom rows, first column) shows a dense concentration of stars, which we will later show corresponds to an over-massive and over-dense stellar component.
\if\dobullet1 \item {\bf Cosmic rays:} \fi
Finally, our cosmic ray simulations are shown in the last three columns of the bottom rows. The first of these three is the non-RT case, where we find an appearance similar to the \iMHDSfFb~simulation, but with a lower stellar mass and a more centrally concentrated gas-rich environment. We will later show that while both the \iMHDSfFb~and \CRiMHDSfFb~simulations follow similar star formation histories, the simulation with cosmic rays is nevertheless able to more efficiently eject gas and prevent star formation. Our most interesting simulations are the full-physics \RTnsCRiMHDSfFb~and \RTCRiMHDSfFb~simulations. These show a more significant gas reservoir as well as a denser CGM. The bottom rightmost panel in Fig.~\ref{fig:FrontPanel} illustrates how most of the gas in \RTCRiMHDSfFb~is outflowing. 
\if\dobullet1 \end{itemize} \fi

\subsection{The stellar mass to halo mass relation}
\label{ss:HMSM}
\begin{figure*}
    \centering
    \includegraphics[width=2.1\columnwidth]{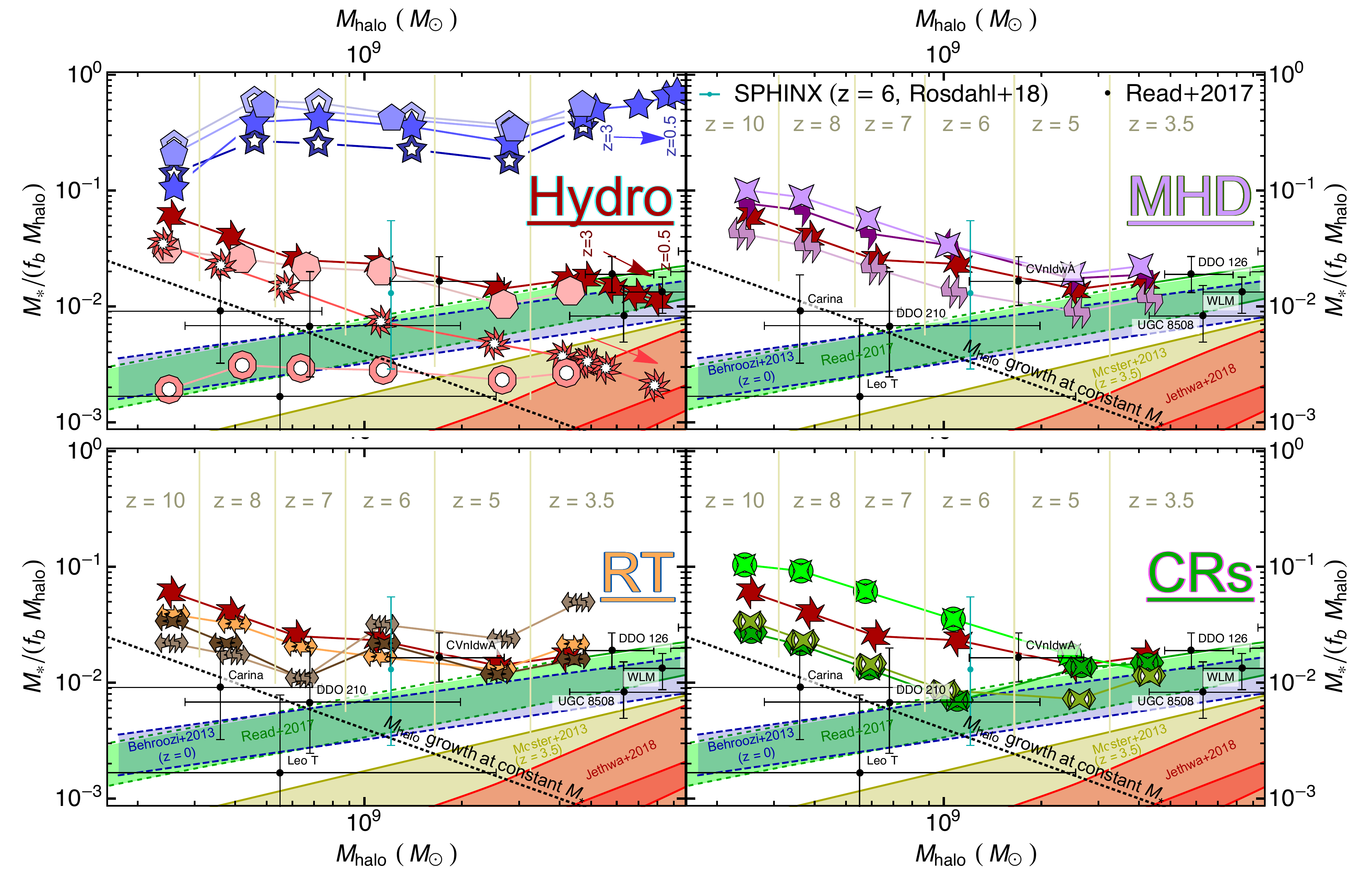}\\
    \DwarfBarLegend\\
    \caption{Baryonic conversion efficiency ($\Mst / (f_b M_\text{halo})$) versus halo mass $M_\text{halo}$ for all our simulations of the dwarf galaxy studied here. Simulations are separated into hydrodynamical simulations comparing star formation and SN feedback models (Hydro, top left), MHD (MHD, top right), radiative transfer with and without MHD (RT, bottom left) and CRMHD with and without radiative transfer (CRs, bottom right), with the \HDSfFb~model included in all panels as reference. For each simulation, we show the conversion efficiency at several redshifts. For the \HDSfNoFb, \HDSfFb, and \HDSfFbBoost~simulations, we include four additional evolution markers corresponding to redshifts $z =$ 0.5, 1, 2, 3 moving from rightmost towards the left of the panel. These illustrate how simulations including SN feedback have negligible star formation between $z = 3.5$ and $z = 0.5$. The blue band shows an extrapolation of the baryonic conversion efficiency predicted by \citet{Behroozi2013} for $z = 0$, the yellow band indicates the $1\sigma$ prediction by \citet{Moster2013} at $z = 3.5$ and the green band corresponds to the abundance matching prediction from \citet{Read2017}. We also include the stellar mass-halo mass relation predicted by \citet{Jethwa2018} for the Milky Way satellites population as the red band in the bottom right corner. We provide additional details regarding their comparison and about additional relations at the end of Section~\ref{ss:HMSM}. The dotted black line indicates an evolutionary track assuming a constant $\Mst$. We include as the cyan data point the result from the {\sc sphinx} simulation \citep{Rosdahl2017} for $M_\text{halo} \sim 1.5 \cdot 10^9 \Msun$ at $z = 6$. Black data points show estimates for LeoT, Carina and isolated dwarf galaxies in the LG \citep{Read2017}. Baryonic conversion efficiency changes considerably depending on the adopted physical models. The `full-physics' simulations at $z = 3.5$ provide the best match to LG dwarf galaxies observed at $z = 0$, the {\sc sphinx} population and evolve along the scalings of \citet{Behroozi2013} and \citet{Read2017} (both for $z = 0$) after $z \sim 6$.}
    \label{fig:HMSM}
\end{figure*}

One of the most studied relations in galaxy formation is the stellar mass $\Mst$ to halo mass $M_\text{halo}$ relation. A physically insightful form of this relation employs the baryonic conversion ratio, $\Mst / (f_b M_\text{halo})$, versus $M_\text{halo}$, where $f_b = \Omega_\text{b} / \Omega_\text{m}$ is the baryon fraction, and $\Omega_\text{b}$, $\Omega_\text{m}$ are the cosmological baryon and matter density parameters, respectively. This form assigns to a specific halo mass an efficiency for the conversion of its baryonic mass into stellar mass. In Fig.~\ref{fig:HMSM} we explore the evolution of the baryonic conversion efficiency for each simulation of our dwarf galaxy. 

In order to contextualise the differences across simulations, we include shaded bands corresponding to various abundance matching relations. As many of these relations only reach halo mass limits as low as $M_\text{halo} \sim 10^{10} \Msun$ we often resort to extrapolating them into the studied regime. Consequently, such extrapolations have to be examined with care. For example, the extrapolation of the \citet{Behroozi2013} and \citet{Read2017} relations predict considerably higher stellar masses than a variety of other models such as \citet{Garrison-Kimmel2014} or \citet{Moster2018}. We include models with stellar masses in between these two regimes, by showing the predictions by \citet{Moster2013} and \citet{Jethwa2018}. Models predicting stellar masses below the presented range \citep[e.g.,][]{Garrison-Kimmel2014, Moster2018} are not shown in Fig.~\ref{fig:HMSM} in order to facilitate visual examination of the differences across our models.

Furthermore, we compare our galaxy with the population of galaxies of similar halo masses ($M_\text{halo} (z = 6) \sim 1.5 \cdot 10^{9} \Msun$) in the {\sc sphinx} simulation (see cyan data point). {\sc sphinx} is a high-resolution galaxy formation simulation including on-the-fly radiative transfer that reproduces well the reionization history of our Universe. This comparison at $z = 6$ may provide us with some insight regarding how our different models compare to those that reproduce observed reionization constraints. Finally, we include in the same figure observational data from \citet{Read2017} as black points for LeoT, Carina and various isolated LG dwarf galaxies.  We first discuss the general evolution of Pandora dwarfs in our no-feedback and SN-only simulations, and then move onto the discussion of the simulations with additional physics.

In the absence of stellar feedback, we find baryonic conversion ratios to be too large ($> 0.1$). Out of the two star formation prescriptions studied, the MTT model yields somewhat lower albeit still unrealistically large values. The return of metals to the ISM by stars seems to only significantly increase the stellar mass of the MTT star formation prescription model. As expected, accounting for SN feedback dramatically reduces the stellar mass of our dwarf galaxy at all redshifts explored. While the effect is comparatively minor shortly after the onset of star formation ($z \sim 10$), SN feedback leads to a large reduction in stellar mass at lower redshifts. In fact, at $z \sim 4$, the \HDSfFb~simulation reaches a conversion ratio of $\sim\!\!2\%$, comparable with the extrapolation of the \citet{Behroozi2013} model to this halo mass. The simulations with standard SN feedback show no large differences between the density threshold and our MTT star formation implementations in terms of the baryonic conversion efficiency. With the classical density threshold star formation implementation, SN feedback has a slightly higher impact. We attribute this to SN events taking place at lower densities and thus providing a more efficient deposition of momentum \citep{Iffrig2015}. In practice, star formation in the MTT simulations typically occurs at densities $\rho_\text{gas} \gtrsim 10^{-21} \gcc$, whereas in the density threshold case, we allow star formation only when $\rho_\text{gas} > \rho_\text{th} \sim 10^{-23} \gcc$. Furthermore, in the density threshold simulations, the earlier onset of star formation suppresses the initial collapse of the gas. However, most gas is re-accreted later on and the simulation reaches baryonic conversion efficiency comparable to \HDSfFb.

A common way of further reducing the stellar mass of simulated galaxies is to boost the strength of SN feedback (motivated by considering that other sources of stellar feedback such as binary stars, hypernovae, top heavy IMF, etc. have been neglected). This is what we do in our \HDSfFbBoost~simulation. This simulation has, as expected, a lower baryonic conversion efficiency that continues to decrease until it reaches $\sim 0.4\%$ by $z = 3.5$. Here, the expulsion of gas is much more efficient and the galaxy maintains a significantly lower stellar mass. We further note that the simulation with boosted SN feedback and density threshold for star formation has the most extreme effect in reducing the stellar mass, due to the powerful SNe exploding in relatively low density environments, to a level that is likely unrealistic. Unfortunately, the high computational cost of the simulations including more complex physical processes impedes evolving all our simulations to $z \sim 0$. However, we extend some of our hydrodynamical simulations to $z = 0.5$ to demonstrate that our dwarf galaxy and its halo do not undergo significant mass evolution after $z \sim 4$. In particular, the halo mass evolves from $M_\text{halo} (z = 3.5) \sim 4 \cdot 10^9 \Msun$ to $M_\text{halo} (z = 0.5) \sim 8 \cdot 10^9 \Msun$, but most importantly, the stellar mass is barely changed, evolving from $\Mst (z = 3.5) \sim 1.0 \cdot 10^7 \Msun$ to $\Mst (z = 0.5) \sim 1.3 \cdot 10^7 \Msun$ for \HDSfFb, and $\Mst (z = 3.5) \sim 2.2 \cdot 10^6 \Msun$ to $\Mst (z = 0.5) \sim 2.3 \cdot 10^6 \Msun$ for \HDSfFbBoost. Consequently, our dwarf galaxy is efficiently quenched by $z \sim 3.5$. We therefore expect most of our results at $z \sim 3.5$ to remain generally valid at later times, facilitating observational comparison with local dwarf galaxies.

\if\dobullet1 \begin{itemize} \item {\bf Magnetic fields:} \fi
Although the simulations including magnetic fields look distinct in Fig.~\ref{fig:Galaxies1Close}, by $z = 4$ their baryonic conversion efficiency is approximately equal to that of our 'default' SN feedback simulations, and only slightly lower for \sMHDSfFb. The time evolution of the simulations with magnetic fields shows some minor differences, which are most notable at early times. In agreement with previous work, we find magnetic pressure delays the onset of star formation \citep[occurring approximately at $z \gtrsim 14$;][]{Martin-Alvarez2020, Koh2021}, where the small-scale magnetic pressure support is also accounted for by our MTT star formation prescription \citep[see appendix B in][]{Martin-Alvarez2020}. The stellar mass attained by $z \sim 10$ depends thereby on our setup for the magnetic field. For the intermediate and injected magnetic field simulations (\iMHDSfFb~and \MHDSfFb), the additional magnetic support and star formation delay early on leads to a higher amount of gas accumulating in the galaxy. In the absence of any significant early stellar feedback due to, for example, stellar radiation or stellar winds, star formation proceeds unimpeded until a significant number of SN events take place. This leads to a higher stellar mass by $z \sim 10$. Contrarily, for our simulation with a strong primordial magnetic field (\sMHDSfFb) we find an initial reduction of the stellar mass. In this simulation, star formation is delayed even further, allowing the first SNe to commence regulating star formation. This sensitivity to the assumed initial magnetic field makes relic dwarf galaxies formed at extremely high redshifts a promising observational window to primordial magnetic fields. 

\if\dobullet1 \item {\bf Radiative transfer:} \fi
The inclusion of stellar radiation and its immediate feedback in the aftermath of star formation results in a small reduction of the stellar mass at $z \sim 10$ for all our simulations with radiative transfer. In the no-magnetic field model (\RTSfFb), this initial suppression does not significantly reduce the stellar mass at lower redshifts, eventually leading to a slightly higher stellar mass by $z = 3.5$. Such positive feedback due to stellar radiation has already been suggested previously by e.g. \citet{Smith2021}. This is caused by a less efficient expulsion of gas from the dwarf galaxy when stellar radiation is included in our simulations.

The combination of radiative transfer and magnetic fields is particularly relevant for star formation: magnetic fields affect the efficiency and spatial distribution of star formation whereas radiation regulates its local suppression through the evaporation of molecular clouds. The {\sc sphinx-mhd} simulation \citep{KMA2021} pioneered the study of this combination of physical processes in high-resolution cosmological simulations, investigating its effects on reionization and as well as the on the properties of the simulated population of galaxies at the end of cosmic dawn ($z \gtrsim 6$). One of the results found by \citet{KMA2021} was that magnetic fields can affect the surface brightness of galaxies. Here we delve deeper into exploring this combination of magnetic fields and radiation, although for a single dwarf galaxy. 

Our simulations with radiative transfer and magnetic fields show a stronger suppression of early star formation than the \RTSfFb~simulation. This is particularly notable for the \RTsMHDSfFb~simulation, which shows the delayed onset and suppression of star formation also observed in \sMHDSfFb. This is expected, as the effects of radiation will not emerge until the formation of the first stars in our simulation. Interestingly, the suppression of the baryonic conversion efficiency in the \RTiMHDSfFb~and \RTsMHDSfFb~simulations is comparable to that of the \HDSfFbBoost~simulation until $z \sim 7$. However, they display a different time evolution at lower redshifts. This could be due to the more concentrated star formation in the presence of strong magnetic fields \citep{Martin-Alvarez2020} leading to higher stellar radiation fluxes. However, this does not lead to an irreversible expulsion of gas from the system. Accretion around $z \sim 7 - 8$ triggers an important peak of star formation at $z \sim 6$. This burst of star formation drives the stellar masses of the \RTiMHDSfFb~and \RTsMHDSfFb~simulations to values comparable to the fiducial \HDSfFb~simulation. These simulations display an additional star formation burst at $z \sim 5$. The positive feedback of the stellar radiation observed in the \RTSfFb~simulation disappears in the \RTiMHDSfFb~model, where $\sim\!\!\muG$ magnetic fields are present in the galaxy from $z \sim 10$. On the other hand, the dwarf galaxy in the \RTsMHDSfFb~simulation has an even higher final stellar mass, most likely due to the high magnetic pressure in the CGM further confining outflows and preventing outflowing gas from escaping the halo.

\if\dobullet1 \item {\bf Cosmic rays:} \fi
In our simulation with cosmic rays and magnetic fields \CRiMHDSfFb, cosmic rays only have a secondary effect at the onset of galaxy formation, behaving similarly to its non-cosmic ray counterpart (\iMHDSfFb), thus reaching a higher initial stellar mass than \HDSfFb~due to the early effects of magnetic fields at $z \gtrsim 10$. The significant effects appear as cosmic ray energy density builds up and becomes comparable to the thermal energy density. As SN feedback events are the exclusive source of cosmic rays in our simulations, their impact is only found at later times. Indeed, from $z \sim 8$ star formation is quenched in the \CRiMHDSfFb~simulation, and its final stellar mass ends up below the \iMHDSfFb~model and slightly below that of the \HDSfFb~simulation. 

Our most interesting models are the two `full-physics' simulations, \RTCRiMHDSfFb~and \RTnsCRiMHDSfFb. Both include magnetic fields, radiative transfer, and cosmic rays, with their only difference being whether they account for cosmic ray streaming. The addition of cosmic rays to radiation and magnetic fields further reduces the early gas content. The star formation peak observed in \RTiMHDSfFb~and \RTsMHDSfFb~occurs at $z \sim 6.3$ in these simulations. \RTCRiMHDSfFb~and \RTnsCRiMHDSfFb~maintain instead a low baryonic conversion efficiency at $z \gtrsim 5$, (coincidentally) comparable to that of our \HDSfFbBoost~simulation. Indeed, \RTCRiMHDSfFb~displays the lowest stellar mass across all our MTT-star formation simulations, even lower than \HDSfFbBoost. The \HDSfFbBoost~simulation employs a comparable feedback boost to that used by the {\sc sphinx} simulation. This may suggest that the inclusion of cosmic rays in this simulation may lead to the same halo mass-stellar mass relation and the same reionization history at $z \gtrsim 5$, in agreement with observations without requiring the calibration boost of SN feedback. On the other hand, \citet{Farcy2022} find the inclusion of cosmic rays to reduce the escape fraction of ionizing radiation in isolated galaxies. Furthermore, they also find cosmic rays generate a smaller stellar mass reduction than their calibrated SN feedback, with a similar boost to the one employed here. The main differences between our setups are the inclusion of radiative transfer in their boosted SN feedback model (as done in {\sc sphinx}), and the fact that the Pandora simulations are evolved in a cosmological context whereas the galaxies simulated by \citet{Farcy2022} are isolated. Consequently, it will be important to revisit the impact of cosmic rays on reionization using {\sc sphinx}-like simulations.

We note that \RTnsCRiMHDSfFb~and \RTCRiMHDSfFb~follow distinct evolutionary tracks in the time interval from $z = 6$ to $z = 4$. After $z \sim 5$, our galaxy commences its final merger with a neighbouring dwarf galaxy. In the \RTnsCRiMHDSfFb~simulation, the dwarf galaxy star formation and associated SN feedback heats the neighbouring dwarf galaxy, preventing significant star formation inside it until both systems merge. On the contrary, SN feedback in \RTCRiMHDSfFb~occurs later, such that the star formation persists in the neighbouring dwarf galaxy for longer. This leads to the higher stellar mass in the \RTCRiMHDSfFb~simulation at around $z \sim 5$. However, both models converge to a similar stellar mass at $z \sim 3.5$.

Finally, we note that in the presence of cosmic rays, the positive feedback effect from stellar radiation \citep{Smith2021} in our \RTSfFb~simulation disappears, and in fact the stellar mass is reduced by $\sim$20\% compared to that in the \HDSfFb~simulation. We will later discuss the interplay of these two physical effects, as the efficiency of cosmic rays in ejecting gas is affected by radiative transfer.
\if\dobullet1 \end{itemize} \fi

The analysis of the stellar mass growth in all our simulations reveals that, due to their shallow potential, dwarf galaxies are extremely sensitive to the inclusion of different physical processes that affect their baryonic conversion efficiency. It is encouraging that a subset of `full-physics' simulations converge to a similar stellar mass when the galaxy is ultimately quenched. However, we stress that all these models assume the same star formation and SN feedback model which may (in part) lead to the apparent final stellar mass `convergence'.

We conclude this section by comparing our simulations with the predictions by \citet{Behroozi2013}, \citet{Moster2013}  and \citet{Jethwa2018}, as well as the observations by \citet{Read2017}. Firstly, we note that these relations (except \citealt{Moster2013}) correspond to $z = 0$ galaxies, whereas our simulations, due to their expensive computational cost, only reach $z \sim 3.5$. Secondly, abundance matching relations are generally uncertain and subject to multiple caveats. In particular, the relations by \citet{Behroozi2013} and \citet{Read2017} yield higher stellar mass per halo mass. They have also been argued to be too high when considering reionization constraints \citep{Graus2019} and observational incompleteness \citep{Garrison-Kimmel2014}. Finally, we also note that the relation obtained by \citet{Jethwa2018} is for satellite galaxies, whereas our simulation corresponds to an isolated system. With these considerations in mind, we find all our simulations with MTT star formation and standard SN feedback have $\Mst (z = 3.5) \sim 10^7 \Msun$, whereas those with boosted SN feedback have $\Mst (z = 3.5) \sim 2\cdot10^6 \Msun$. These boosted SN feedback simulations have a stellar mass that may be too low when compared with both our extrapolation of the \citet{Behroozi2013} scaling relation and the estimates for LG dwarfs, but in better agreement by those predicted by \citet{Jethwa2018} from Milky Way satellites (note that our simulated galaxy is a field dwarf). Nonetheless, we caution that there are still significant observational uncertainties and large data scatter in this low galaxy mass regime preventing us to draw any firm conclusions. Contrarily, our simulations with standard SN feedback only appear to provide a good match to both abundance matching and observations. Furthermore, due to its positive feedback, \RTSfFb~has a stellar mass that may be somewhat high for this halo mass range. \RTsMHDSfFb~appears to be in tension with both observations and predictions \citep{Behroozi2013, Jethwa2018}. We attribute this to the extremely strong primordial magnetic field used. While our results based on a single simulated galaxy should be taken with caution, we note that $B_0 < 3 \cdot 10^{-11}$~G leads to more reasonable stellar masses. Finally, note that our \RTCRiMHDSfFb~and \RTnsCRiMHDSfFb~simulations provide the best match to comparable observed galaxies by \citet{Read2017} for all halo masses.

While reproducing the observational estimates of the $\Mst / (f_b M_\text{halo})$ vs $M_\text{halo}$ relation is an important test, this is not a sufficient condition to appropriately simulate dwarf galaxies as many other of their properties may be unrealistic. We proceed to review how these other properties are affected in the following sections.

\subsection{Dynamical masses and sizes of stars and gas}
\label{ss:sizes}

\begin{figure*}
    \centering
    \includegraphics[width=2.1\columnwidth]{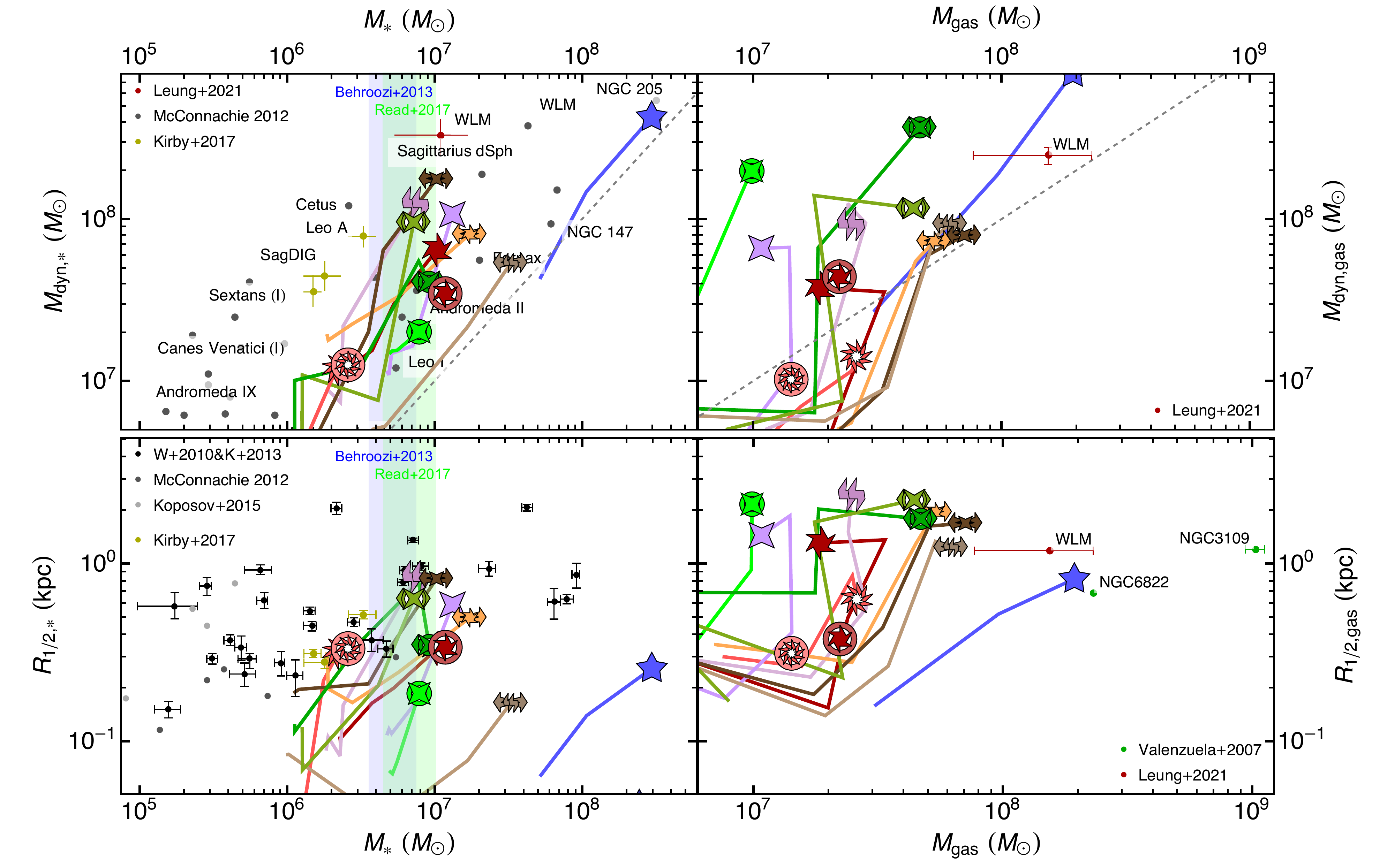}\\
    \DwarfBarLegendFrom{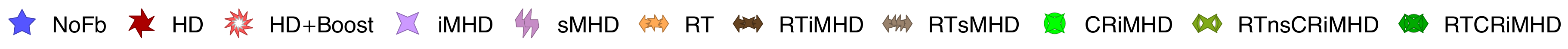}\\
    \caption{Dynamical mass (top row) and half-mass radius (bottom row) as a function of stellar mass (left column) and gas mass (right column). Dynamical masses are estimated from stars (left) and gas (right), and computed according to Equation~(\ref{eq:Mdyn}). Data from different observations at $z = 0$ is included for comparison (see legends). In order to aid the comparison of our $z \sim 3.5$ results with observations, we include $z = 0.5$ values where available as encircled symbols of the corresponding models. Dashed gray lines (top row) correspond to the one-to-one relation between dynamical and component mass. On the left column, vertical shaded bands indicate the stellar mass range predicted for the simulated galaxy halo mass ($M_\text{halo} (z = 3.5) \sim 4.1 \cdot 10^{9}\,\Msun$) according to \citet[shown in blue]{Behroozi2013} and \citet[shown in green]{Read2017}. We only show a subset of our simulations: the representative models that differ significantly from the \HDSfFb~run, with segmented lines representing their evolution from redshift $z = 7$ down to $z = 3.5$ (see text for more details). Additional physical processes affect non-linearly the dynamical mass and galaxy size, with most notable `outliers' being runs with strong primordial field (both with and without radiation) and boosted SN simulations. In general, simulations with radiation lead to more gas rich and extended galaxies. Overall, a number of simulations provide a good match to recent observations of isolated dwarfs, which are reasonable analogues to our simulated system.}
    \label{fig:StellarMassStellarRadius}
\end{figure*}

In this section we review how our different physical models affect the kinematics and morphology of our simulated galaxy and compare with the corresponding properties of observed galaxies. The panels of Fig.~\ref{fig:StellarMassStellarRadius} show the dynamical mass (top row) and half-mass radius (bottom row) as a function of stellar (left) and gas mass (right) for a representative subset of our simulations that differ significantly from the \HDSfFb~simulation. We include evolutionary tracks as segmented lines, sampling the median values of each simulation during $z = (7, 6, 5) \pm 0.5$ and $z \in [3.4,3.7]$, with symbols shown for the final redshift interval. Note that the dynamical mass is more accessible observationally than the `true' halo mass. The dynamical mass is calculated separately for the stars and gas as follows
\begin{equation}
    M_\text{dyn,i} = \frac{\sigma_{\mathrm v, \text{i}}^2 (2 R_\text{1/2,i})}{G}\,,
    \label{eq:Mdyn}
\end{equation}
where $\sigma_{\mathrm v, \text{i}}$ is the mass-weighted 3D velocity dispersion and $R_\text{1/2,i}$ is the half-mass radius of either stars or gas (with i$=$`*' or `gas', respectively). We limit the radial extent of all our mass measurements as well as the computation of $\sigma_{\mathrm v, \text{i}}$ to twice the half-mass radius of either stars or gas. The half-mass radius is in turn obtained for each component employing the mass within a 0.2 $r_\text{halo}$ sphere centred on the galaxy.

The $\Mdynst$ vs $\Mst$ relation shows the expected correlation, with our simulations clustering in three main groups. These are in order of decreasing dynamical mass: no feedback simulations, standard SN feedback, and boosted SN feedback. Amongst the simulations without the boosted SN feedback, the inclusion of magnetic fields appears to have a relatively small effect, possibly driving a slight increase of $\Mdynst$ especially in the case with strong primordial magnetic fields. This is likely the consequence of a higher central density concentration, most pronounced for the \sMHDSfFb~simulation. Accounting for radiative stellar feedback generally increases the dynamical mass, and results in larger sizes and a larger baryonic mass. The most interesting effect is that of cosmic rays, which decrease the dynamical mass. Unsurprisingly, simulations with boosted SN feedback have a lower stellar mass as well as a lower dynamical mass. As \HDSfFb~and \HDSfFbBoost~evolve to $z = 0.5$, they undergo a negligible increase of $\Mst$ and a small change in $\Mdynst$, which appears to 
 be set by the total mass of the galaxy (see Section~\ref{ss:compareKinematics}).

We compare our simulations with the large sample of observations compiled by \citet{McConnachie2012}. Note that several of these observed systems correspond to satellites, they may have undergone significant dark matter-stripping while preserving their stellar mass \citep{Penarrubia2008}. On the other hand, our simulated dwarf galaxy is evolving in a `field' environment without experiencing any substantial stripping. We also include data for Leo A, SagDIG and Aquarius by \citet{Kirby2017}, which may represent the closest present-day analogues of our simulated galaxy being also isolated systems. We use the mass model for WLM presented by \citet{Leung2021} to estimate a dynamical mass for that galaxy.  An important caveat regarding our comparison with observations throughout this Section is that our main simulated models only reach $z \sim 3.5$, whereas the observations correspond to $z \sim 0$. Therefore to further reinforce our analysis, we include all simulation models that reach $z = 0.5$. These are labelled in Fig. \ref{fig:StellarMassStellarRadius} and the rest of the manuscript as encircled versions of the corresponding model, and show only small displacements of the dynamical masses. Overall, most of our simulations provide a good match to the observations within their considerable scatter. Bearing in mind that WLM and LeoA are the systems that best resemble the Pandora dwarf, the \CRiMHDSfFb~and boosted SN feedback simulations show the largest tension with this data.

The $\Rhalfst$ versus $\Mst$ plot shows that the sizes of our simulated galaxy are in good agreement with the broad distribution of the observations \citep{Wolf2010, McConnachie2012, Kirby2013, Koposov2015}\footnote{Note that here observational data shows the half-light radius rather than the half-mass radius we measure in our simulations, and that observational measurements are projected.} for most of our simulations with SN feedback. Despite a large stellar mass reduction, boosting SN feedback only has a minor impact on the stellar half-mass radius. Both for \HDSfFb~and \HDSfFbBoost~ we find negligible changes to their stellar half-mass radii when considering their evolution down to $z \sim 0.5$. While inclusion of magnetic fields tends to slightly increase our $\Rhalfst$ compared with their non-MHD counterparts, for stronger primordial fields than those explored here, Sanati et al. (in prep) find a trend towards a mild size increase with increasing $B_0$, followed by a considerable shrinkage. However, when including radiation with our extreme primordial magnetic field (\RTsMHDSfFb), we find the most considerable shrinkage. We attribute this to the combination of a lower efficiency of SN feedback due to stellar radiation (see also \citealt{Rosdahl2015b, Smith2021}), and outflow confinement by the strong CGM magnetic pressure generated by the primordial fields. For the other models, the inclusion of radiation leads to higher $\Rhalfst$ values, due to flattened and more extended stellar density radial profiles. We attribute these to a more extended distribution of the gas, as discussed later on in this section. 

Furthermore, cosmic rays somewhat reduce $\Rhalfst$, particularly in the absence of radiation (\CRiMHDSfFb). This is due to the non-thermal ISM support from cosmic rays \citep{Dashyan2020}, which inhibits the number and mass of star forming clumps \citep{Farcy2022}. With less star forming clumps where gravity dominates over the gas support at large distances from the galaxy centre, the galaxy features a lower $\Rhalfst$ measurement. Such size reduction has also been reported for simulations of larger galaxy masses \citep{Buck2020}. The reduction of the stellar half-mass radius appears less significant for the simulations combining radiation and cosmic rays, with sizes comparable to that of \RTiMHDSfFb. \RTCRiMHDSfFb~exhibits a decrease of $\Rhalfst$ at $z \sim 3.5$, which we attribute to its recent merger event\footnote{Note that this merger event takes place at slightly different times in each of our models, featuring also somewhat different mass ratios and impact parameters.} (see Section~\ref{ss:CuspCore}).

The right column of Fig.~\ref{fig:StellarMassStellarRadius} reviews the same relations for the gas component. We compare our dynamical masses with an observational estimate generated using the WLM mass profiles from \citet{Leung2021}. Our simulations cluster in three groups. From higher to lower gas mass content these are: no feedback, stellar radiation and SN feedback runs. While, as expected, no feedback yields unrealistic $\Mgas$ and $\Mdyngas$ values for this halo, the inclusion of SN feedback drives our simulations to the lower values of the two quantities. In particular, the boosted feedback simulation has the lowest $\Mdyngas$ while maintaining a comparable gas content to the fiducial \HDSfFb~case. As we evolve \HDSfFb~and \HDSfFbBoost~to $z = 0.5$, their dynamical masses measured through the gaseous component remain relatively unchanged, and their gas masses remain around $\Mgas \sim 2 \cdot 10^7 \Msun$. Simulations with magnetic fields do not significantly affect the gas mass content of our galaxies, but increase dynamical mass estimates based on the gas, consistent with higher concentrations \citep{Martin-Alvarez2020}. Including stellar radiation leads to a consistent increase of gas mass in the studied galaxy by $\sim\!0.3 - 0.5$~dex, in agreement with predictions from isolated galaxy simulations \citep{Smith2021}, but maintains the same approximate $\Mdyngas/\Mgas$ ratio as the SN feedback simulations. Cosmic ray simulations have a mild reduction in gas mass, which we attribute to cosmic ray-driven galactic outflows. Additionally, the \RTCRiMHDSfFb~simulation shows an increased dynamical mass estimate from the gas dynamics at $z = 3.5$, likely due to a recent merger.

The gas mass-size relation shows that the boosted SN feedback produces the smallest sizes. This is likely due to the small amount of remaining gas, as the majority of gas is effectively removed from the system by the SN feedback already at high redshift. As the \HDSfFb~and \HDSfFbBoost~simulations evolve to $z \sim 0.5$, their gaseous extent decreases in size due to a reduction in the frequency of SN feedback. On the other hand, stellar radiation leads to a slightly more extended gas distribution. Due to their additional non-thermal support of the ISM, the inclusion of cosmic rays also increases $\Rhalfgas$. We include data by \citet{Valenzuela2007} and \citet{Leung2021} for comparison. All our simulations with stellar feedback have somewhat lower gas contents than observed galaxies with the same gas half-mass radius, favouring our models with radiative transfer. We note however that the stellar masses of some of the observed systems we compare to \citep{Valenzuela2007} are higher than that of our simulated dwarf galaxy\footnote{Masses are $\Mst \sim 7.6 \cdot 10^7 \Msun$ for NGC6822 \citep{Read2017}) and $\Mst \sim 7.6 \cdot 10^7 \Msun$ for NGC3109 \citep{McConnachie2012}, while our dwarf galaxy has $\Mst \sim 10^7 \Msun$.}, and that our system is not evolved to $z = 0$.

\subsection{Integrated and resolved kinematics of stars and gas}
\label{ss:kinematics}
\subsubsection{Stellar kinematics}
\begin{figure}
    \centering
    \includegraphics[width=\columnwidth]{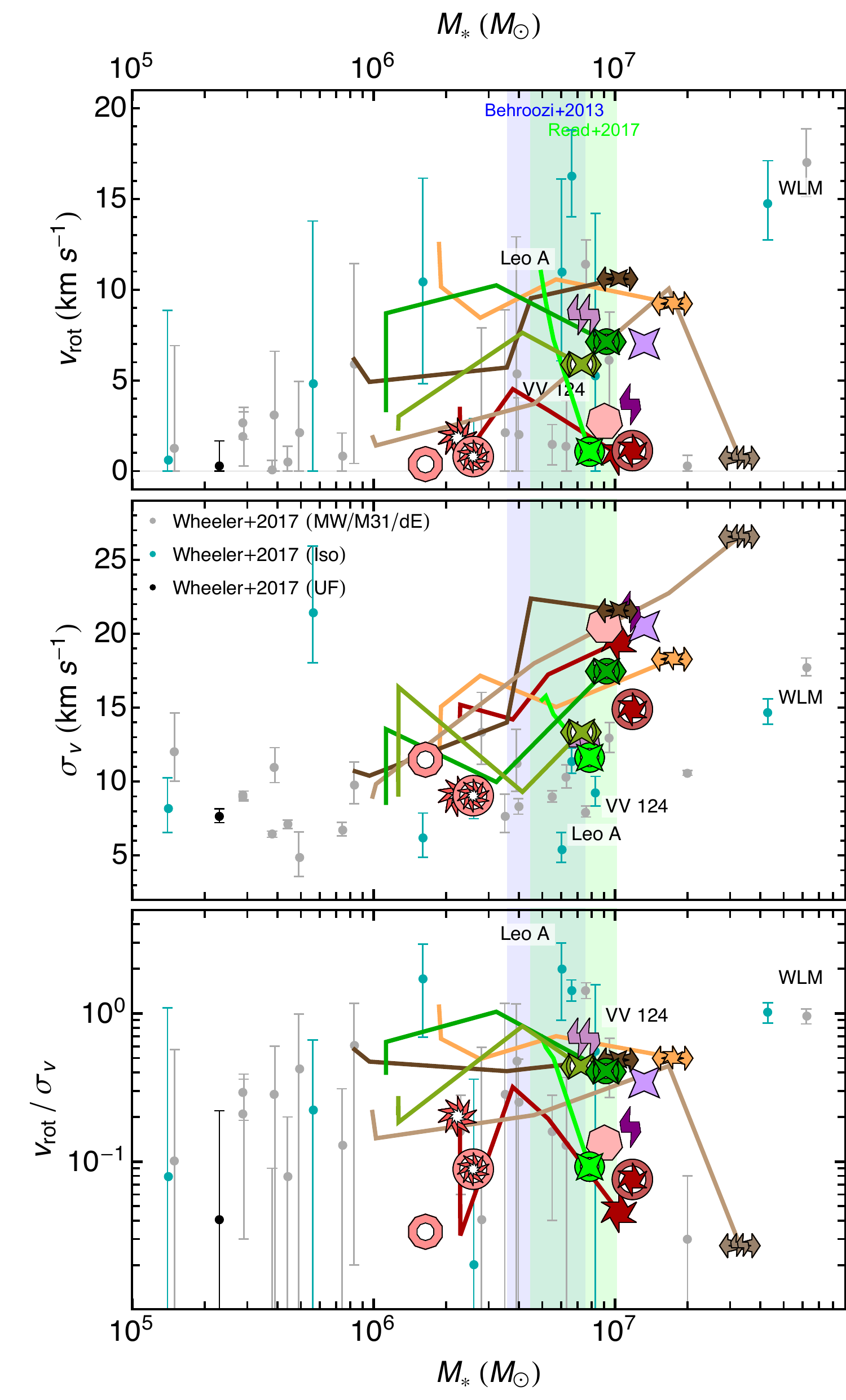}\\
    \HalfDwarfBarFrom{Images/DwarfLegendHalf.pdf}\\
    \caption{Integrated stellar kinematic properties versus $\Mst$ for our simulated galaxy at $z \sim 3.5$, compared against available observations of local dwarf galaxies \citep{Wheeler2017} at $z = 0$. From top to bottom, panels show mass-weighted rotational velocity $v_\text{rot}$, velocity dispersion $\sv{*}$ and $v_\text{rot}/\sv{*}$, all measured within $2 \Rhalfst$. Vertical shaded bands show, as in Fig.~\ref{fig:StellarMassStellarRadius}, the expected stellar mass range for our simulated dwarf galaxy based on abundance matching. We only show evolution tracks for some representative models, displaying the evolution from $z = 7$ until $z = 3.5$. Most of our simulations attain $\sv{*}$ values generally located at the upper end of the observations distribution or above it. Amongst them, our models with cosmic rays have the lowest values of $\sv{*}$ at a given $\Mst$.}
    \label{fig:StKinematics}
\end{figure}

The study of the stellar kinematics serves as an important further diagnostic when comparing simulations with observations. Fig.~\ref{fig:StKinematics} shows the integrated stellar kinematics of our simulated dwarf galaxy along with observational data for local dwarf galaxies as compiled by \citet{Wheeler2017}\footnote{Observational source manuscripts are: \citet{Mateo1998, Koch2007, Simon2007, Walker2009, Fraternali2009, Leaman2009, Geha2010, Koposov2011, Leaman2012, Tollerud2012, McConnachie2012, Frinchaboy2012, Ho2012,  Collins2013, Martin2013, Kirby2014, Martin2014, Salomon2015, Walker2015}.}. We include evolutionary tracks sampling median values during $z = (7, 6, 5) \pm 0.5$ and $z \in [3.4,3.7]$ for the \HDSfFb, \RTSfFb, \RTiMHDSfFb, \RTsMHDSfFb, \CRiMHDSfFb, \RTnsCRiMHDSfFb~and \RTCRiMHDSfFb~simulations. Each of the kinematic quantities is computed as a mass-weighted average within $2 \Rhalfst$. From top to bottom, panels show the stellar rotational velocity ($\mathrm{v}_\text{rot}$), stellar velocity dispersion ($\sv{*}$) and the ratio of rotational velocity to velocity dispersion ($\mathrm{v}_\text{rot}/\sv{*}$). For all three quantities, our {\it NoFb} simulations are in tension with observations (not shown, due to their values being significantly above the ranges displayed for $\mathrm{v}_\text{rot}$ and $\sv{*}$). The inclusion of SN feedback provides better agreement. Rotational velocities with fiducial SN feedback are of order $\sim\!\!1-5\km\,\s^{-1}$, and become somewhat lower when the strength of the SN feedback is boosted. During their time evolution to $z = 0.5$, \HDSfFb~and \HDSfFbBoost~maintain low rotational velocities. Their evolution suggests that other models may maintain similar $\mathrm{v}_\text{rot}$ values or undergo a slight reduction. With additional physics included, rotational velocities of our simulated dwarf are in the $\sim\!\!7-10\km\,\s^{-1}$ range, found in reasonable agreement with observations of isolated dwarfs (shown as cyan data points).

Most of our fiducial SN feedback simulations are located at the upper end or slightly above the $\sv{*}$ observational data. The boosted SN feedback simulations, with their lower stellar masses, show a lower velocity dispersion that is in good agreement with observations. When evolved to $z = 0.5$, the stellar velocity dispersion of \HDSfFb~is reduced by $\sim 5 \kmps$, which brings the model in closer agreement with observations. This suggests that simulations with additional physics may experience a similar reduction. \HDSfFbBoost~maintains an approximately constant value for $\sv{*}$ during its evolution. Simulations with stellar radiation have particularly large stellar velocity dispersion, whereas the models with cosmic rays (with or without radiation) have the lowest values of $\sv{*}$ for a realistic $\Mst$. \RTnsCRiMHDSfFb~in particular yields a reasonable match to observations of isolated dwarfs in the same stellar mass range. While \RTCRiMHDSfFb~is within the range of observations prior to its merger at $z \sim 5$, this event is the cause of the higher final $\sv{*}$ values than the \RTnsCRiMHDSfFb~case. The ratio of rotational velocity to velocity dispersion is within the scatter of observations for all our simulations. While some minor variation is observed when evolving \HDSfFb~and \HDSfFbBoost~to $z = 0.5$, their qualitative properties remain unchanged. The only notable trend for this quantity is that the inclusion of stellar radiation leads to a more rotationally-dominated support of the stellar component, which appears to best resemble the isolated dwarf galaxies from the \citet{Wheeler2017} sample.

\subsubsection{HI properties}
\label{sss:HIproperties}

\begin{figure}
    \centering
    \includegraphics[width=\columnwidth]{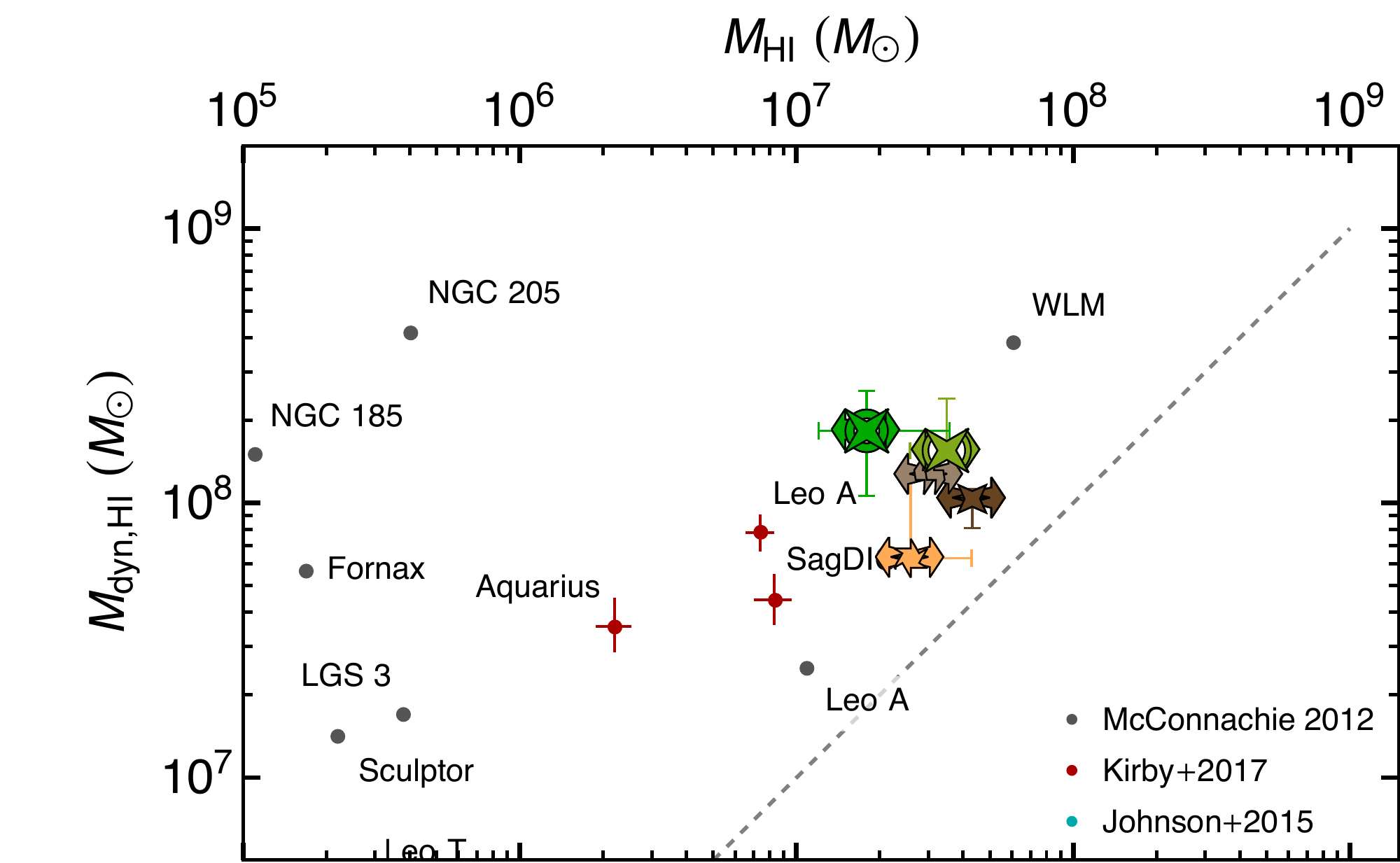}\\
    \includegraphics[width=\columnwidth]{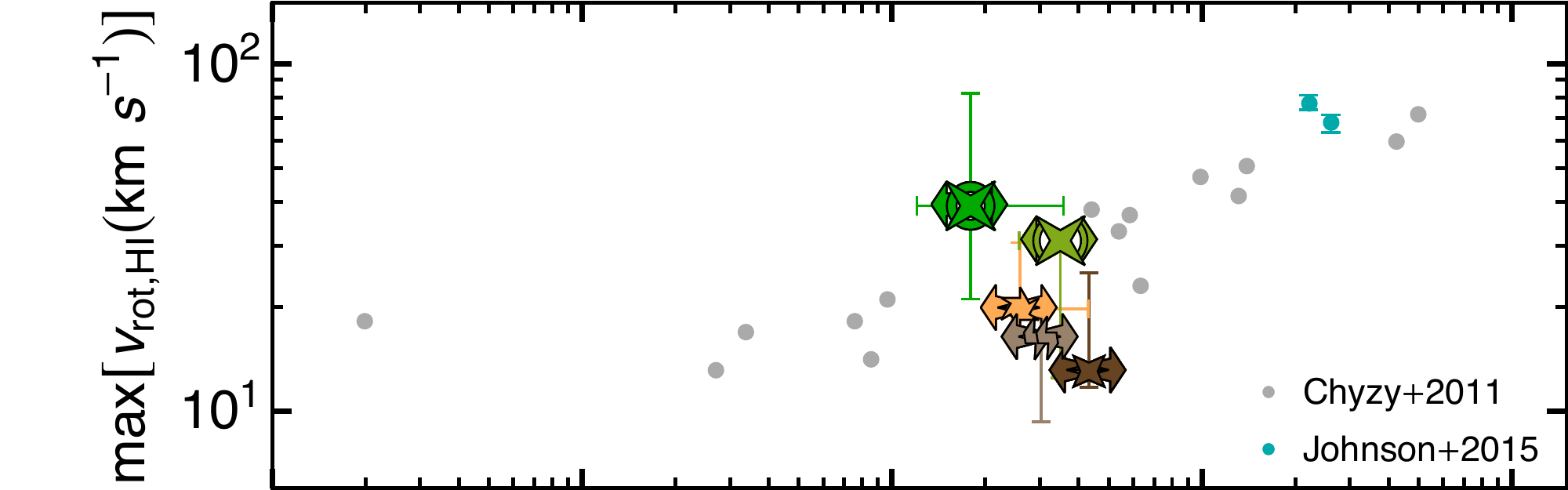}\\
    \includegraphics[width=\columnwidth]{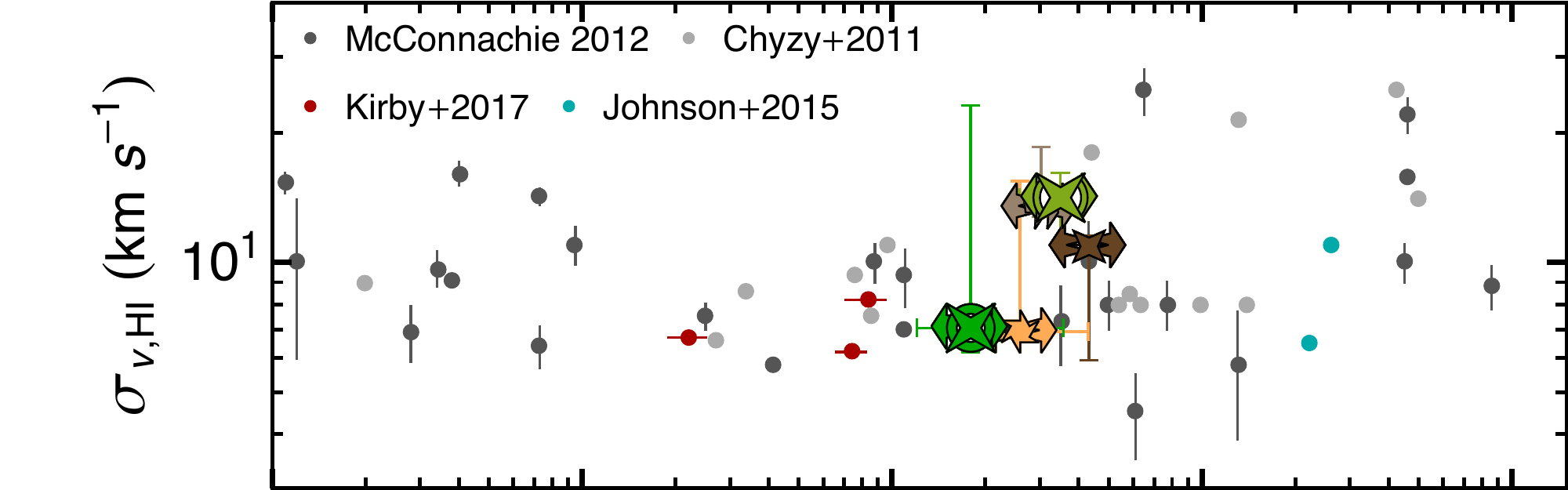}\\
    \includegraphics[width=\columnwidth]{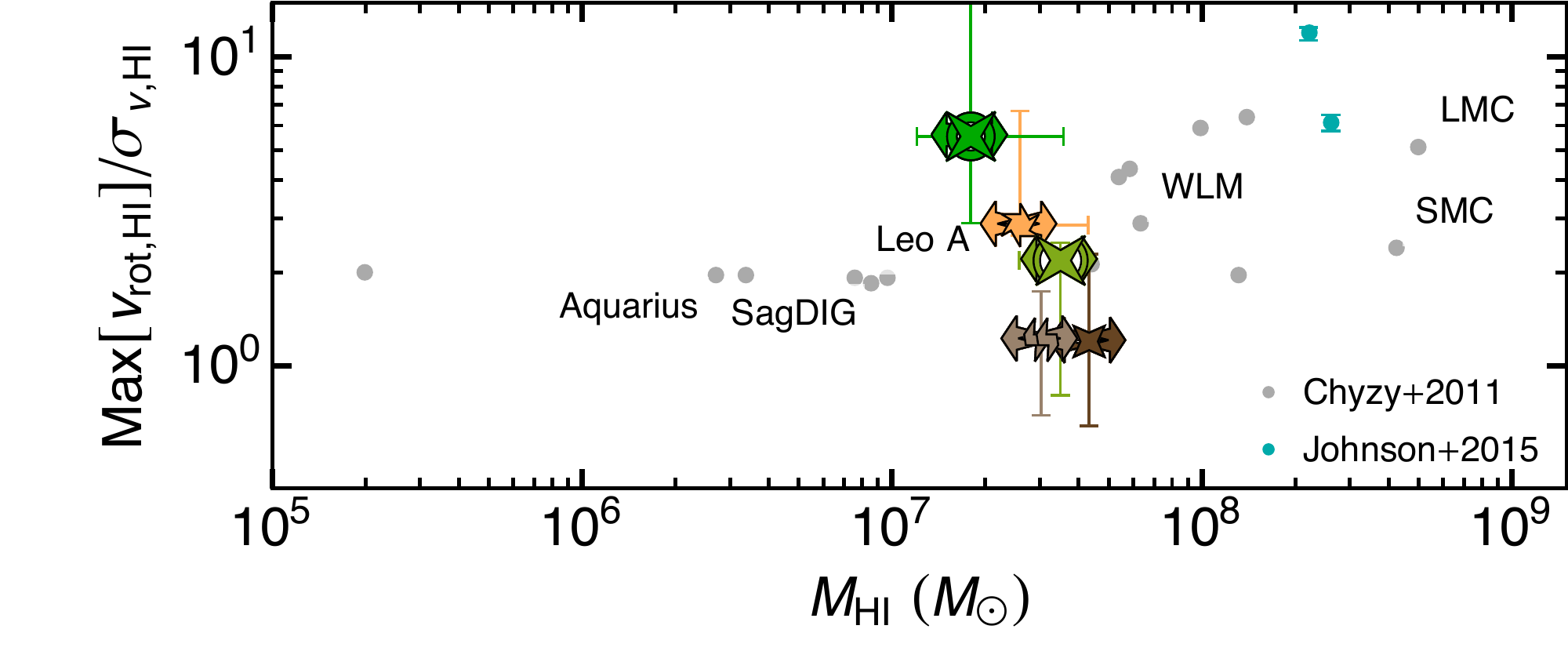}\\
    \DwarfBarLegendRT\\
    \caption{HI dynamical mass and kinematics vs $\MHI$ for our simulations featuring radiative transfer (i.e. self-consistently modelling HI). From top to bottom, panels display the dynamical mass estimate using HI ($\MdynHI$; Equation~(\ref{eq:Mdyn})), maximum HI rotational velocity ($\text{max}(v_\text{rot,HI})$), HI velocity dispersion ($\sv{HI}$) and the $\text{max}(v_\text{rot,HI}) / \sv{HI}$ ratio. Error bars show median quartiles during the redshift interval $z \in [3.4, 3.7]$. Overall, all our radiative transfer simulations models at $z = 3.5$ are in good agreement with HI observations at $z \sim 0$, with the \RTSfFb, \RTnsCRiMHDSfFb~and \RTCRiMHDSfFb~simulations providing the best match to the local isolated dwarfs reported by \citet{Kirby2017}.}
    \label{fig:HIkinematics}
\end{figure}

Including radiative transfer leads to a more self-consistent modelling of the HI content in our dwarf galaxy. This allows us to explore the subset of radiation-hydrodynamics simulations and compare their HI properties to observations. Fig.~\ref{fig:HIkinematics} revisits the kinematic properties of the studied galaxy, but now for the HI component, comparing with HI observations for similar mass galaxies. We compute each of these quantities adopting the same procedure as done for Figs.~\ref{fig:StellarMassStellarRadius} and \ref{fig:StKinematics}, except for the maximal rotational velocity $\text{max}(v_\text{rot,HI})$, which corresponds to the maximum value of the rotational component of the velocity profiles within $2 \RhalfHI$. We compare our models at $z \sim 3.5$ with HI observational results for similar galaxies at $z \sim 0$ by \citet{Chyzy2011}, \citet{McConnachie2012}, \citet{Johnson2015} and \citet{Kirby2017}. 

In agreement with previous studies \citep[e.g.,][]{Rey2022}, we find a higher time-variability for measurements associated with the gaseous components, such as the HI values presented here. This variability is illustrated by the included error bars, where the median values remain relatively robust once the final merger event has taken place. We also find these shown quantities to remain relatively unchanged in the \HDSfFb~and \HDSfFbBoost~simulations
\footnote{We have quantified the changes in the kinematic quantities of the \HDSfFb~and \HDSfFbBoost~models between $z \sim 3.5$ and $z = 0.5$. Note however that this measurement is done for the gas component: with no radiative transfer, these simulations do not provide a self-consistent modelling of hydrogen ionization. Keeping this caveat in mind, we find that their maximal rotational velocity remains within $\sim 20 - 30 \kmps$, the $\sv{gas}$ within $\sim 5 - 10 \kmps$, and their ratio remains at $ \sim 3$. The evolution of $\Mdyngas$~is reviewed in Fig.~\ref{fig:StellarMassStellarRadius}.}.
Our measurements of the dynamical mass using HI (top panel) as a tracer are more robust than those using the total gas, with their scatter reduced by $\sim\!0.5$~dex and all our models matching well the comparable galaxies from \citet{Kirby2017}. Our simulations spread around the expected values for $\text{max}(v_\text{rot,HI})$ (second panel), with the simulations combining radiative transfer and MHD falling slightly below the observations. For the HI velocity dispersion $\sv{HI}$ (third panel), all our simulations have values of $\sv{HI} \sim 10 \kmps$ and the same approximate scatter, providing a good match to the distribution of observations. Finally, the ratio of $\text{max}(v_\text{rot,HI}) / \sv{HI}$ (fourth panel) is in best agreement with observations for the \RTSfFb~and \RTnsCRiMHDSfFb~simulations, dominated by the value measured for $\sv{HI}$. Nonetheless, all our radiative transfer simulations predict reasonable HI content and properties of our simulated dwarf galaxy. 

\subsubsection{Comparing HI and stellar kinematics}
\label{ss:compareKinematics}

\begin{figure}
    \centering
    \includegraphics[width=\columnwidth]{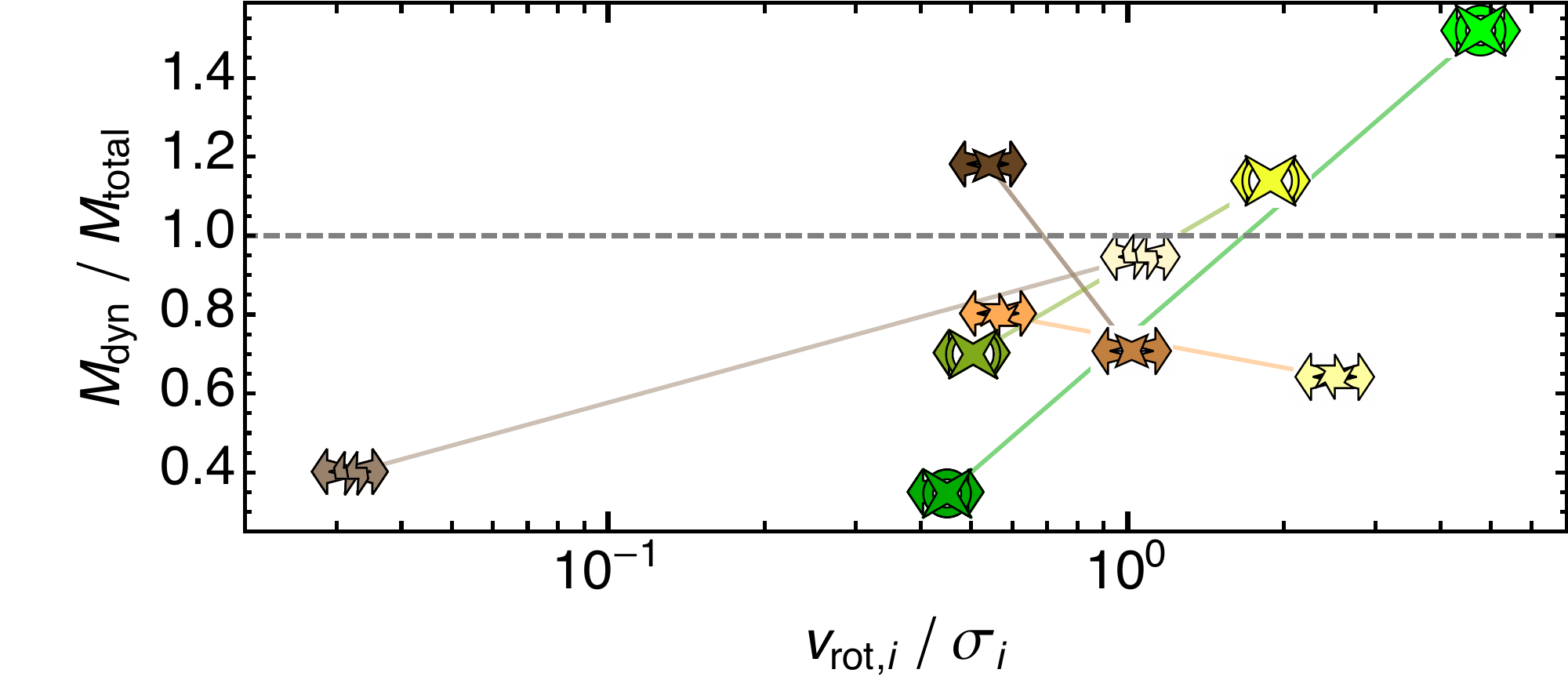}\\
    \DwarfBarLegendRT\\
    \caption{Ratio of the dynamical mass with respect to the total mass, inferred through the stellar (standard symbols) and HI (lighter symbols) components. The two estimates of the dynamical mass encompass well the total mass, except for the model with an extreme primordial magnetic field (\RTsMHDSfFb).}
    \label{fig:compareKinematics}
\end{figure}
With different components of the galaxy subject to different dynamics, we expect some variation in their kinematic properties. This is reflected in our overall results, where the stellar component displays larger velocity dispersion ($\mathrm{v}_\text{rot} / \sv{*} < 1$) while the HI is dominated by rotation ($max[\mathrm{v}_\text{rot,HI}] / \sv{HI} > 1$), in agreement with their collisional (HI) and collisionless (stellar) nature.
Rotational velocities for the stars are $\lesssim 10\,\kmps$ whereas the maximum HI rotational velocity tends to be within $10 - 30\,\kmps$. In terms of velocity dispersion, HI displays $\sv{HI} \lesssim 10\,\kmps$, whereas $\sv{*}$ tends to be within $14 - 20 \,\kmps$. Notably, employing the maximum velocity as done by the observations will bias the measurement towards higher rotational support estimates. It is also important to consider that the stellar velocity dispersion may be subject to some degree of dynamical heating due to particle mass resolution \citep{Wheeler2019, Ludlow2023}.

Another comparison worth exploring is that of the estimated dynamical masses from each component as the dynamical mass is expected to trace the total mass of the galaxy. We show in Fig.~\ref{fig:compareKinematics} the ratio of the stellar dynamical mass ($\Mdynst$; standard symbols) and the gaseous dynamical mass ($\MdynHI$; lighter symbols) with respect to the total mass. Both components tend to provide relatively accurate estimates for the total mass of the galaxy. The HI measurement tends to yield ratios $> 1$, whereas the stellar component tends towards somewhat lower values. However, combining the two estimates provides a good approximation to $M_\text{total}$, with the exception of the \RTsMHDSfFb~model with an extreme primordial magnetic field. The \RTCRiMHDSfFb~model shows the largest separation from the two components, likely as a consequence of its recent `wet' merger.

\subsubsection{Spatially resolved kinematics}
\label{sss:manga}

\def \thiswidth{0.34} 
\begin{figure*}
    \centering
    \includegraphics[width=2.1\columnwidth]{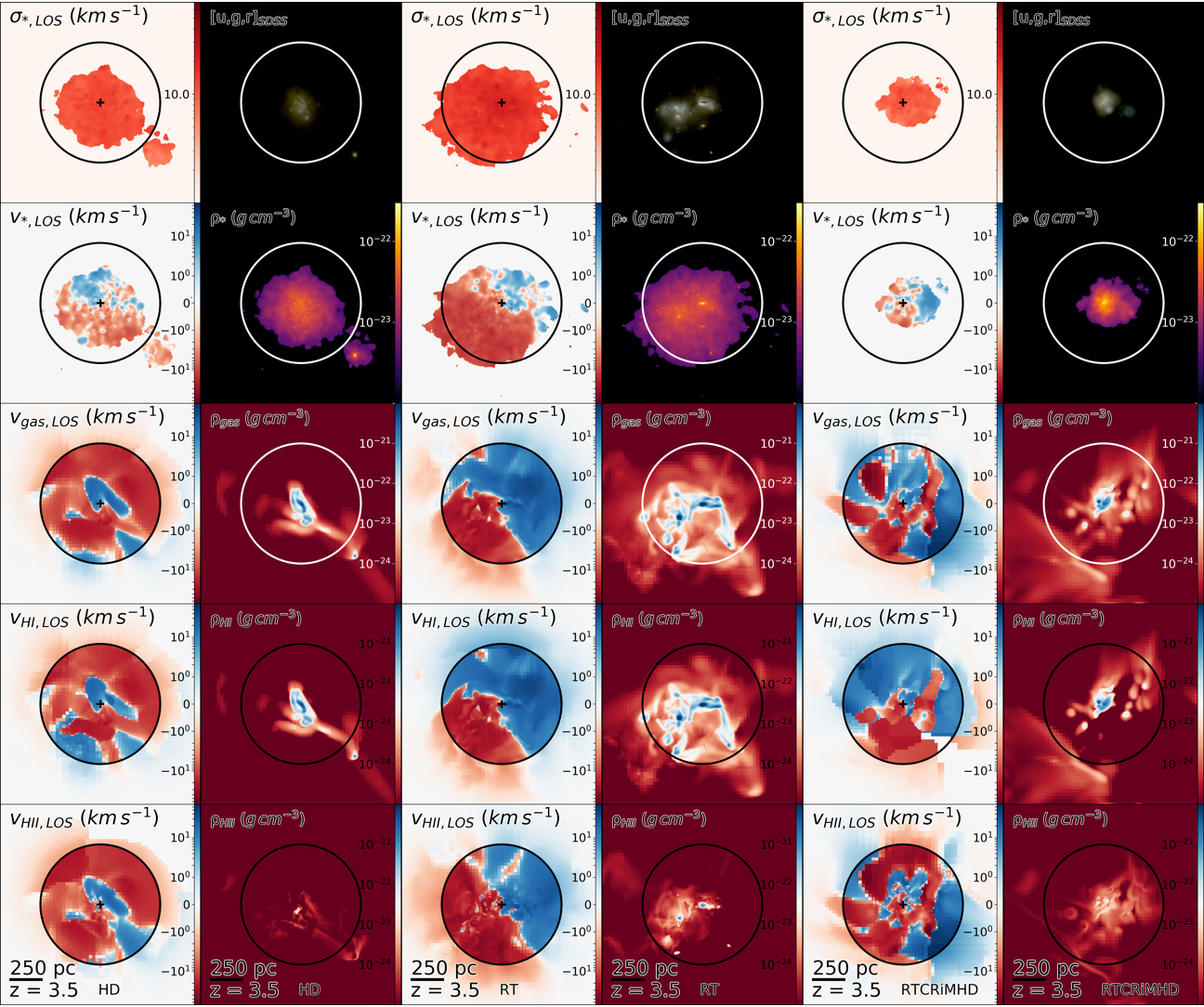}\\
    \caption{For the subset of simulations shown (as labelled), the left column displays line-of-sight stellar velocity dispersion (top row) and line-of-sight velocities for stars (second row), gas (third row), neutral hydrogen (fourth row) and ionized hydrogen (bottom row). Panels in right column show a synthetic SDSS mock image (top row), and the corresponding density maps next to each line-of-sight velocity field. All projections are computed along the x-axis of the box. Only gas cells with total densities higher than 0.1 $\mH\,\cm^{-3}$ are employed in the line-of-sight computation.}
    \label{fig:LOSstars}
\end{figure*}

Our review of various kinematic quantities revealed important differences between our models featuring different physical processes. These differences will be relevant for observations with access to spatial velocity information, employing integral field spectroscopy, such as CALIFA \citep{Garcia-Benito2015} or MaNGA \citep{Penny2016}. In Fig.~\ref{fig:LOSstars} we show multiple kinematics projections of our studied dwarf system at $z = 3.5$, separated into its baryonic components. Each simulation is presented by a pair of columns. In the left column we show the line-of-sight stellar velocity dispersion (top row), the velocity along the line-of-sight for the stars (second row), the gas (third row), neutral hydrogen (fourth row) and ionized hydrogen (fifth row). These two latter components are obtained either using the ionization fractions computed self-consistently in simulations with radiative transfer, or using the gas thermodynamical properties for our simulations without. Stellar velocity projections are density-weighted, whereas gas velocity projections employ a $\rho^2$ weighting, to more closely mimic the emission-weighted maps. For the velocity projections, we display only gas cells with densities $\rho > 0.1 \mH\,\cm^{-3}$ to avoid a bias due to the motion of gas in the lowest density regions. We have verified that such a cut does not affect and of the main features of the projections. We smooth out gas velocity fields outside the central circle for clarity. In the right column, we show from top to bottom a synthetic SDSS mock image (following the same procedure employed in Fig.~\ref{fig:Galaxies1Close}), stellar density, gas density, neutral hydrogen density and ionized hydrogen density. We analyse the projections for the redshift interval $z \sim 5$ to $z \sim 3.5$ and describe below the main results that remain consistent through this period. 

While significantly dispersion supported, as found from our integrated kinematics analysis, the \HDSfFb\ simulation displays a level of coherent rotation in stars and gas, with the gas being more centrally concentrated and largely in the neutral state. Including radiative transfer in our simulations leads to a dwarf galaxy which is more gas rich. This is exemplified by the \RTSfFb~simulation, which exhibits a more extended HI distribution and various clear dust bands. Our simulated dwarf for this particular simulation is embedded within an extended gaseous distribution. This model also exhibits a somewhat higher degree of coherence of rotation both in gas and stars, but it is worth noting that the line-of-sight velocity field and velocity dispersion are affected by a close-by gas-rich companion at this particular redshift. Gas within the dwarf contains a mixture of neutral and ionized hydrogen, and interestingly these two components have somewhat different line-of-sight velocity fields. 

Turning now to our full physics run, \RTCRiMHDSfFb, both the stellar and gaseous distribution is more centrally concentrated than in the \RTSfFb\ model, with the dwarf being somewhat more gas poor. In the synthetic SDSS image it is apparent that the morphology of the dwarf is more amorphous. Importantly, all line-of-sight velocity fields contain a number of kinematically distinct `clumps', which are (partly) caused by significant cosmic ray-driven outflows, also causing kinematically misaligned and decoupled motions in stars, ionized hydrogen as well as neutral hydrogen. Performing the same analysis on the \RTCRiMHDSfFb\ simulation at different times (not shown here) we further find that kinematic misalignment between the stellar component and ionized hydrogen are particularly prevalent during or shortly after significant star formation events that trigger cosmic ray-driven outflows.

\subsection{Colour-magnitude relation}
\label{ss:colmagnitude}
\begin{figure}
    \centering
    \includegraphics[width=\columnwidth]{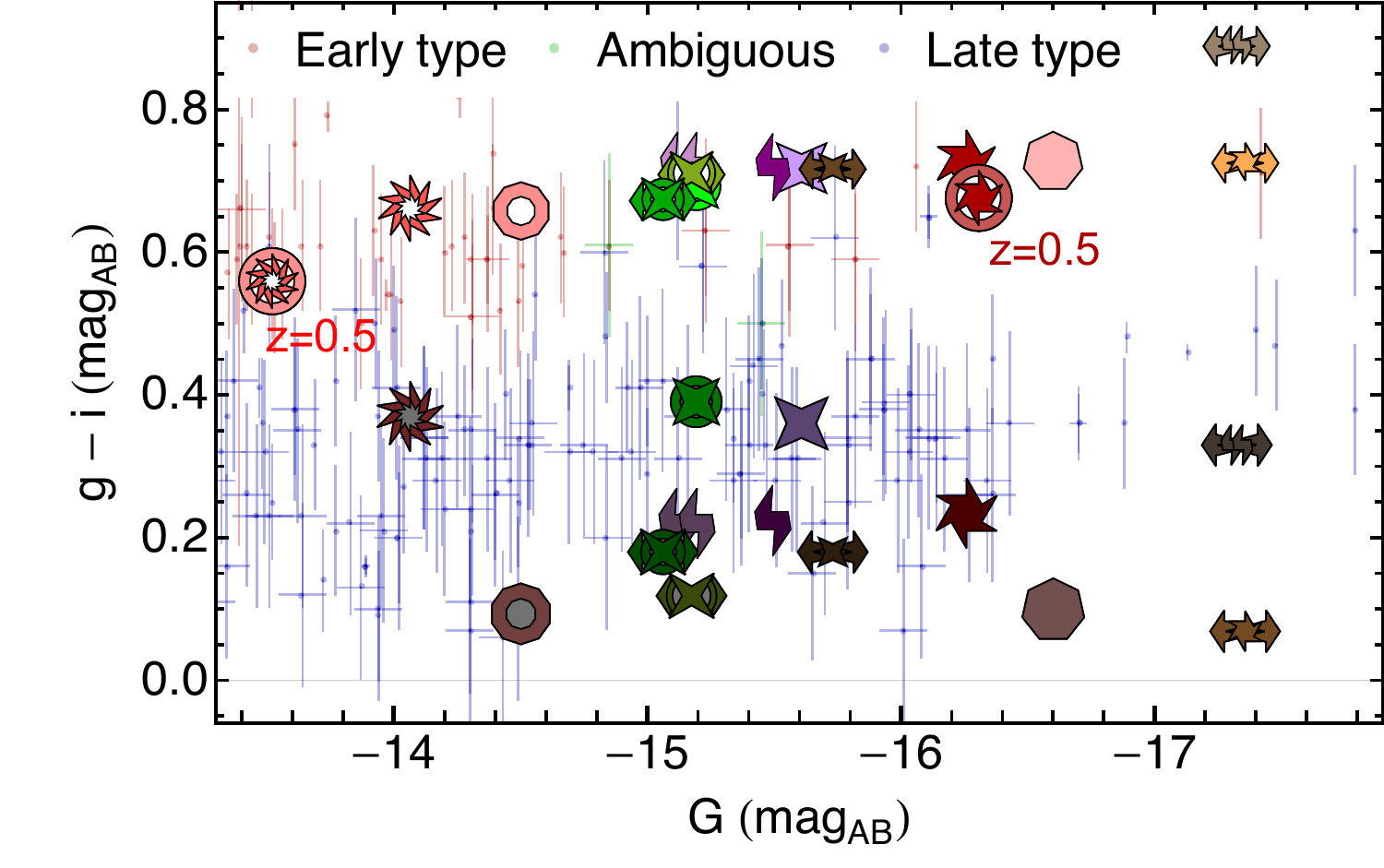}\\
    \includegraphics[width=\columnwidth]{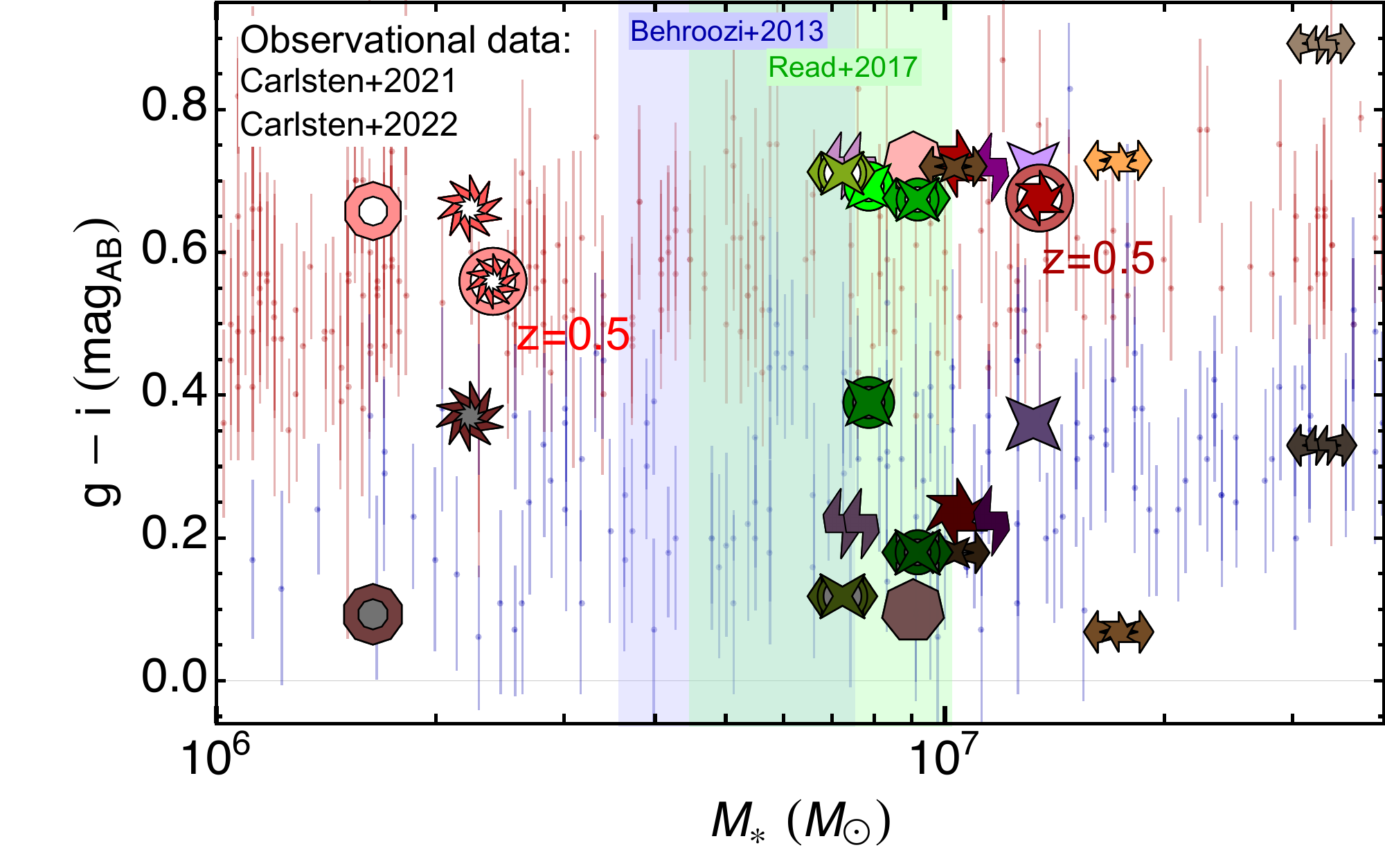}\\
    \HalfDwarfBarFrom{Images/DwarfLegendHalf.pdf}\\
    \caption{$g - i$ colour vs $G$ absolute magnitude (top) and stellar mass (bottom) for the simulated dwarf galaxies accounting for dust obscuration compared with ELVES dwarfs at $z \sim 0$ \citep{Carlsten2021, Carlsten2022}. These observations are divided into `Early type' (red), `Late type' (blue) and those with an ambiguous classification (green). Points show the galaxy at $z = 3.5$ (shaded symbols) and the colour upper limit estimate for $z \sim 0.5$ (standard symbols; see text for details), respectively. We also include as encircled symbols the exact measurements at $z \sim 0.5$ for the \HDSfFb~and \HDSfFbBoost~simulations, which we evolved down to $z = 0.5$. The coloured bands in the bottom panel indicate the range of stellar masses predicted from the abundance matching analysis as labelled.}
    \label{fig:CMD}
\end{figure}

As part of our analysis, we review the optical emission properties of our simulated dwarf galaxy. Fig.~\ref{fig:CMD} shows the $g - i$ colour versus absolute magnitude $G$ (top) and versus stellar mass within 2 $\Rhalfst$ (bottom), compared with the observed properties of the ELVES dwarf galaxies \citep{Carlsten2021, Carlsten2022}. In order to measure the emitted magnitudes for each of the filters, we produce synthetic observations accounting for dust obscuration. For these, we assume each stellar particle to be a single stellar population with its emission following \citet{Bruzual2003} depending on its age and metallicity. We adopt a simple dust obscuration model through a 3D dust absorption screen \citep[with metal-to-dust ratio 0.4;][]{Kaviraj2017}. Magnitudes are obtained by integrating the emission within the central $1.25$~kpc. The measurement for our galaxies at $z =3.5$ is shown with shaded symbols. Since for the majority of our simulations we only evolve the dwarf galaxies to $z = 3.5$, their stellar populations are still relatively young in comparison to the ELVES sample of local dwarfs. Hence, to provide a more adequate range for comparison, we also show the same measurement for our galaxies assuming their stellar populations have evolved passively down to $z = 0.5$ (i.e. we increase all particle ages by $\sim\!6.8$~Gyr) in their colour computation. The selected value of $z = 0.5$ redshift for this passive evolution is based on the availability of the \HDSfFb~and \HDSfFbBoost~simulations to allow for direct comparison. Further evolution to $z = 0$ would increase the $g - i$ colour by an additional $(0.04 \pm 0.02)\,\text{mag}_\text{AB}$. Therefore, the resulting data points are obtained assuming that no additional star formation will occur during this period. Including these data points provides us with an upper limit for the colour attainable by the simulated galaxies. To further review these estimates, we also include the exact measurements at $z = 0.5$ for the simulations that reached that redshift (i.e. for the \HDSfFb~and \HDSfFbBoost\ models, at $z = 0.5$). These are shown as encircled datapoints. We compare this `manual' ageing with the actual simulation data for the \HDSfFb~and \HDSfFbBoost\ runs which reached $z = 0.5$ to examine the accuracy of this approach. For \HDSfFb, the approximation is remarkably accurate, with errors in the estimated colour and absolute magnitude of $\sim 0.05\,\textrm{mag}_\text{AB}$ and $0.05\,\textrm{mag}_\text{AB}$, respectively. These estimates experience larger deviations for the absolute magnitude in \HDSfFbBoost, with only $\Delta(g - i) \sim 0.05\,\textrm{mag}_\text{AB}$ but $\Delta G \sim 0.5\,\textrm{mag}_\text{AB}$. 

We conclude that the `passive ageing' procedure leads to reasonable estimates of the observed $g - i$ colour, with a possible overestimation of the absolute magnitude $G$ of up to $\sim 1
\,\textrm{mag}_\text{AB}$, at least for those two models. If our other models followed reasonably similar evolutions, we would expect the majority of them to be relatively close to the estimated colour upper limit. Nonetheless, we note that some degree of additional star formation may take place in Pandora models, and some could even undergo a mild re-ignition of star formation \citep{Rey2020}.

Unsurprisingly, at $z \sim 3.5$, our galaxies are relatively blue compared with the overall observed ELVES population. By $z \sim 0.5$ however, we expect most of our simulated dwarfs to be in good agreement with the `Early type' observed sample, and possibly at the upper colour end of the `Late type' if they underwent some star formation. By `manually' ageing the simulated galaxies we have assumed that they remain completely quenched from $z = 3.5$ to $z =0.5$, hence our $g - i$ colours should be considered only as a conservative upper limit, and the stellar masses and absolute $G$ magnitude as a conservative lower limit. If the trend displayed by the \HDSfFb~and \HDSfFbBoost~models is reproduced by our other simulations, this could potentially imply that all of our feedback models would lead to a population of early-type dwarfs in the local Universe. Hence alternative feedback models, assembly histories or star formation re-ignition are needed to produce the observed sample of `Late type' dwarfs. To determine this, simulation to low redshifts and larger samples of simulated dwarfs are needed.

\subsection{Constraints on magnetic field strength}
\label{ss:magnetism}

\begin{figure*}
    \centering
    \includegraphics[width=\columnwidth]{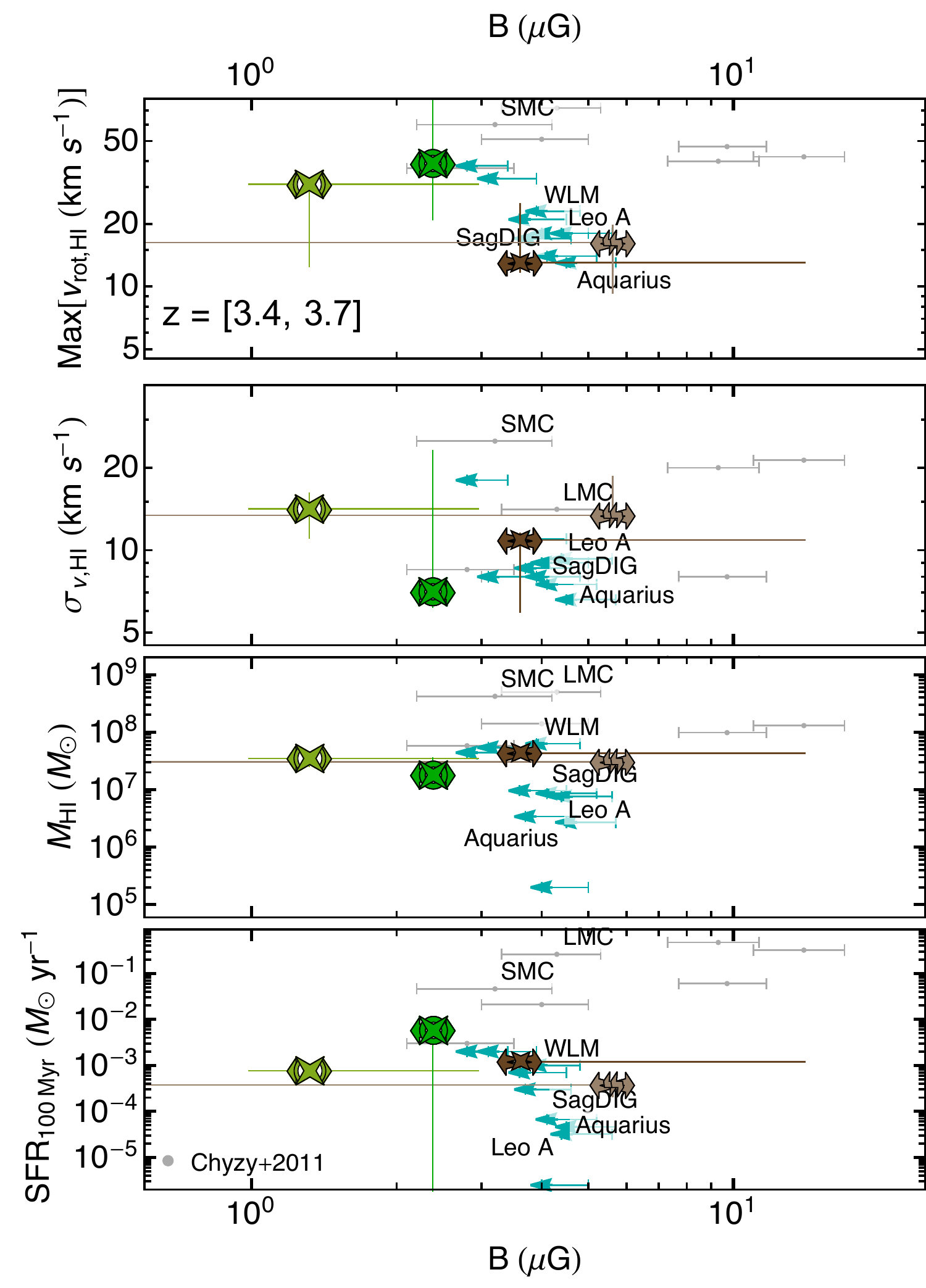}%
    \includegraphics[width=1.035\columnwidth]{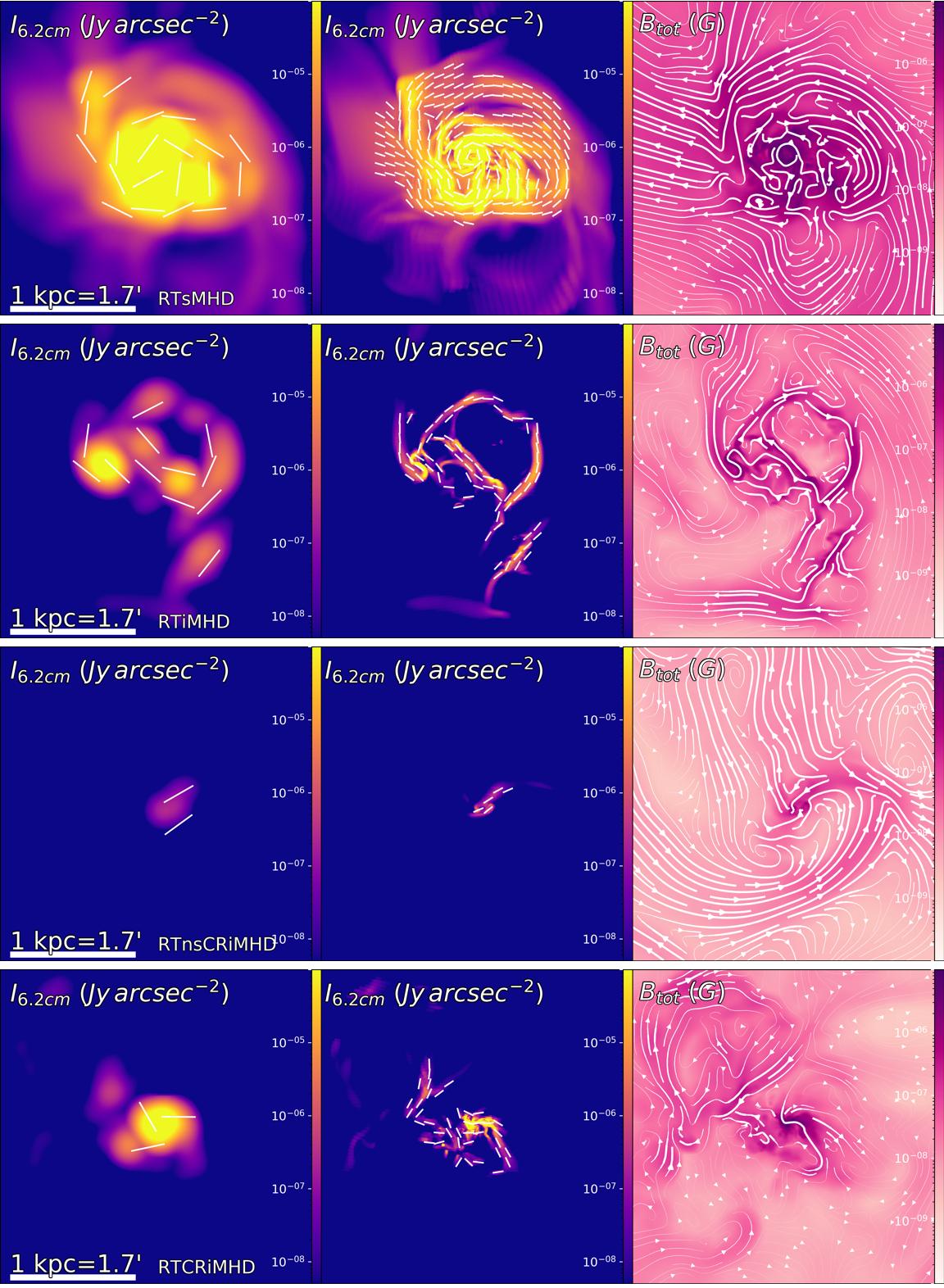}\\
    \vspace{-0.15cm}\includegraphics[width=0.95\columnwidth]{Images/DwarfLegendRT.png}\hspace{\columnwidth}\vspace{-0.2cm}\\
    \caption{{\bf (Leftmost panels)} From top to bottom, the $y$-axis of various panels displays maximum rotational velocity of HI, $\text{max}(v_\text{rot,HI})$, total HI mass, $\MHI$, HI velocity dispersion, $\sv{HI}$, and the star formation rate averaged over the last 100~Myr, $\text{SFR}_\text{100 Myr}$ versus the average magnetic field strength $B$. Our simulated dwarf galaxies have magnetic fields at $z \sim 3.5$ that are in agreement with observations by \citet{Chyzy2011} of local dwarf galaxies at $z \sim 0$. {\bf (Right panels)} Comparison of the synchrotron synthetic views (for $\lambda = 6.2$ cm) for our simulated dwarf galaxy at $z = 3.5$ generated using {\sc polaris} (see text for details). We assume the galaxy to be observed at a distance of 2 Mpc, convolving the images with the corresponding telescope resolution. White dashes display the local linear polarisation orientation rotated 90 degrees to align with the local magnetic field. Each row displays from left to right: VLA-like synthetic map, SKA-like synthetic map and magnetic field strength (streamlines represent magnetic field lines). From top to bottom, rows show the \RTsMHDSfFb, \RTiMHDSfFb, \RTnsCRiMHDSfFb~and \RTCRiMHDSfFb~simulations. Emission traces regions of stronger magnetic field. Due to the turbulent nature of magnetic field lines, polarisation only captures large-scale magnetic field orientation features. Our simulations with cosmic rays and radiative transfer have a more concentrated synchrotron emission.}
    \label{fig:Bfield}
\end{figure*}

Galaxy formation simulations simultaneously featuring radiative transfer and magnetic fields are still rare. The leftmost panels in Fig.~\ref{fig:Bfield} show how our simulations compare with observational relations between various HI properties and magnetic fields in local dwarf galaxies by \citet{Chyzy2011}. Our average magnetic field at galactic scales is on the order of $\sim\!2 - 5 \muG$, with slightly higher field strengths in the \RTsMHDSfFb~simulation due to a rather high value of its primordial magnetic seed. Encouragingly, for all of the HI quantities examined, our galaxies are either consistent with current observations or fall within the range where \citet{Chyzy2011} can only provide upper limits for the magnetic field. Simulated galaxies have magnetic fields which are also compatible with measurements at higher $\MHI$ values, where these observations suggest $B \sim 3 - 15 \muG$.

Detailed simulations of galaxies modelling the ionization state of hydrogen as well as MHD, as is the case for part of our Pandora suite studied here, will help our understanding of observations in the radio \citep{Heald2022, Lopez-Rodriguez2022b} and far-infrared \citep{Borlaff2021} wavelength ranges. These simulations will be crucial to aid both current and upcoming surveys such as the Survey of extragALactic magnetiSm with SOFIA (SALSA) using the Stratospheric Observatory for Infrared Astronomy (SOFIA; \citealt{Lopez-Rodriguez2022a}) or SKA. These surveys aim to probe magnetisation down to dwarf galaxies. We showcase this capability for our simulations on the right-hand side of Fig.~\ref{fig:Bfield}, where we present synthetic synchrotron observations for the simulations combining radiative transfer and magnetic fields.

To produce these synthetic observations we employ the publicly available code {\sc polaris}\footnote{\href{https://portia.astrophysik.uni-kiel.de/polaris/}{https://portia.astrophysik.uni-kiel.de/polaris/}} \citep{Reissl2016, Reissl2019}. We extract from the simulations an adaptive grid centred on the galaxy, resolved with double the local AMR resolution, interpolating all native {\sc ramses} quantities required by {\sc polaris}. Due to their limited spatial resolution, our simulations do not fully reproduce the entire magnetic energy inverse-cascade \citep{Martin-Alvarez2018}, thus failing to capture additional magnetic energy expected to reside below the local grid size \citep{Schekochihin2002}. As we are only interested in a qualitative comparison across our models, we account for the loss of additional magnetic energy by a simple boost of the magnetic field strength. To estimate the approximate amount of energy boost required, we expand a Kazantsev-like spectrum from our simulation resolution to the scales where turbulence should be converged. \citet{Martin-Alvarez2022} show the peak of the magnetic energy spectrum $k_\text{peak}$ to scale with the finest simulation resolution $\Delta x_\text{min}$ following $k_\text{peak} \propto \Delta x_\text{min}^{-1}$. \citet{Kortgen2017} suggest resolutions $\Delta x_\text{min} < 0.1 \pc$ are required to approach turbulence convergence, finding the average turbulence to stabilize at values of $\Delta x_\text{min} \lesssim 0.03 \pc$. Extrapolating the \citet{Martin-Alvarez2022} scaling of $k_\text{peak}$ from our suite resolution $\Delta x_\text{min,sim} \sim 7 \pc$ to $\Delta x_\text{min,exp} = 0.005 \pc$ yields a ratio of $k_\text{peak,exp} / k_\text{peak,sim} \sim 40.1 $. Expanding a Kazantsev-like spectrum ($\propto k^{3/2}$) by such ratio provides an increase of the average magnetic field strength of approximately 2 dex. As we are only interested in a qualitative exploration of these synthetic observations, we leave the review of alternative models to capture the sub-grid magnetic energy (e.g. \citealt{Reissl2019}) to future work.

The use of {\sc polaris} requires additional quantities not modelled by our version of {\sc ramses}. We follow a similar approach to that used by \citet{Reissl2019} to determine these quantities on the grid. We bound the Lorentz factor of electronic cosmic rays between a fixed $\gamma_\text{min} = 4$ minimum \citep{Webber1998} and a $\gamma_\text{max} = 300$ maximum (i.e. $\gamma_\text{max} \gg 1$). We also follow \citet{Reissl2019} and fix the power-law index to $p = 3$ \citep{Miville2008}, related to the spectral index by $\alpha = (p - 1) / 2$. In order to obtain the distribution of thermal electrons, we compute their number density as a function of gas number density according to \citet{Pellegrini2020}. Finally, we model the local energy density of electronic cosmic rays $e_{e^{-} \text{CR}}$ following the CR2 model also presented by \citet{Reissl2019}, which assumes energy equipartition between electronic cosmic rays and the local magnetic field (without any magnetic field increase). The produced synthetic radio observations show the resulting intensity map at $\lambda = 6.2$ cm. We overlay on these maps white dashes oriented perpendicularly (i.e. rotated 90 degrees) to the local linear polarisation (accounting for Faraday depolarisation) in order to align with local magnetic field. These dashes are only shown in regions where the linearly polarised intensity is higher than $5\%$ of the average intensity of the entire map. Finally, we assume the observed galaxy to be at a distance of 2 Mpc, convolving our images with a 2D gaussian profile with a full width at half maximum equal to the resolution of the corresponding telescope (12.6 arcsec for VLA and 0.1 arcsec for SKA).

In all maps, synchrotron emission traces regions of strong magnetic field. Due to its extreme primordial magnetic field, \RTsMHDSfFb~features the brightest emission, extending well beyond the galaxy. The observed polarisation traces the large-scale structure of the magnetic field at the outskirts of the galaxy and into its halo. Contrarily, models with SN-injected magnetic fields have weaker and more turbulent magnetic fields outside the galaxy. In these simulations, the polarisation vectors seemingly match only features at and below galactic scales. Our simulations with cosmic rays have a less extended emission, which is seen as a single gaussian in the VLA observations. While magnetic field lines are comparably turbulent in the three SN-injected scenarios, the polarisation vectors appears more coherent in the models combining radiative transfer and cosmic rays, particularly at scales of the VLA resolution.

\subsection{Halo density profiles}
\label{ss:CuspCore}
\begin{figure}
    \centering
    \includegraphics[width=0.9\columnwidth]{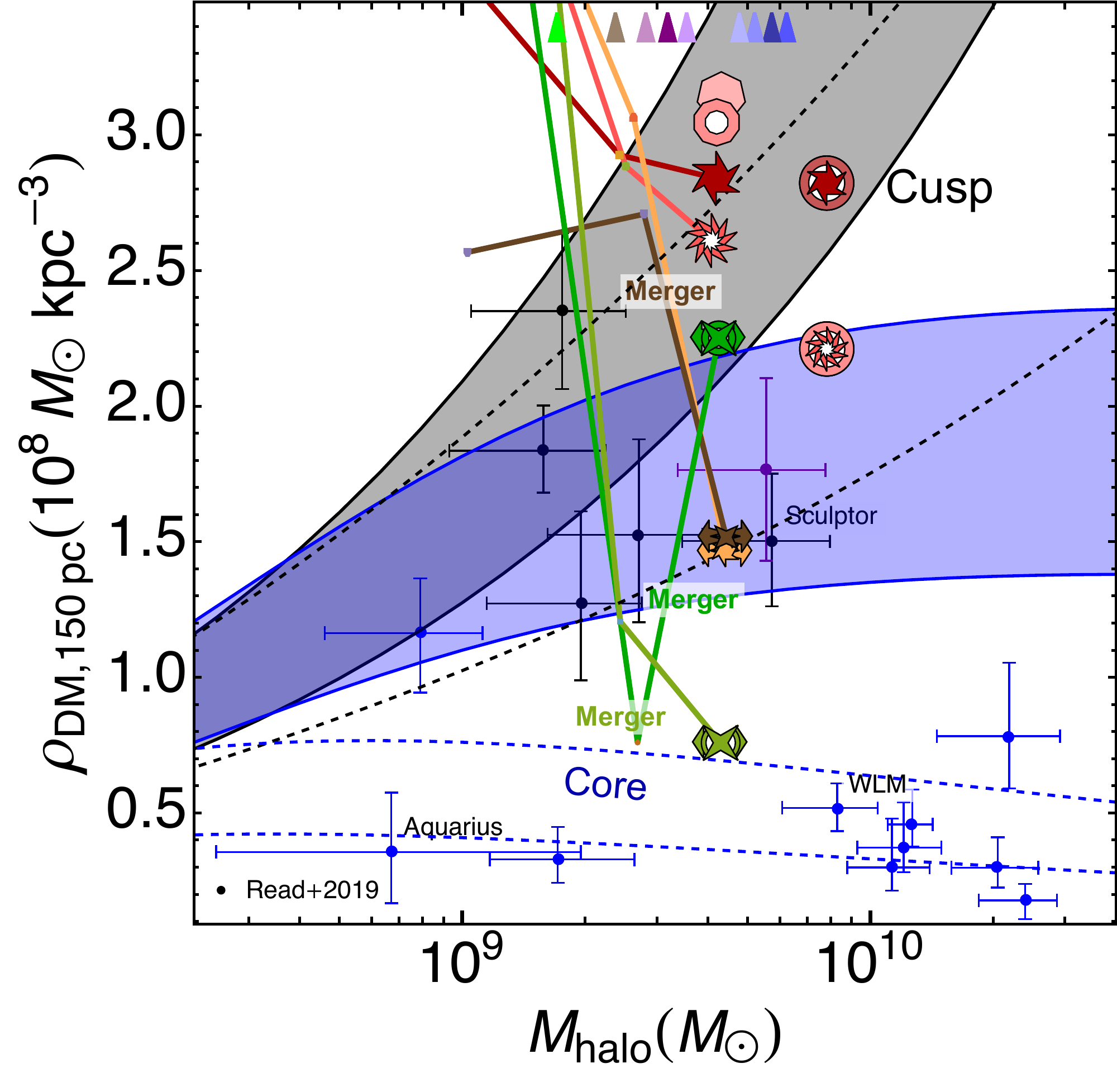}\\
    \HalfDwarfBarLegend\\
    \caption{Average dark matter density within the inner $150$~pc of the simulated dwarf galaxy versus $M_\text{halo}$, where the models \HDSfFb~and \HDSfFbBoost~are also shown at their later redshift $z = 0.5$ with encircled symbols. Our results are compared with Local Group dwarf galaxies data by \citet{Read2019a}. We follow \citet{Read2019a} and include bands for the expected density at $z = 3$ (shaded bands) and $z = 0$ (dashed bands). The latter provide a good approximation to the bands expected at $z = 0.5$, to be compared with the encircled symbols. We only show evolutionary tracks for a number of representative models, with coloured lines displaying the evolution from $z = 7$ until $z = 3.5$, and symbols showing the simulated data at $z = 3.5$. Simulations with very high central densities are indicated with arrowheads at the top of the graph, separated by artificial displacements along $x$ axis for visibility. Following \citet{Read2019a}, we colour the observational points corresponding to galaxies that stopped forming stars more than $6$~Gyr ago as black, between $6$ and $3$~Gyr as purple, and less than $3$~Gyr ago as blue. Colour bands show the prediction by \citet{Read2019a} for the central density in haloes with cusps (black) or cores (blue). While the majority of our runs have cuspy central dark matter distributions, some of our simulations with radiation and cosmic rays fall within the `core' band.}
    \label{fig:HaloFits}
\end{figure}

Dark matter density profiles in dwarf galaxies estimated from observations frequently suggest shallower central slopes than that of the theoretical expectation from dark matter assembly in a cold dark matter cosmology \citep{Navarro1997}. Such cores may be carved out of the high density NFW cusps through the action of baryonic feedback sufficiently violent to induce rapid variations in the gravitational potential \citep[see e.g.][]{Navarro1996, Pontzen2012}. Cores may be particularly prevalent in dwarf galaxies due to their shallower gravitational potentials, which make their dark matter more susceptible to the SN-driven outflows \citep{Read2016a}. Hence, in this section, we analyse the density distribution of our simulated dwarfs using their density profiles, and compare these with observations.

We centre our profiles in the galaxy rather than using the centre of the dark matter halo, motivated by our aim to resemble observational measurements. We review the effects of centring the measurements on the dark matter component in Appendix~\ref{ap:CentralDensHalo}. We find an overall density increase of $\sim 0.3 \cdot 10^8 \Msun$ for this choice, with the global trends for each set of models preserved. Density profiles may be at times sensitive to the selected galaxy centre, which is not always unambiguously defined. To obtain the galaxy centre for the radial profiles, we commence our calculation using the centre of the host dark matter halo, and select an initial sphere of the baryonic mass extending up to 0.2 $r_\text{halo}$. We recursively apply the shrinking spheres method \citep{Power2003} to the baryonic mass inside this sphere until a centre is obtained.

Fig.~\ref{fig:HaloFits} compares the average central dark matter density within the inner $150$~pc of our dwarfs with data from \citet{Read2019a}. Following \citet{Read2016a} and \citet{Read2019a}, we compute the bands of expected central densities for cusps and cores according to the cosmology of our simulations and employing the mass - concentration relation as a function of redshift provided by \citet{Maccio2007}. These bands indicate their density expectation for haloes that have (blue bands) or have not (black bands) undergone a transition from cusps to cores, at redshifts $z = 3$ (shaded bands) and $z = 0$ (dashed lines). The vast majority of our simulations have central densities in agreement with cusp profiles (or even higher). At face value, the lack of cores could be attributed to the fact that our simulations have been evolved only to $z \sim 3.5$, and thus cannot capture additional SN feedback cycles. While there is the possibility of additional star formation, based on the few simulations evolved to lower redshift, we do not expect significant evolution for the Pandora dwarf after $z \sim 4$. When evolved to $z = 0.5$, both the \HDSfFb~and \HDSfFbBoost~simulations remain within the $z = 0$ cuspy halo band (dashed black lines), which is approximately equal to the band at $z = 0.5$ (not shown for clarity). \HDSfFb~has no change in central density, whereas it is somewhat reduced in \HDSfFbBoost. Considering that Pandora does not undergo any additional mergers beyond $z \sim 5$ (see Appendix~\ref{ap:HaloGrowth}), we expect all our other models with SN feedback to either maintain their central densities or further reduce them as they evolve to $z = 0$. \citet{Read2019a} suggest that additional star formation at latter redshifts would lead to lower central densities. This is in qualitative agreement with our models displaying significant star formation after $z \lesssim 6$. On the other hand, the models where most of the star formation takes place at earlier times (e.g. the \iMHDSfFb, \HDSfNoFb~models) have considerably higher central densities. Although cusp destruction is hypothesized to be driven by SN feedback, we note that boosting the SN specific energy has only a moderate effect in our simulations by $z = 3.5$, even though this run is very efficient at quenching the dwarf and was driving a significant gas fraction out of the halo. The inclusion of radiative transfer leads to lower central dark matter densities. This effect is intensified when stellar radiation is combined with cosmic rays, as reflected by the tracks of \RTCRiMHDSfFb\ and \RTnsCRiMHDSfFb\ runs at $z \sim 5$. While these lower densities persist in \RTnsCRiMHDSfFb~down to our final studied redshift, the central dark matter density in the \RTCRiMHDSfFb~simulation increases after $z \sim 4$. We attribute this to a `wet' merger, which erases the cored profile in the \RTCRiMHDSfFb\ simulation and leads to a central density that resembles more the fiducial \HDSfFb~case. Nonetheless, we note that the final central density measured for \RTnsCRiMHDSfFb~increases by a non-negligible amount when measured using the dark matter centre (reaching $\sim 1.5 \cdot 10^8 \Msun$).

These trends are further quantified in Fig.~\ref{fig:DensityProfiles}, which shows density profiles for the gaseous, stellar and dark matter components. The leftmost column shows some representative simulations (\HDSfNoFb, \HDSfFb, \HDSfFbBoost~and \RTiMHDSfFb) at $z = 3.5$. The vertical solid line indicates $r = 0.15$~kpc, used in Fig.~\ref{fig:HaloFits} to calculate the average central dark matter density. Dotted lines spanning from $r = 0.15 \kpc$ up to $r = 2 \kpc$ show the NFW density scalings $\rho_\text{DM} \propto r^{-1}$ (black) and $\rho_\text{DM} \propto r^{-3}$ (gray). Gray and blue bands within $r = 0.15 \kpc$ denote the values of 'cuspy' and 'cored' dwarf profiles, respectively. These are computed as in Fig.~\ref{fig:HaloFits}, and are shown at $z = 3$ (shaded bands) and $z = 0$ (dashed horizontal lines). The $y$ range shown corresponding to the halo mass that matches our simulated dwarfs at $z = 3.5$. 

Compared to the no feedback simulations, the inclusion of SN feedback and radiative transfer (see bottom left panel) leads to a significantly shallower central dark matter density profile that persists to up to a few $100$~pc. However, this shallower dark matter density profile still falls within the `cusp' category as defined by \citet{Read2019a}, which is indicated by the grey coloured bands. In terms of the baryonic contribution to the total density profile, it is interesting that due to SN feedback and local radiation feedback from young stars, the central stellar distribution is diffuse with gas dominating in the centre.  

The central and right-hand columns of Fig.~\ref{fig:DensityProfiles} show the redshift evolution of the density profiles from $z = 5$ to $z = 3.5$ for the `full-physics' \RTCRiMHDSfFb~and \RTnsCRiMHDSfFb~simulations, respectively. Here, focusing first on the dark matter density profiles, the central profile is somewhat shallower than the $\rho_\text{DM} \propto r^{-1}$ scaling and there is a range of redshifts where the central density distribution is more in line with a `core'. As discussed previously, due to the merging event taking place at $z \sim 4$ both the baryonic and dark matter density profiles steepen in the central region for the \RTCRiMHDSfFb~simulation, while this is not the case for \RTnsCRiMHDSfFb. This indicates that, while radiative transfer and cosmic rays contribute to the formation of dark matter cores, `wet' galaxy mergers may transform these cores to cusps again. Furthermore, it is interesting to contrast the distribution of the baryonic component when the cosmic rays are included as well. While in the SN feedback only simulation as well as the \RTiMHDSfFb~model, gas is the dominant component in the centre of the galaxy, this is not the case in the simulations with cosmic rays. Now, at all redshifts explored, the stellar component dominates the total density profile within the central region with gas being significantly depleted (apart from the aftermath of the merger at $z=3.5$ in the \RTCRiMHDSfFb~simulation). This is caused by efficient cosmic ray-driven outflows, pushing the gas out of our simulated dwarfs.

\def \thiswidth{0.7} 
\begin{figure*}
    \centering
    \includegraphics[width=\thiswidth\columnwidth]{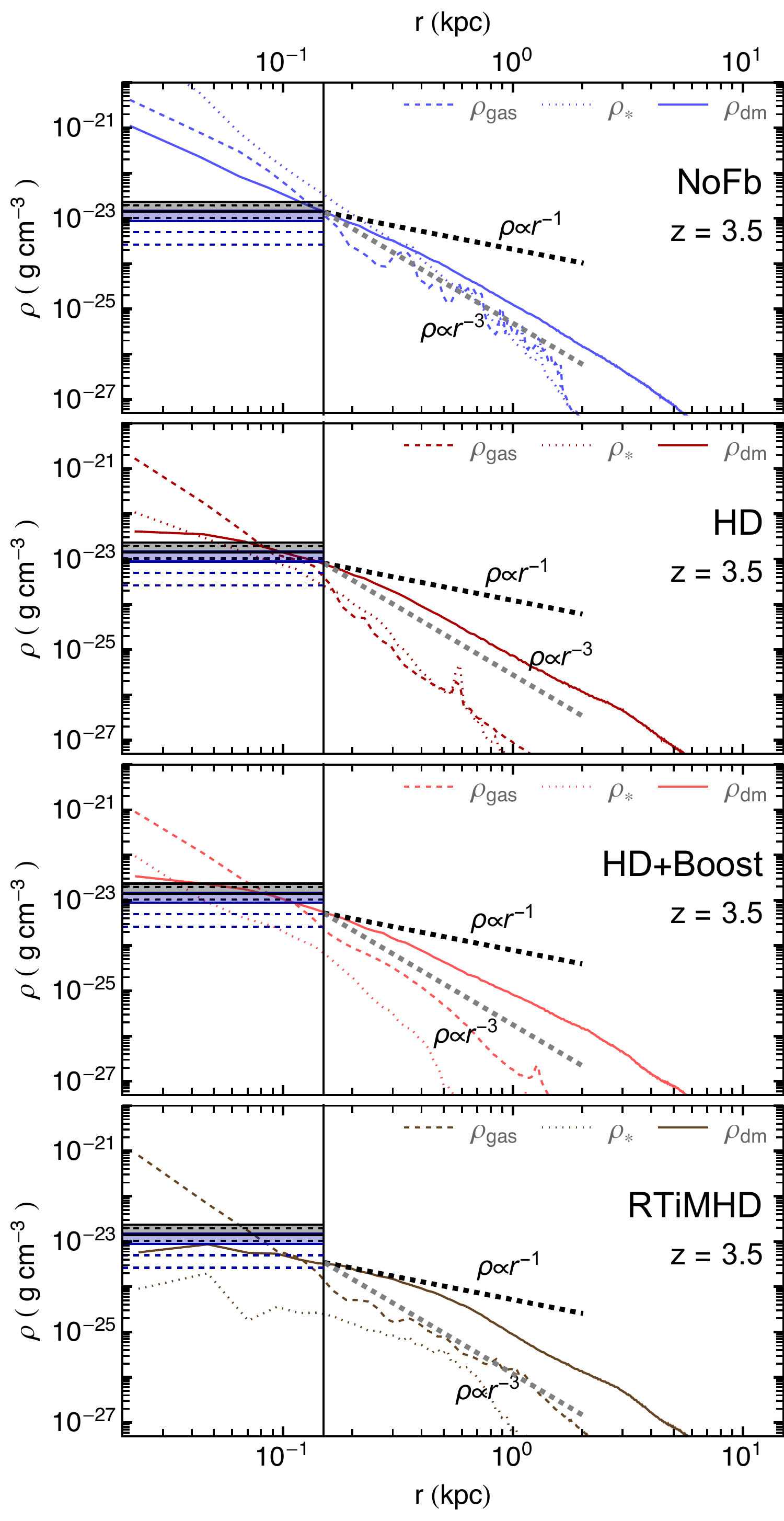}%
    \includegraphics[width=\thiswidth\columnwidth]{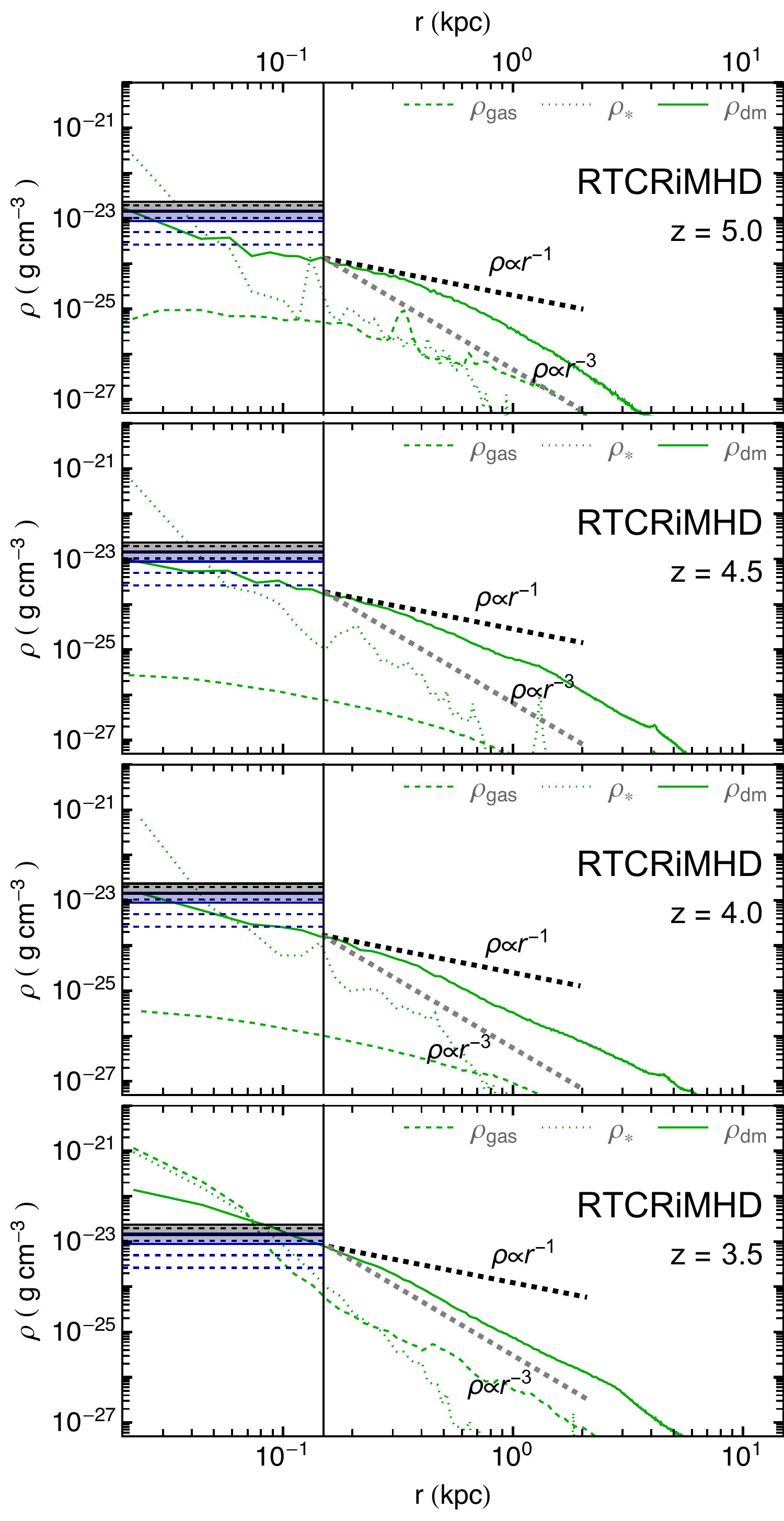}%
    \includegraphics[width=\thiswidth\columnwidth]{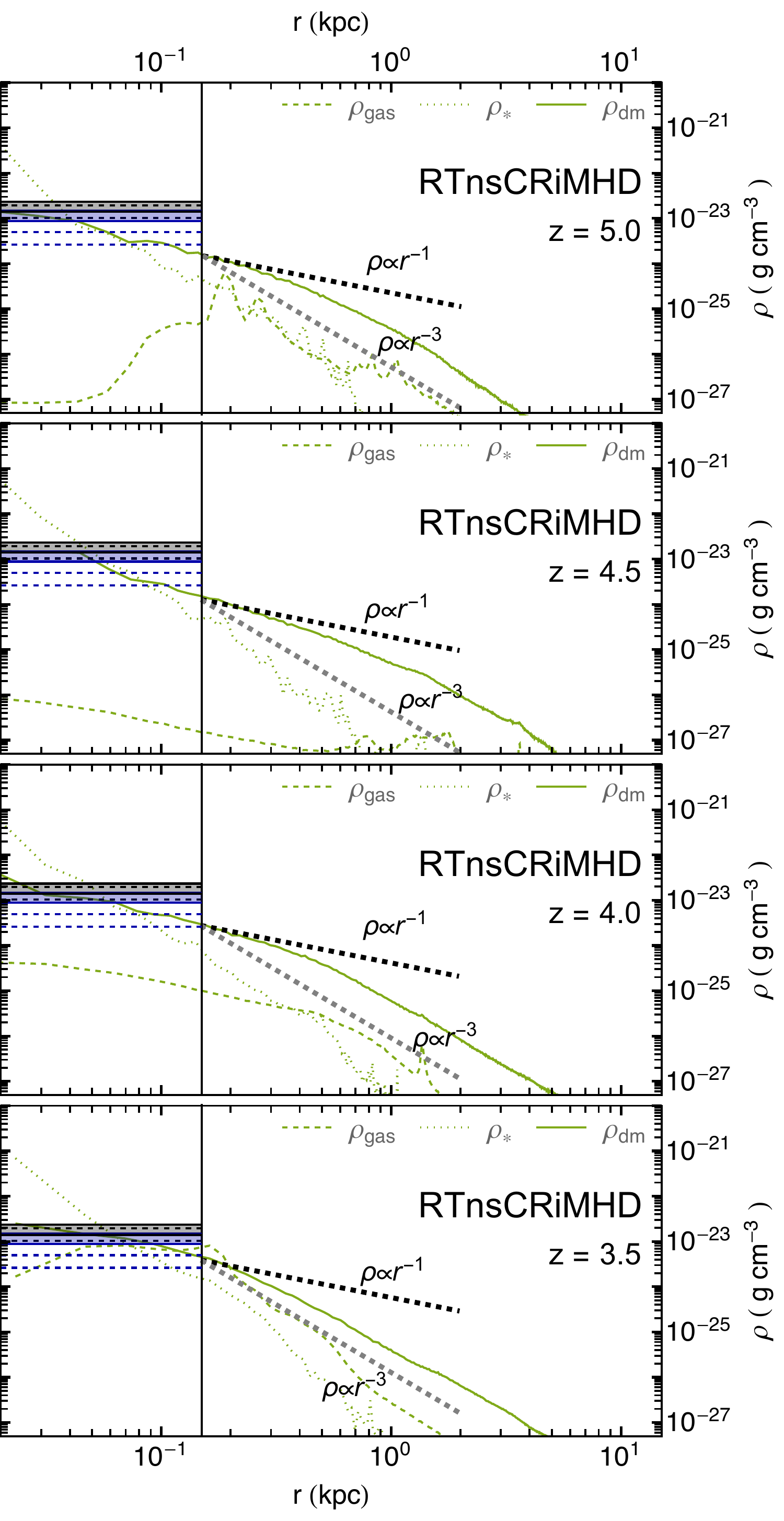}\\
    \caption{Radial density profiles of the gaseous (dashed), stellar (dotted) and dark matter (solid) components for various representative models. The vertical solid line indicates $r = 0.15 \kpc$, used in Fig.~\ref{fig:HaloFits} and \citet{Read2019a}. Colour bands show the prediction for the central density in haloes with cusps (gray) or cores (blue) for $M_\text{halo} = 4.5 \cdot 10^{9} \Msun$ at $z = 3$ (shaded bands) and $z = 0$ (dashed lines) computed in an analogous manner to that of \citep{Read2019a}. Finally, we include two additional lines spanning from $r = 0.15 \kpc$ up to $r = 2 \kpc$ showing the NFW density scalings, $\rho_\text{DM} \propto r^{-1}$ (gray) and $\rho_\text{DM} \propto r^{-3}$ (black). {\bf (Left column)} From top to bottom, the panels show \HDSfNoFb, \HDSfFb, \HDSfFbBoost~and \RTiMHDSfFb at $z = 3.5$, illustrating how SN feedback and radiation decrease the central dark matter density and flatten the radial slope of the dark matter profile. {\bf (Central column)} Radial density profile evolution for the full-physics \RTCRiMHDSfFb~simulation from $z = 5$ (top) to $z = 3.5$ (bottom). {\bf (Right column)} Same as the central column but for the \RTnsCRiMHDSfFb~simulation. These two columns illustrate that cosmic ray feedback together with radiation lead to a central dark matter profile more in line with `cores', which is transformed back to a cusp at $z \sim 3.5$ in \RTCRiMHDSfFb, due to a `wet merger'.}
    \label{fig:DensityProfiles}
\end{figure*}

\section{Conclusions}
\label{s:Conclusions}
In this paper, we introduce the Pandora suite of high resolution cosmological zoom-in simulations of a dwarf galaxy (with halo mass $M_\text{vir} (z = 0) \approx 10^{10} \Msun$) combining magneto-hydrodynamics, radiative transfer and cosmic rays. This set of simulations is generated using our own modified version of the {\sc ramses} code \citep{Teyssier2002}, employing multiple extensions by \citet[][MHD]{Fromang2006}, \citet[][RT]{Rosdahl2015a} and \citet[][CRs]{Dubois2019}. Our suite of simulations builds up from a fiducial model (\HDSfFb) comprised of hydrodynamics, a magneto-thermo-turbulent star formation model and a mechanical SN feedback scheme. For a subset of our simulations we also explore an alternative star formation model based on a gas density threshold and vary the SN feedback strength. We investigate multiple configurations for these models, gradually increasing the complexity of physical mechanisms, ultimately leading to two {\it `full-physics'} simulations, \RTnsCRiMHDSfFb~and \RTCRiMHDSfFb, which simultaneously account for magnetic fields, stellar radiation and SN-generated cosmic rays.

The simulated dwarf galaxy is a gas-rich system at very high redshift, but does not evolve significantly after $z \sim 4$, as it resides in a field environment. Consequently, we evolve all our simulations, summarised in Table \ref{table:setups}, down to $z \sim 3.5$. We also evolve a small subset of simulations all the way to $z = 0.5$ confirming the largely passive evolution since $z  \sim 3.5$, when the dwarf is quenched. We compare in detail the properties of our simulated dwarf with a wealth of observational data, with the aim to place constraints on the most likely physical mechanisms regulating the evolution of field dwarfs. Our main findings are as follows:
\begin{enumerate}

\item Unsurprisingly, due to their shallow potential well, the properties of dwarfs are very sensitive to the physical processes included in the simulation. SN feedback remains one of the key processes regulating their final stellar masses. With our fiducial SN feedback, simulated dwarf galaxies are approximately located on the extrapolated stellar mass - halo mass relation of \citet{Behroozi2013}, and agree with observations of isolated dwarf galaxies of similar halo masses \citep{Read2017}. While the combination of radiation, cosmic rays and magnetism has only a moderate effect on the final stellar mass of our galaxy, both radiation and cosmic rays can significantly delay the growth of stellar mass in dwarf galaxies.

\item  In terms of the overall morphology and spatial gas distribution, models with strong SN feedback produce an over-quenched, amorphous and compact system. The inclusion of stellar radiation leads to a much more extended and gas-rich dwarf galaxy, with some central dust lanes observed in our mock optical images. Models that incorporate cosmic rays in addition to radiation, lie somewhere in between of these two extremes, due to cosmic ray-driven outflows that deplete the gas reservoir in the ISM and can more effectively quench star formation.

\item Consequently, strong SN feedback simulations are a poor match to observed dwarfs with a similar estimated halo mass (LeoA, WLM, SagDIG) in the dynamical mass - stellar mass relation, while our `full physics' simulations provide much more reasonable predictions for the mass-size and dynamical mass - mass relations. All of this implies that `dialling' up SN feedback strength in simulations to match (some) observational constraints cannot realistically account for the effects of radiation and cosmic rays.
    
\item While a number of models with stellar radiation lead to integrated rotational velocities and velocity dispersion for stars and HI in good agreement with the kinematics of local isolated dwarf galaxies, spatially resolved kinematics reveals that cosmic ray-driven outflows can induce more realistic and diverse kinematics, with distinct `clumps' and misaligned motions in stars, ionized and neutral hydrogen as observed with IFU surveys in some dwarfs. 

\item Our fiducial SN models display cusp-like dark matter profiles, where increasing the SN feedback strength only leads to a minor reduction of profile cuspiness. This is the case even for the overquenched models, as the majority of the SN feedback takes place in a single burst. However, episodic removal of gas in our `full physics' models leads to more core-like dark matter profiles. We note that this cusp-core transformation and its longevity is further compounded by the dwarf galaxy merger history, with a single `wet merger' able to re-establish a cusp.   
\end{enumerate}

While our simulations explore additional physical processes frequently omitted in many numerical studies (due to their complexity and/or their computational cost), we note that our models are still far from complete. Some of the physical processes that we are missing and which may affect dwarf galaxy formation are stellar winds \citep{Agertz2021}, higher accuracy modelling of ISM turbulence either by alternative refinement strategies \citep{Martin-Alvarez2022} or subgrid turbulence models \citep{Semenov2016, Kretschmer2020}, and more realistic cooling prescriptions \citep{Katz2022}. Furthermore, the possibility of AGN activity in dwarf galaxies has gained traction in recent years \citep[e.g.][]{Pardo2016}, with numerical studies suggesting that AGN feedback has the potential to affect the formation and evolution of dwarf galaxies \citep{Dashyan2018, Koudmani2019, Koudmani2022}.

Our numerical simulations have unveiled the complexity of physical processes needed to more realistically model dwarf galaxies. `SN feedback-only' models struggle to match realistic masses, sizes and kinematics of observed dwarf galaxies, either leading to over-quenched objects (for their halo mass) or too centrally concentrated baryons. 

Inclusion of local stellar radiation sources and SN-driven cosmic rays leads to more extended, rotationally-supported systems, where star formation and feedback is more spatially distributed and better able to regulate dwarf properties. Detailed resolved kinematics of dwarf galaxies from IFU surveys together with the upcoming JWST constraints on the multi-phase nature of outflows in dwarf galaxies working in conjunction with detailed simulation models such as attempted here, will help us to unravel the surprisingly complex nature of our Universe's smallest building blocks.

\section*{Acknowledgements}
The authors kindly thank the referee for their careful consideration of this manuscript, and their insightful comments and suggestions, which have highly improved the quality of this manuscript. The authors would like to thank M. Smith for sharing his initial conditions. This work was supported by the ERC Starting Grant 638707 ``Black holes and their host galaxies: co-evolution across cosmic time''. SMA, DS and MGH acknowledge support from the UKRI Science and Technology Facilities Council (grant number ST/N000927/1). This work used the DiRAC@Durham facility managed by the Institute for Computational Cosmology on behalf of the STFC DiRAC HPC Facility (www.dirac.ac.uk). The equipment was funded by BEIS capital funding via STFC capital grants ST/P002293/1, ST/R002371/1 and ST/S002502/1, Durham University and STFC operations grant ST/R000832/1. DiRAC is part of the National e-Infrastructure. This work was performed using resources provided by the Cambridge Service for Data Driven Discovery (CSD3) operated by the University of Cambridge Research Computing Service (www.csd3.cam.ac.uk), provided by Dell EMC and Intel using Tier-2 funding from the Engineering and Physical Sciences Research Council (capital grant EP/P020259/1), and DiRAC funding from the Science and Technology Facilities Council (www.dirac.ac.uk).

\section*{Data availability}
The data employed in this manuscript is to be shared upon reasonable request contacting the corresponding author.
\bibliographystyle{mnras}
\bibliography{references}
\appendix
\section{Halo mass growth and the re-ignition of star formation}
\label{ap:HaloGrowth}
\begin{figure}
    \centering
    \includegraphics[width=\columnwidth]{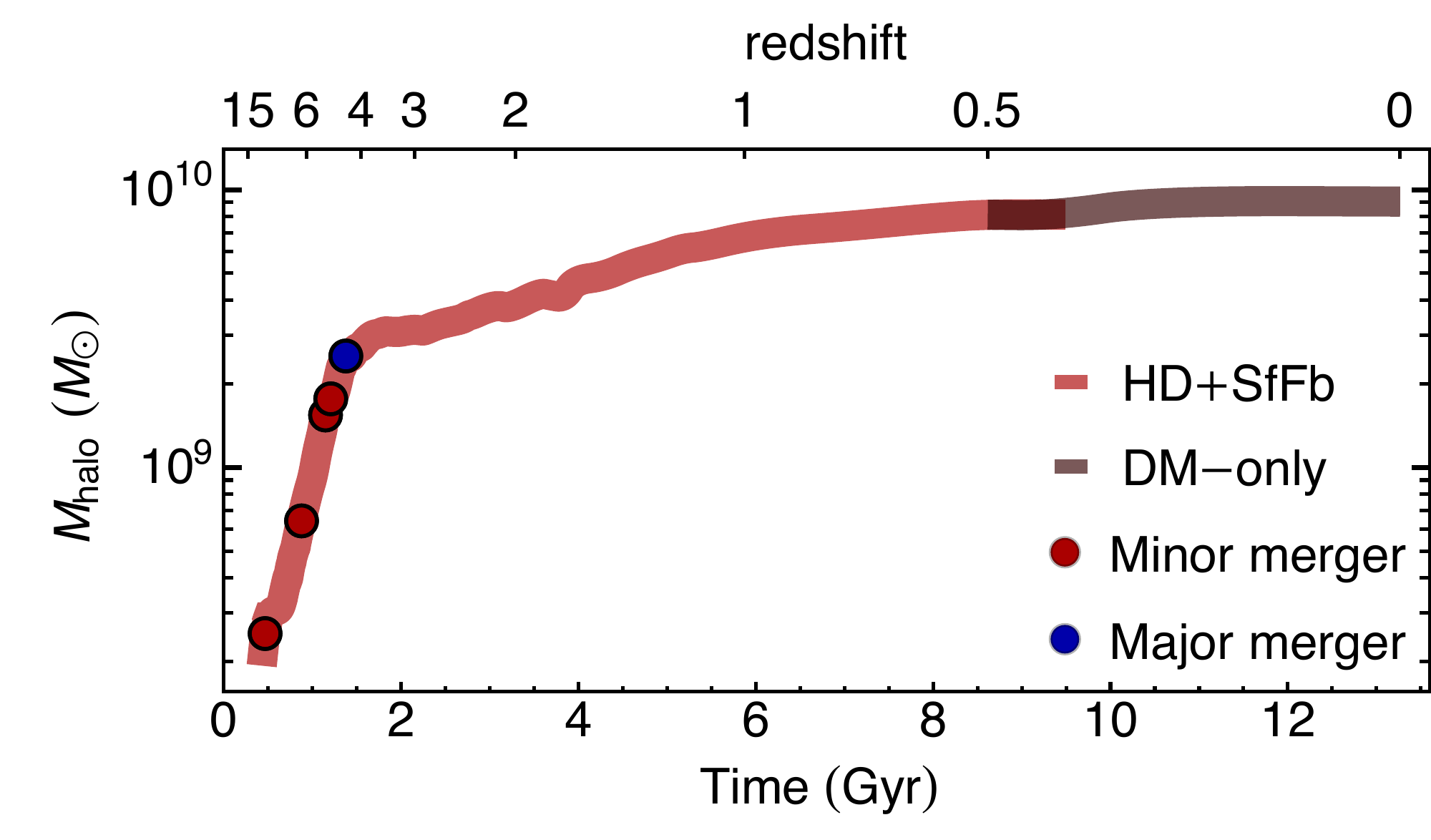}\\
    \caption{Pandora halo mass growth down to $z = 0$ for the \HDSfFb~simulation parsed with the halo mass growth from a dark matter-only simulation beyond $z = 0.5$. We overplot all major (blue circles) and minor (red circle) mergers experienced by the halo. The studied halo forms rapidly at high redfshift, with no mergers beyond $z \sim 4$. After that time, the halo only grows in a gradual and secular manner.}
    \label{fig:mhalo}
\end{figure}

There is an integral connection between the evolution of galaxies and their hosting dark matter haloes. Exploring the evolution of the halo for the studied galaxy thus provides further insight on the fate of the galaxy. Events such as mergers, sustained gas accretion or encounters with gaseous structures have the potential to re-ignite star formation \citep{Wright2019, Rey2020, Gutcke2022}. In order to further illustrate the lack of significant evolution for the Pandora galaxy after the simulated period (i.e. $z = 3.5$), we show in Fig.~\ref{fig:mhalo} the halo mass growth. This is shown for the \HDSfFb~simulation until $z = 0.5$, and down to $z = 0$ for a dark matter-only simulation. We include mergers with systems down to 0.1 of the mass of the galaxy, as red (minor mergers) and blue (major mergers) circle markers, respectively. The studied halo and galaxy only undergo significant mergers prior to $z = 4$, and evolve secularly afterwards. We also verify the absence of any significant gas accretion or encounters with any gaseous structures in \HDSfFb~down to $z = 0.5$ (see Fig.~\ref{fig:gasevo}). While there is some progressive halo growth during the $z = 3.5$ to $z = 0$ interval, the absence of any further galaxy mergers leads to no star formation re-ignition through a merger-driven channel. Combined with the lack of significant gas accretion, we predict a very low probability for star formation re-ignition in the presence of SN feedback. This is in agreement with the expected evolution of a very isolated dwarf galaxy.

\def \gasevowidth{0.45}
\def \gasevowidth{0.33}
\begin{figure}
    \centering
    \includegraphics[width=\gasevowidth\columnwidth]{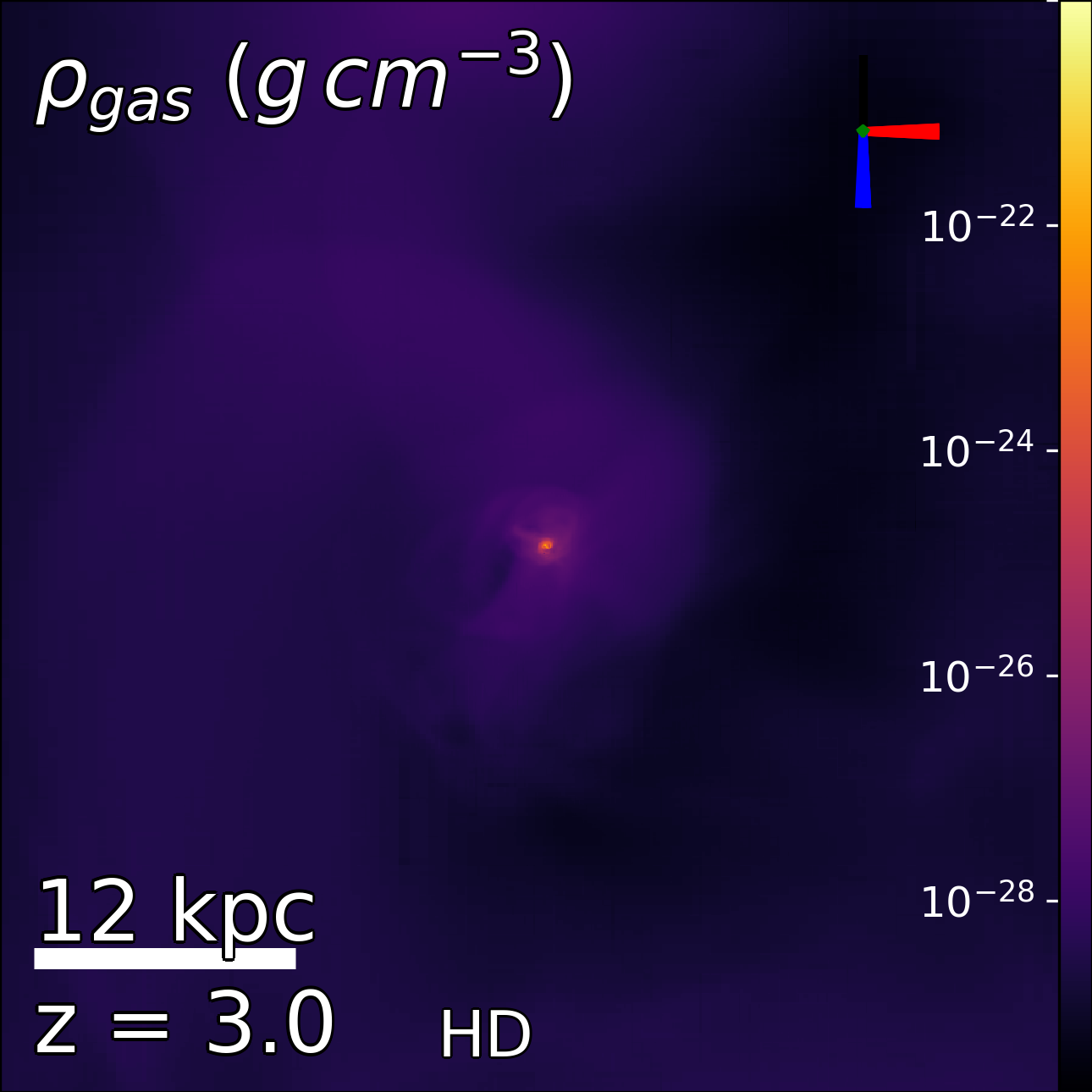}%
    \includegraphics[width=\gasevowidth\columnwidth]{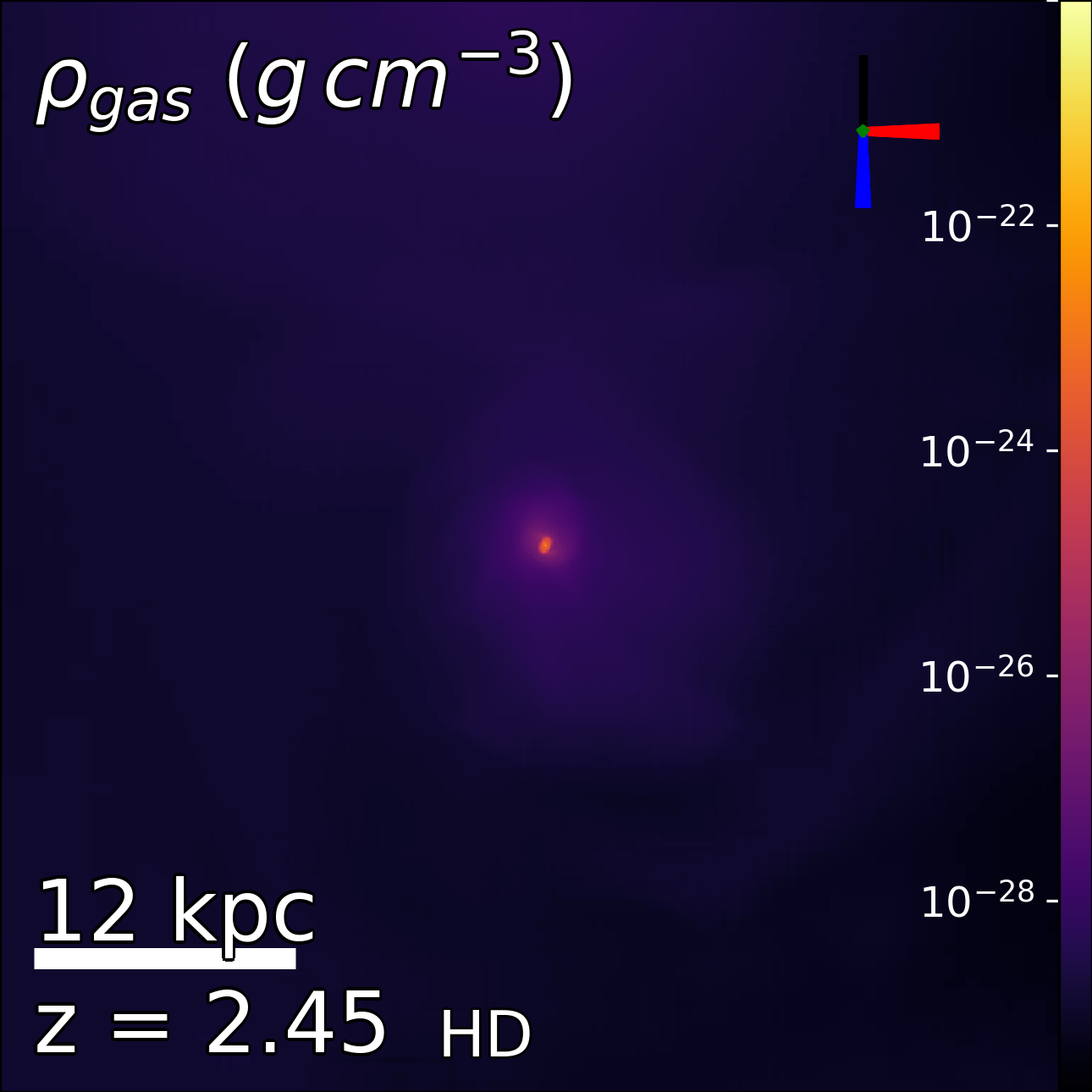}%
    \includegraphics[width=\gasevowidth\columnwidth]{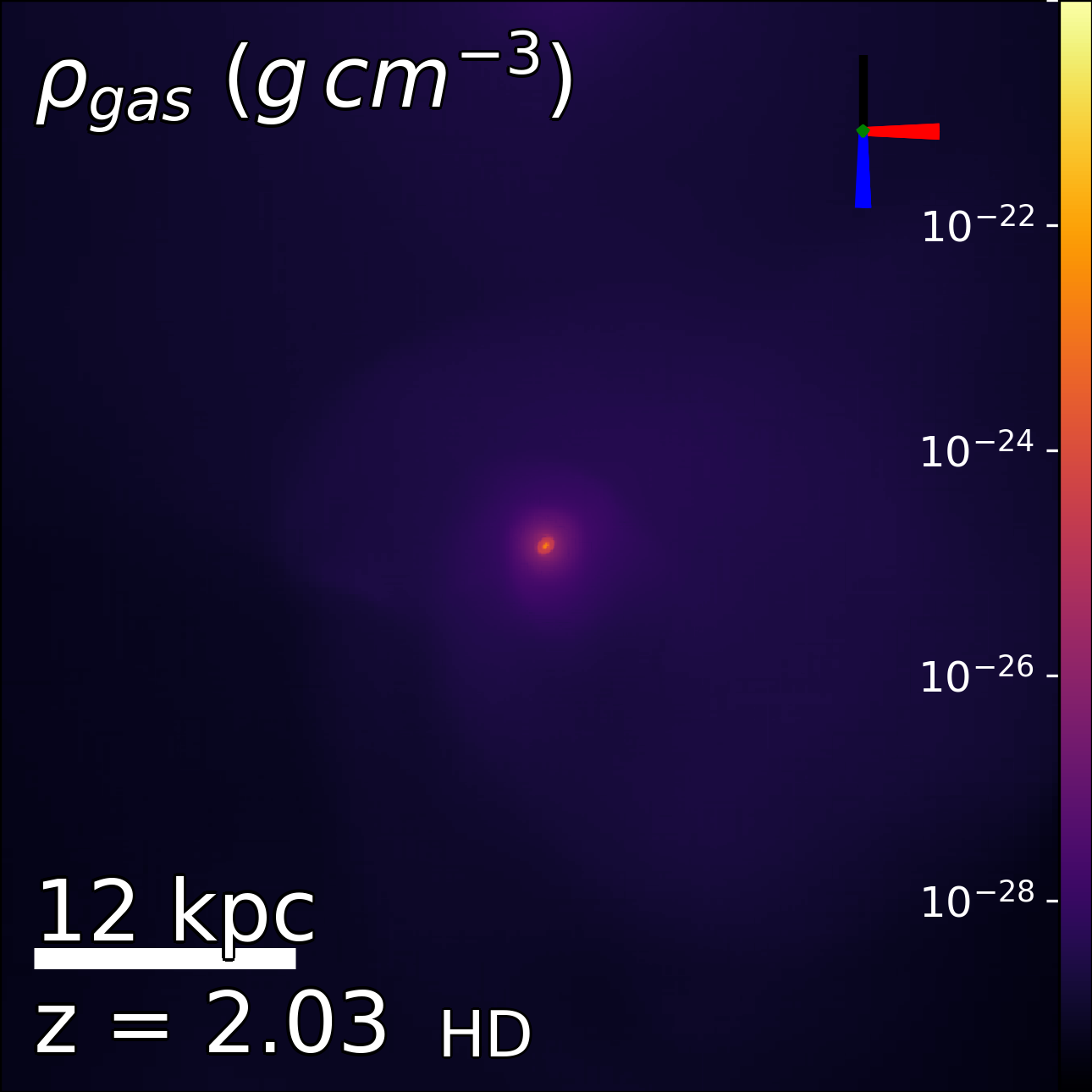}\\
    \includegraphics[width=\gasevowidth\columnwidth]{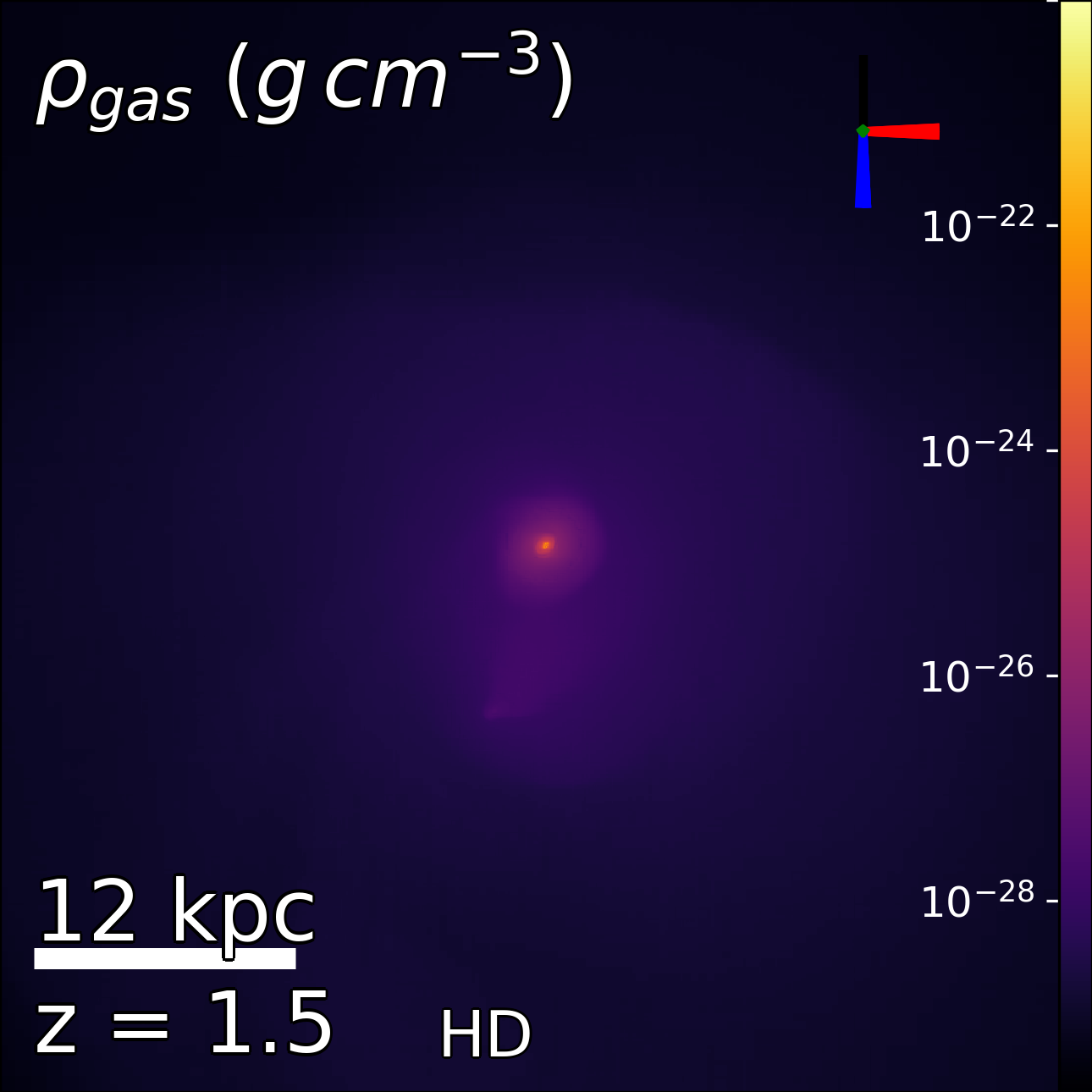}%
    \includegraphics[width=\gasevowidth\columnwidth]{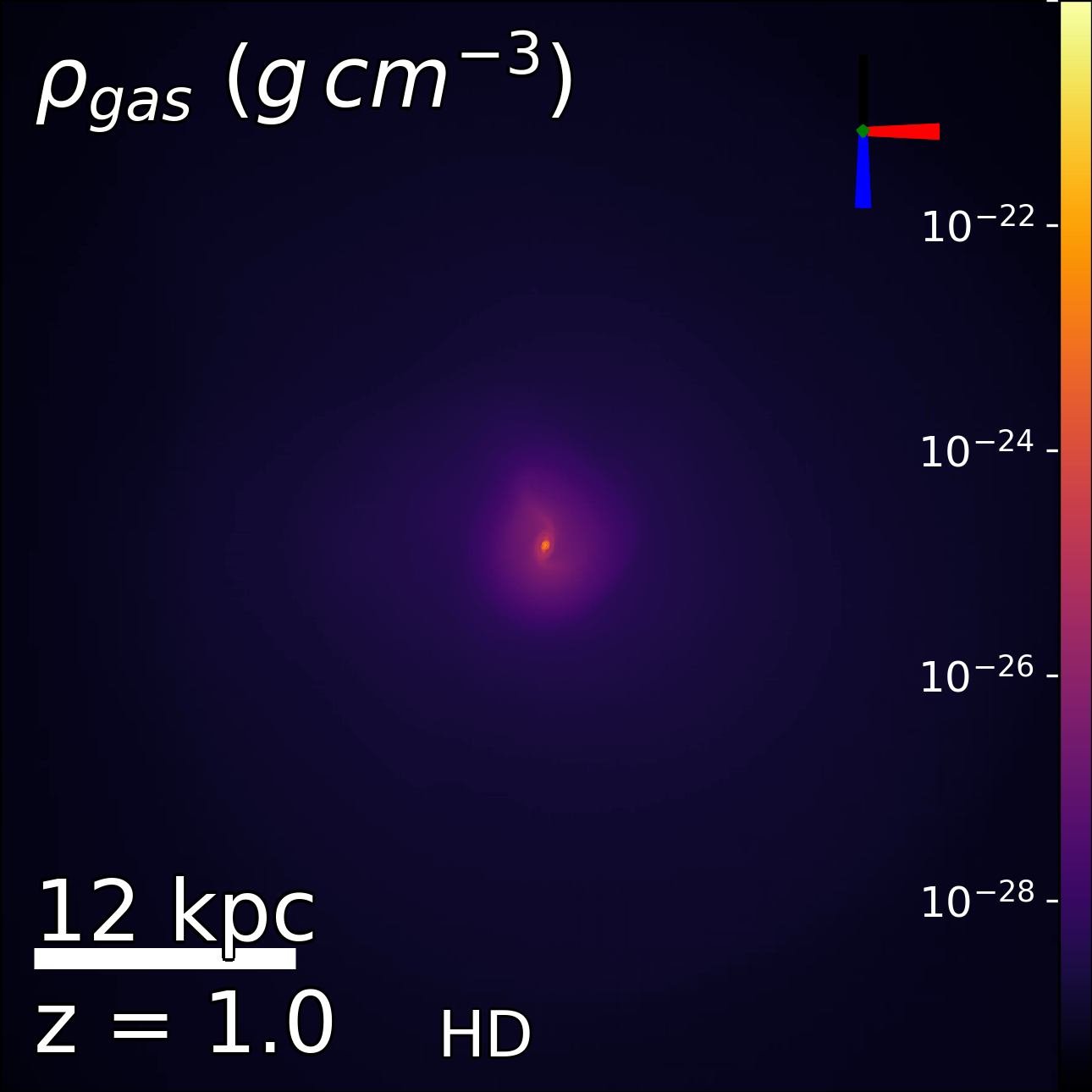}%
    \includegraphics[width=\gasevowidth\columnwidth]{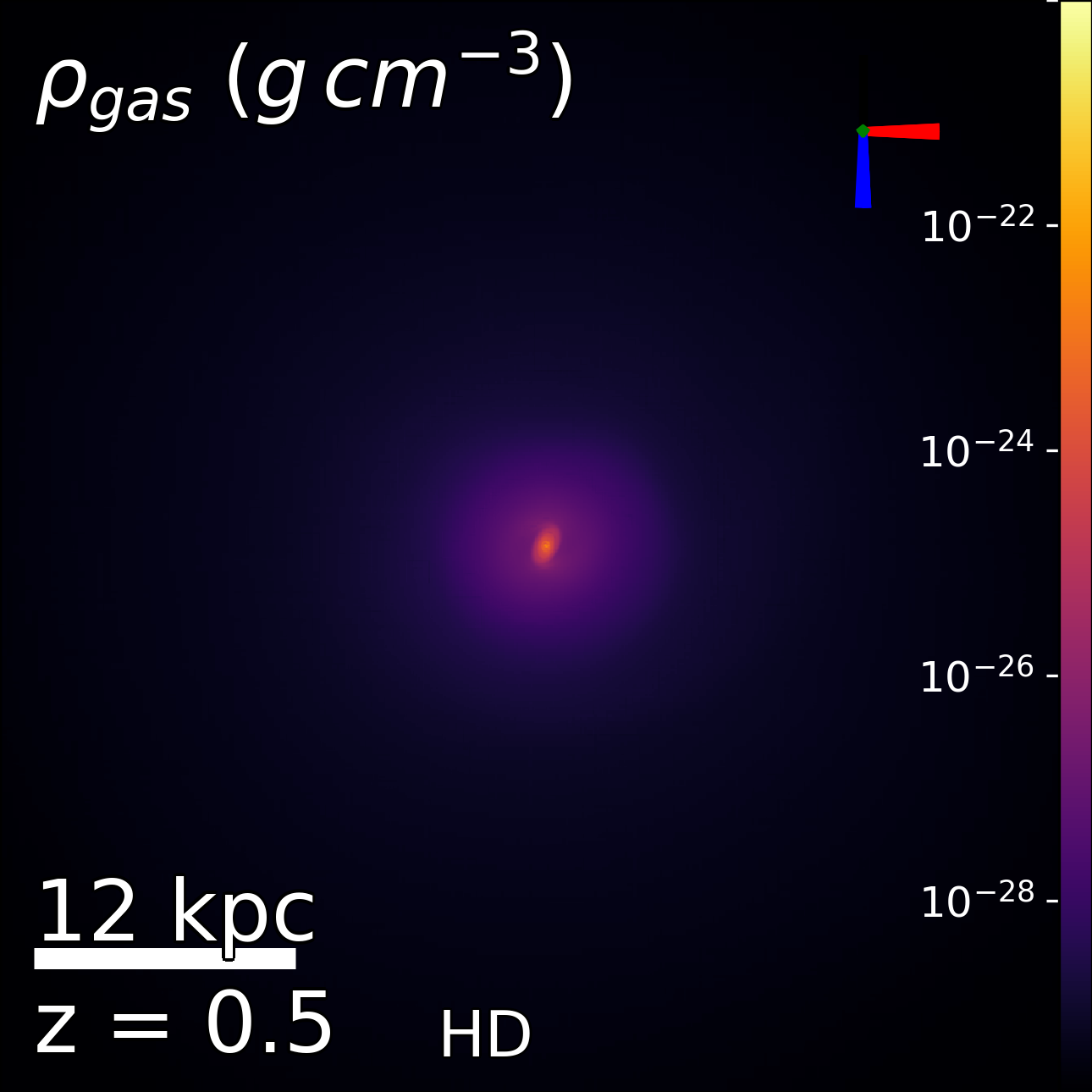}\\
    \caption{Gas density evolution of the studied halo in the \HDSfFb~simulation. No significant gas accretion nor encounters with gaseous structures occur between $z = 3.5$ and $z = 0.5$ due to the considerable spatial isolation of the simulated galaxy.}
    \label{fig:gasevo}
\end{figure}

\section{Central densities of the galaxy vs the halo}
\label{ap:CentralDensHalo}
\begin{figure}
    \centering
    \includegraphics[width=0.9\columnwidth]{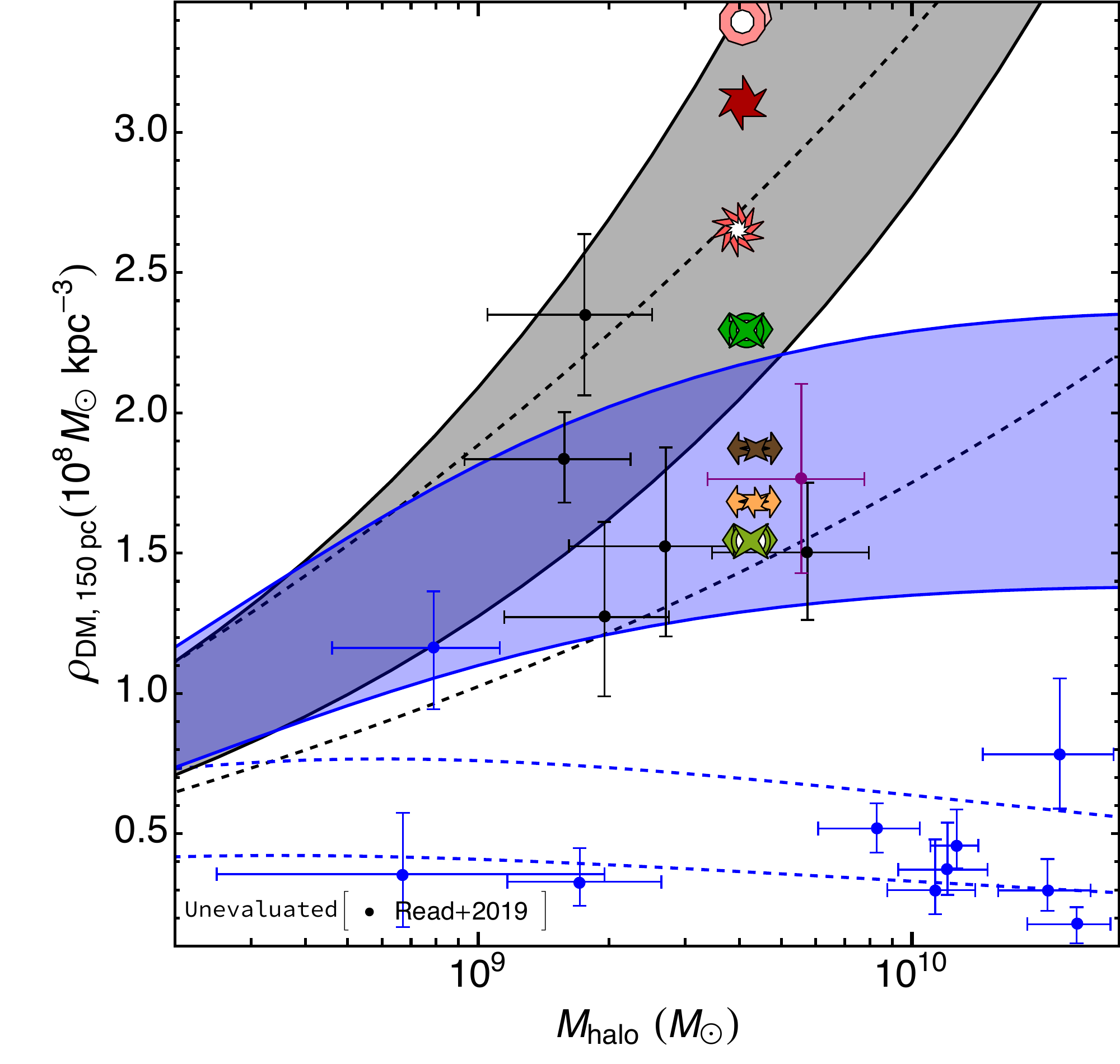}\\
    \includegraphics[width=0.8\columnwidth]{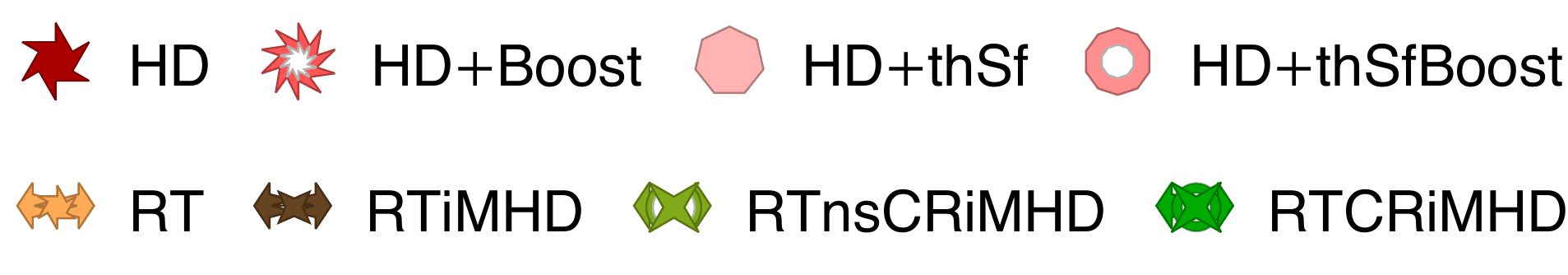}\\
    \caption{Same as Fig.~\ref{fig:HaloFits}, but now centred on the dark matter-only centre of the halo. Average dark matter density within the inner $150$~pc of the simulated dwarf galaxy versus $M_\text{halo}$ at $z = 3.5$. Selecting the dark matter-only centres leads to an increase of the central density measurement with respect to the galaxy centres.}
    \label{fig:xhaloDens}
\end{figure}

In Section~\ref{ss:CuspCore} we measured the central density of the Pandora galaxy to explore how different physics affect the formation of dark matter cores. In order to better reproduce observations, we performed this measurements centred on the galaxy. While galaxies such as Pandora are generally located at the centre of their hosting dark matter halo, we measure small separations of $\sim 20 - 30$ pc, with somewhat larger deviations during disruptive events such as mergers. Studies that focus on dark matter core formation also opt to centre their measurements on the dark matter component \citep[e.g.,][]{Orkney2021}. To explore the impact of our centring choice, we show the central densities of Pandora in Fig.~\ref{fig:xhaloDens}, now using the halo centre obtained with our shrinking spheres algorithm applied exclusively to the dark matter. This selection results in an approximate increase of the average central density by $\sim 0.3 \cdot 10^8 \Msun \kpc^{-3}$, with a somewhat larger effect on \RTnsCRiMHDSfFb~and \HDSfFb. Importantly, the described trends for the different physical models are preserved when employing this alternative dark matter-only centering.

\bsp	
\label{lastpage}
\end{document}